\documentclass[prx,twocolumn,superscriptaddress,showpacs,amsmath,amssymb]{revtex4-1}

\usepackage{amssymb}
\usepackage[dvips]{epsfig,graphics,color}
\usepackage{graphicx}

\usepackage{color}

\begin{document}

\title{Partial long-range order in antiferromagnetic Potts models}

\author{M.~P.~Qin}
\affiliation{Institute of Physics, Chinese Academy of Sciences, P.O. Box
603, Beijing 100190, China}

\author{Q.~N.~Chen}
\thanks{Current address: Department of Physics, Beijing Normal
University, Beijing 100875, China.}
\affiliation{Institute of Theoretical Physics, Chinese Academy of Sciences,
P.O. Box 2735, Beijing 100190, China}

\author{Z.~Y.~Xie}
\affiliation{Institute of Physics, Chinese Academy of Sciences, P.O. Box
603, Beijing 100190, China}

\author{J.~Chen}
\affiliation{Institute of Physics, Chinese Academy of Sciences, P.O. Box
603, Beijing 100190, China}

\author{J.~F.~Yu}
\affiliation{Institute of Physics, Chinese Academy of Sciences, P.O. Box
603, Beijing 100190, China}

\author{H.~H.~Zhao}
\affiliation{Institute of Physics, Chinese Academy of Sciences, P.O. Box
603, Beijing 100190, China}

\author{B.~Normand}
\affiliation{Department of Physics, Renmin University of China, Beijing
100872, China}

\author{T.~Xiang}
\affiliation{Institute of Physics, Chinese Academy of Sciences, P.O. Box
603, Beijing 100190, China}
\affiliation{Collaborative Innovation Center of Quantum Matter, Beijing,
China}

\begin{abstract}

The Potts model plays an essential role in classical statistical mechanics,
illustrating many fundamental phenomena. One example is the existence of
partially long-range-ordered states, in which some degrees of freedom remain
disordered. This situation may arise from frustration of the interactions,
but also from an irregular but unfrustrated lattice structure. We study
partial long-range order in a range of antiferromagnetic $q$-state Potts
models on different two-dimensional lattices and for all relevant values of
$q$. We exploit the power of tensor-based numerical methods to evaluate the
partition function of these models and hence to extract the key thermodynamic
properties -- entropy, specific heat, magnetization, and susceptibility
 -- giving deep insight into the phase transitions and ordered states
of each system. Our calculations reveal a range of phenomena related to
partial ordering, including different types of entropy-driven phase
transition, the role of lattice irregularity, very large values of the
critical $q_c$, and double phase transitions.

\end{abstract}

\pacs{64.60.Cn, 05.50.+q, 75.10.Hk, 64.60.F-}

\maketitle

\section{Introduction}

The Potts model \cite{Wu1982_RMP54-235} is a cornerstone of
classical statistical physics. First appearing in Potts' Ph.D. thesis
\cite{Potts_PHD_thesis} as a generalization of the Ising model, it is
a simple but highly nontrivial model. Indeed, the family of $q$-state
antiferromagnetic (AF) Potts models displays a rich and complex range
of behavior, providing many examples of different phase transitions,
critical phenomena, ordered states, and universality classes. Although
the $q = 2$ Potts model is equivalent to the Ising model, and thus has
exact solutions for all planar lattices with nearest-neighbor interactions
\cite{domb_green}, including the square \cite{square_exact}, triangular
\cite{trian_honey_exact}, and honeycomb \cite{trian_honey_exact} geometries,
exact results for $q > 2$ are rare. Many other problems in statistical
mechanics are closely related to the Potts model, including vertex models
\cite{Baxter_vertex}, bond and vertex coloring problems \cite{Baxter_coloring},
and loop models \cite{Jacobsen_loop}.

The behavior of the AF Potts model is dictated by the interplay between
$q$, the number of states per site, and the lattice geometry. When $q$ is
small compared to the average coordination number ${\bar z}$ of the lattice,
at low temperatures the limited number of degrees of freedom will in general
be fixed, and ordered, by geometrical and interaction requirements. However,
when $q$ is similar to or greater than ${\bar z}$, the entropy is such that
the system may not order at any temperature \cite{Salas1997_JSP86-551}. In
addition to the conventional zero-temperature limits of complete order or
disorder, AF Potts models show two further possibilities. One is that the
ground state is genuinely critical, the result of an arrested
``zero-temperature phase transition'' to an ordered state; this type of
physics is known in the $q = 3$ AF Potts model on the square
\cite{Salas1998_JSP92-729} and kagome \cite{kagome_qc} lattices and
in the $q = 4$ AF Potts model on the triangular lattice \cite{triangular_qc}.
The other is that some, but not all, of the degrees of freedom of the system
may form a state of partial long-range order
\cite{x,Kotecky2008_PRL101-030601}; partial or complete ordering processes
occur at different types of ``finite-temperature phase transition.''

At a conventional phase transition, the order parameter becomes finite
everywhere in the system to minimize the free energy, and the result is
a state of complete order at zero temperature. In the example of the
ferromagnetic Ising model, all spins are oriented either upwards or
downwards in the ground state. However, in systems with sufficiently
many degrees of freedom (sufficiently large entropy, as in a Potts model
with sufficiently high $q$), the crossover to a ground state optimizing
the resulting entropic contribution may occur in a step-wise fashion. The
minimization of energy may be achieved in many different ways, and may
involve only some of the lattice sites. The remaining degeneracy, and the
type of order, is then determined by the maximization of entropy. The result
is an ``entropy-driven'' transition, usually occurring at a finite temperature,
and to a state of partial order. On cooling to zero temperature, if the system
retains a nonzero ``residual'' entropy then a ground state with order on only
a subset of the lattice sites can be achieved. The best-known example of the
physics of extensive ground-state degeneracy is found in ice \cite{Pauling}.

The majority of prior work on partial order has concerned frustrated
systems, where the energy cannot be minimized locally, meaning for all
bonds simultaneously \cite{Lipowski}. The AF Ising model on the triangular
lattice \cite{trian_honey_exact} is an archetypal frustrated system,
because no spin configuration can minimize all three bonds on a triangle
simultaneously. Frustrated systems share the same property of highly
degenerate ground states, arising from their frustrated interactions,
and the formation of partially ordered states offers one avenue for
partial frustration relief and partial entropy reduction.

When partial order arises in unfrustrated systems \cite{x}, its origin
lies only in configurational entropy effects. In 2008, Kotecky and
coauthors \cite{Kotecky2008_PRL101-030601} found partial long-range order
in the $q = 3$ AF Potts model on the diced lattice, performing both
analytical and numerical studies of the accompanying finite-temperature
phase transition. In 2011, we \cite{Chen2011_Phys.Rev.Lett.107-165701} traced
their result to the extensive zero-temperature entropy (residual entropy per
site) of this lattice, which arises because it is ``irregular'' in the sense
of having differently coordinated sites (we defer a discussion of lattice
types to Sec.~II). From this insight we demonstrated the existence of
finite-temperature phase transitions and partial order in the $q = 4$
Potts models on the Union-Jack and centered diced lattices. Partial order
is the result of a partial symmetry-breaking, and we found that, depending
on the Potts model in question, the singularity associated with this breaking
of symmetry may be either almost as strong as a full symmetry-breaking, or
may be remarkably weak and difficult to detect.

Partial order in the ground state is known exactly in a number of models.
In the spin-$S$ AF Ising model on the triangular lattice, for sufficiently
large $S$ the ground state is partially ordered on two of the hexagonal
sublattices but disordered on the third \cite{Nagai1993_PRB47-202}.
Also in two dimensions, the ground states of the AF Ising model on
the Union-Jack lattice \cite{Vaks1966_JETPLetters22-820}, kagome
lattice \cite{Azaria1987_PRL59-1629--1632}, dilute centered square
lattice \cite{Debauche1992_PRB46-8214--8218}, anisotropic triangular
lattice \cite{Diep1991_Phys.Rev.B43-8759--8762}, and Villain lattice
(anisotropic square lattice) \cite{Villan} are all partially ordered.
Partial order also exists for some frustrated systems, such as the
$q = 3$ Potts model on the Villain
lattice \cite{Foster2001_JPA34-5183,Foster2004_PRB70-014411}.
Three-dimensional classical models with partial order are mostly
frustrated, including the Ising model on the accumulated
triangular \cite{Blankschtein1984_PRB29-5250--5252} and body-centered
cubic (BCC) \cite{Azaria1989_EPL9-755} lattices, the classical Heisenberg
model on the BCC lattice \cite{Santamaria1997_JAP81-5276-5278}, models
on the simple cubic lattice \cite{Blankschtein1984_PRB30-1362--1365,
Diep1985_JPC18-1067}, the $q = 4$ AF Potts model on the diamond lattice
\cite{Igarashi2010_JPCS200-022019} and the XY model on the checkerboard
lattice \cite{Boubcheur1998_PRB58-400--408}. Experimentally, partial order
has been observed in the frustrated AF material Gd$_2$Ti$_2$O$_7$
\cite{Stewart2004_JPCM16-L321}. Partial order is also predicted
for the periodic Anderson model on the triangular lattice
\cite{Hayami2011_JPSJ80-073704,Hayami2011_ArXive-prints-1107.4401}
and the Heisenberg model on the BCC lattice
\cite{Quartu1997_PRB55-2975--2980,Santamaria1998_JAP84-1953-1957}.

Although the Potts model is one of the simplest in statistical physics,
its analytical study has been restricted by the limited number of
exactly known results beyond $q = 2$. Methods including height mapping
\cite{Kotecky2008_PRL101-030601} have some general utility, while mappings
to related coloring problems \cite{Baxter_coloring} are useful in specific
cases. Previous numerical studies of Potts models have made use of Monte
Carlo \cite{WSK_1,WSK_2} and transfer-matrix \cite{Trans_Mat_1,Trans_Mat_2}
techniques. Monte Carlo simulations are accurate, and can study large but
finite lattice sizes, while transfer-matrix methods are infinite in one
spatial dimension but finite in the other(s), and can be used to study
fractional values of $q$.

In this paper we introduce (Sec.~III) a set of tensor-based numerical methods,
which are quite generally applicable in classical statistical mechanics.
The partition function is written as the trace over a network of tensors
representing the states of the system on an infinite lattice, and in its
evaluation the truncation is performed systematically in the tensor dimension.
Because it evaluates the partition function, this calculational approach
gives access in principle to all thermodynamic quantities, and is not very
resource-intensive in comparison with other numerical techniques.

We apply the tensor-network approach to perform a detailed analysis of
partial order in AF Potts models \cite{qiaoni_thesis}. We consider a number
of irregular lattices in two dimensions, and calculate thermodynamic
quantities including the entropy, specific heat, magnetization, and magnetic
susceptibility. We use these qualitatively to investigate the partial
order or partial breaking of symmetry, which is shown by all the models,
and quantitatively to characterize the phase transitions and partially
ordered states. We find lower bounds for the critical values of $q$ on
each lattice and illustrate the phenomenon of double phase transitions in
particular models. Our results show the power of tensor-based numerical
methods for gaining fresh insight into long-standing problems in
classical statistical mechanics.

The structure of this paper is as follows. In Sec.~II we review the Potts
model, the classes of lattice we consider in two dimensions, and some
known results concerning $q$, regular geometries, and phase transitions.
Section III describes in detail the tensor-based numerical techniques we
employ to compute the partition function and thermodynamics of each model.
In Sec.~IV we focus on the entropy-driven phase transition, using the
entropy and specific heat to compare and contrast its form on a number
of Laves lattices. We calculate in Sec.~V the magnetization of the models
studied in Sec.~IV, in order to characterize the partial order through
its order parameter and susceptibility. In Sec.~VI we expand our discussion
to models showing two successive phase transitions with an intermediate
state of partial order occurring for entropic reasons. For completeness,
in Sec.~VII we examine Potts models on two lattices, which have a high
ground-state degeneracy and do display partial order, but where the
entropy is sub-extensive, i.e.~the residual entropy per site is $0$.
Section VIII contains a brief summary and conclusion.

\section{\label{sec:level1}Models}

\subsection{Potts Model}

In a $q$-state Potts model, the local variable at site $i$ may take one of
$q$ different states, which we label as $\sigma_i = 0, 1, \dots \; q-1$.
The Hamiltonian
\begin{equation}
{\cal H} = J \sum_{<i,j>} \delta_{\sigma_{i}\sigma_{j}} - H \sum\limits_{i \in
\cal{L}} \delta_{\sigma_i, 0}
\label{epm}
\end{equation}
consists of two terms, one for interactions between nearest-neighbor local
variables for every bond of the lattice, and one for an external field $H$
coupled to one of the $q$ states and for one sublattice $\cal{L}$. In the
ferromagnetic Potts model, $J < 0$, a negative energy contribution is
obtained if the neighboring sites are in the same state, while in the AF
case, $J > 0$, neighboring sites tend to occupy different states.

In the case $J < 0$, long-range order is favored and the ground state always
has ferromagnetic order. It has been proven in two dimensions that the
finite-temperature phase transition to a disordered state is continuous if
$q \le 4$ and is first-order if $q > 4$ \cite{Baxter_q_4_order}. Although
there is no exact solution in three dimensions, numerical results
\cite{3D_ferr_Potts,rwxcnx} indicate that a first-order phase transition
occurs for $q \ge 3$.

The AF case is far richer and more complex. If the different local states
are denoted by different colors, at zero temperature the neighboring sites
should not have the same color. Thus the AF Potts model at $T = 0$ is
equivalent to a vertex coloring problem. By using the Dobrushin
Uniqueness Theorem \cite{Dobrushin1968_Theor.Prob.Appl.13-197-224,
Dobrushin1970_Theor.Prob.Appl.15-458-486}, Salas and Sokal
\cite{Salas1997_JSP86-551} proved that for sufficiently large $q$ the
correlation function exhibits exponential decay at all temperatures,
including $T = 0$. The model is therefore disordered even in the ground
state, and no phase transition occurs. For small $q$, by contrast, an ordered
(or, from above, partially ordered) ground state is likely. Based on this
insight, it is thought that for every lattice there exists a value $q_{c}$
for which the system is disordered at all temperatures if $q > q_{c}$. For
$q = q_{c}$ the system is critical at zero temperature, a situation we
discuss below. Any behavior is possible if $q < q_{c}$, and typically one
expects a phase transition of first or second order to a type of
long-range-ordered state \cite{Salas1998_JSP92-729}.

\begin{table}
\caption{\label{tab:table1} Values $q_{c}$ at which the $q$-state
AF Potts model is critical at zero temperature for different two-dimensional
lattices. For regular lattices, ${\bar z}$ denotes the coordination number,
while it represents the average coordination for irregular lattices. For the
square, kagome, and triangular lattices, the value of $q_c$ is exact, while
for the honeycomb lattice it is derived by conjecture. All other values are
deduced from our calculations. The dilute centered diced lattices are
introduced in Sec.~VII.}
\begin{ruledtabular}
\begin{tabular}{lcr}
Lattice & ${\bar z}$ & $q_{c}$ \\
\hline
Decorated Honeycomb \cite{decorated_AF_F} & 2.4 & $ < 3$ \\
Decorated Square \cite{decorated_AF_F} & 2.667 & $ < 3$ \\
Honeycomb \cite{hexagonal_qc} & 3 & 2.618 \\
Square \cite{square_qc} & 4 & 3 \\
Kagome \cite{kagome_qc} & 4 & 3 \\
Diced & 4 & $ > 3$ \\
Triangular \cite{triangular_qc} & 6 & 4 \\
Union-Jack & 6 & $ > 4$ \\
Centered Diced & 6 & $ > 5$ \\
Generalized Decorated Square \cite{gdsl} & 5.333 & $ \ge 4$ \\
Dilute Centered Diced IA & 5 & $ > 4$ \\
Dilute Centered Diced IIA & 5 & $ > 4$ \\
\end{tabular}
\end{ruledtabular}
\label{q_c}
\end{table}

In Table I we use $q_c$ to organize a number of lattices in two
dimensions (Sec.~IIB), including all of those to be discussed in the
remainder of this paper. The results for the $q_c = 3$ on the square
\cite{Salas1998_JSP92-729} and kagome lattices \cite{kagome_qc}, and $q_c$
 = 4 for the triangular lattice \cite{triangular_qc} are obtained from exact
solutions at zero temperature, in which they are proved to be critical at
these values of $q$. On the honeycomb lattice, the fractional value of the
critical $q$ is determined by conjecture \cite{hexagonal_qc}. The results
for the decorated square and honeycomb lattices are obtained by mapping
the AF model to a ferromagnetic model whose critical value of $q$ is
known \cite{decorated_AF_F}. Results for all other lattices are based on
our calculations. Table I shows a very close relationship between the
lattice coordination number and $q_c$, with larger values of ${\bar z}$
requiring larger $q_c$. Beyond this coordination number, however, it is
clear that site equivalence also plays a key role. Although the diced
lattice has the same average coordination number as the square and kagome
lattices, it shows a finite-temperature transition to a partially ordered
ground state \cite{Kotecky2008_PRL101-030601}, and therefore $q_c > 3$. The
crucial difference is that, while all sites in the regular square and kagome
lattices are equivalent, the diced lattice is irregular, being composed of
two inequivalent sublattices of three-fold and six-fold coordinated sites.

\subsection{\label{sec:level2}Archimedean and Laves Lattices}

\begin{figure}[t]
\includegraphics[width=2.1cm]{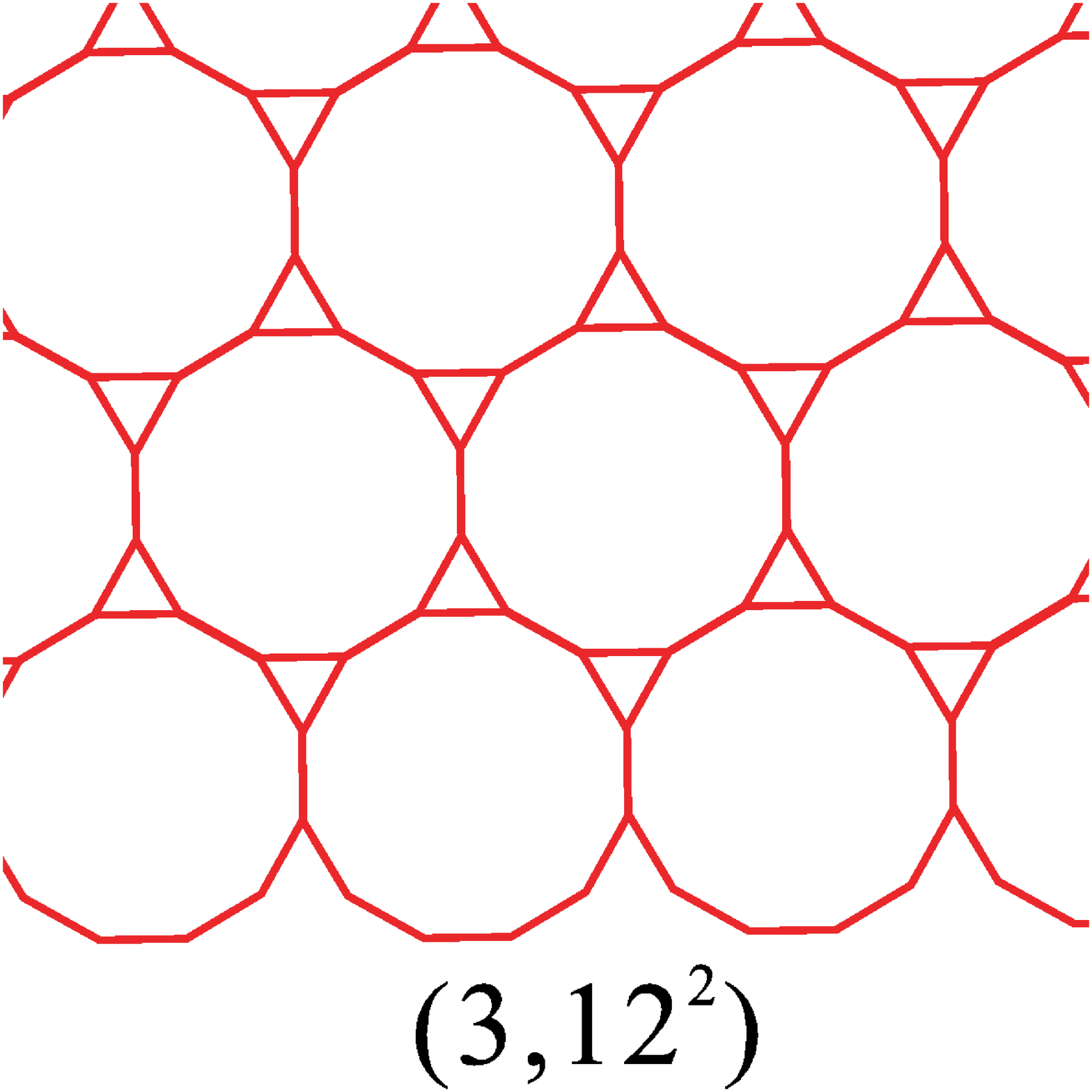}\hspace{0.8cm} \vspace{0.6cm}
\includegraphics[width=2.1cm]{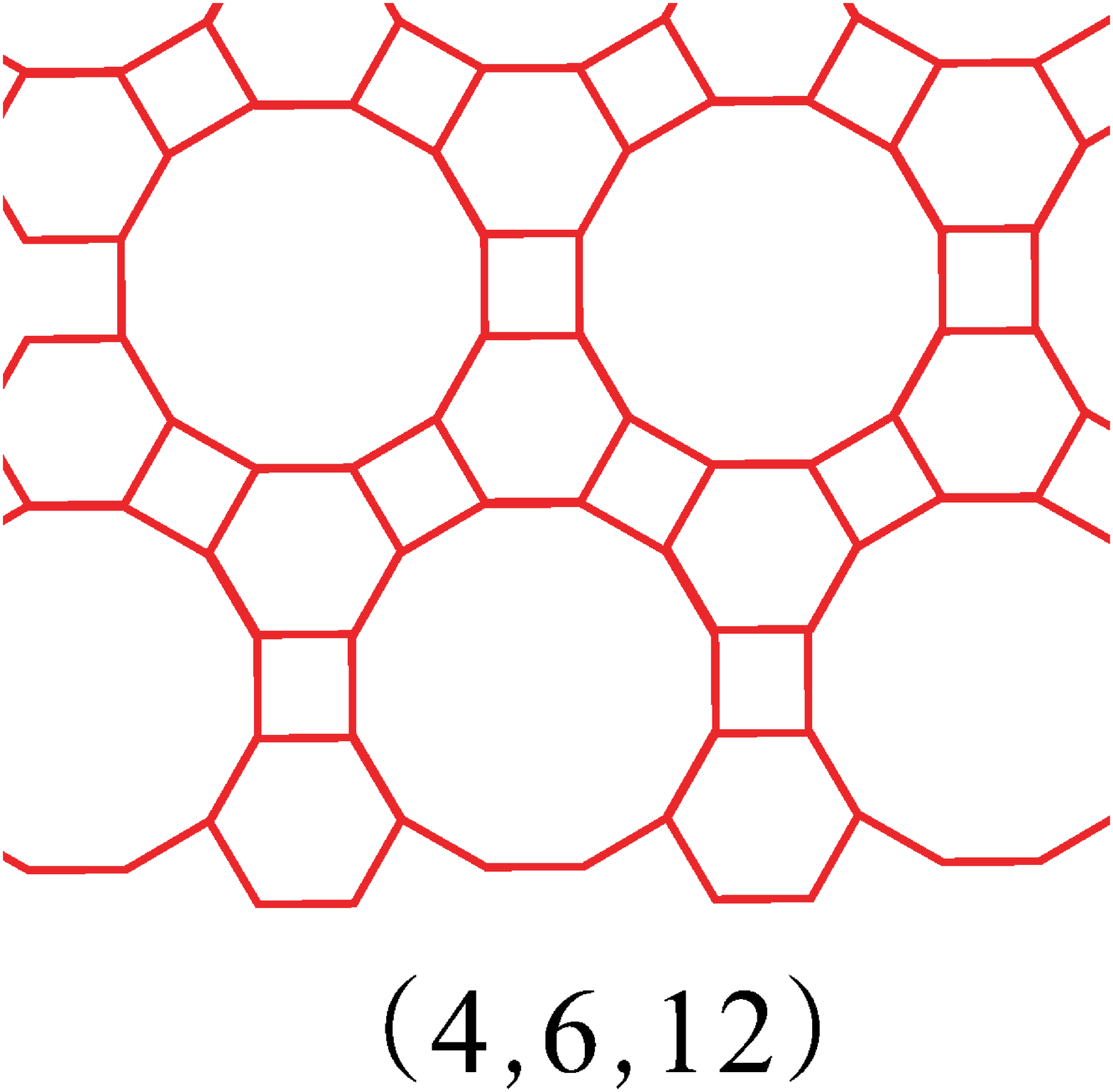}\hspace{0.8cm}
\includegraphics[width=2.1cm]{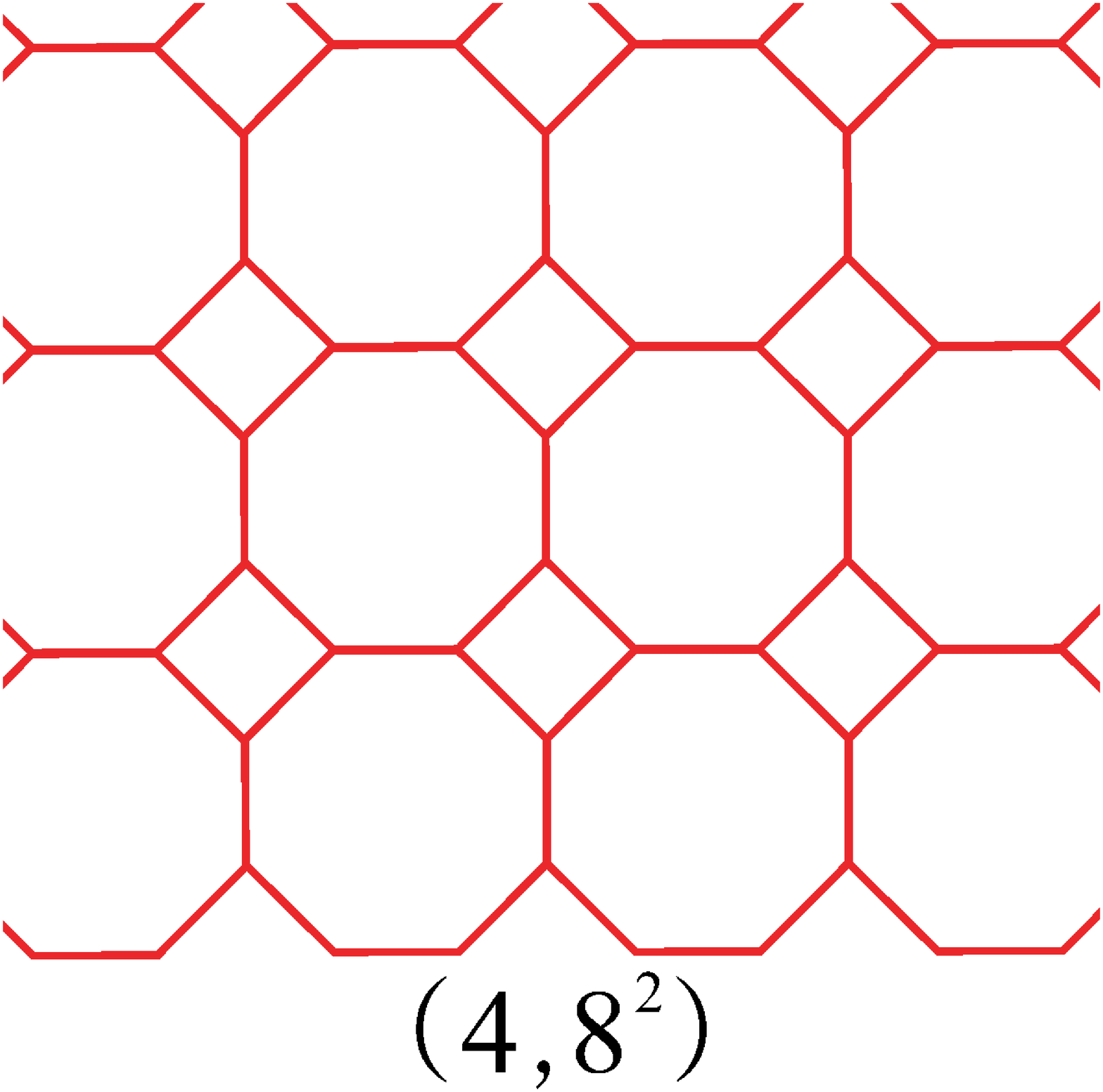} \hspace{0.8cm}
\includegraphics[width=2.1cm]{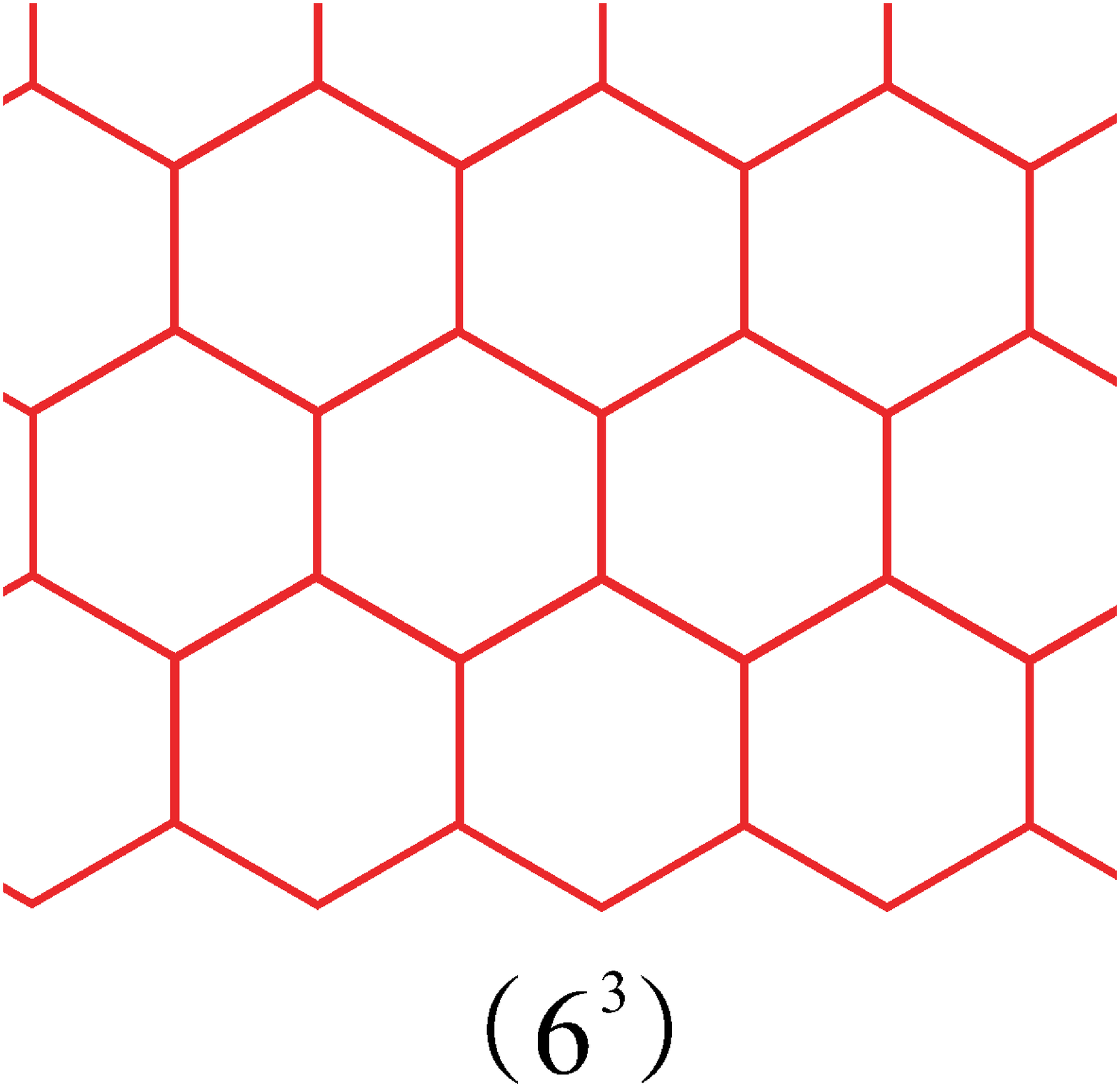}\hspace{0.8cm}\vspace{0.6cm}
\includegraphics[width=2.1cm]{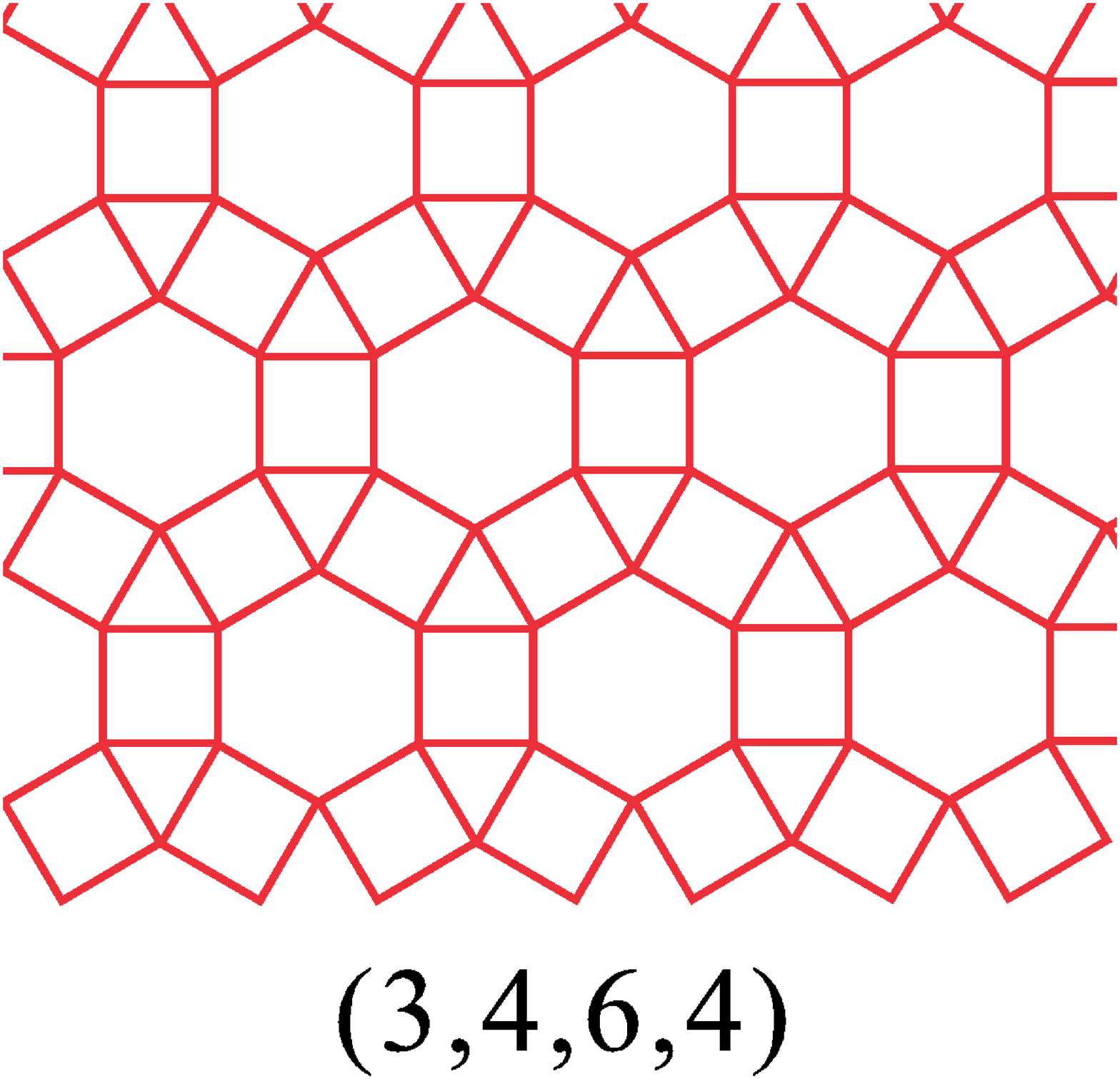}\hspace{0.8cm}
\includegraphics[width=2.1cm]{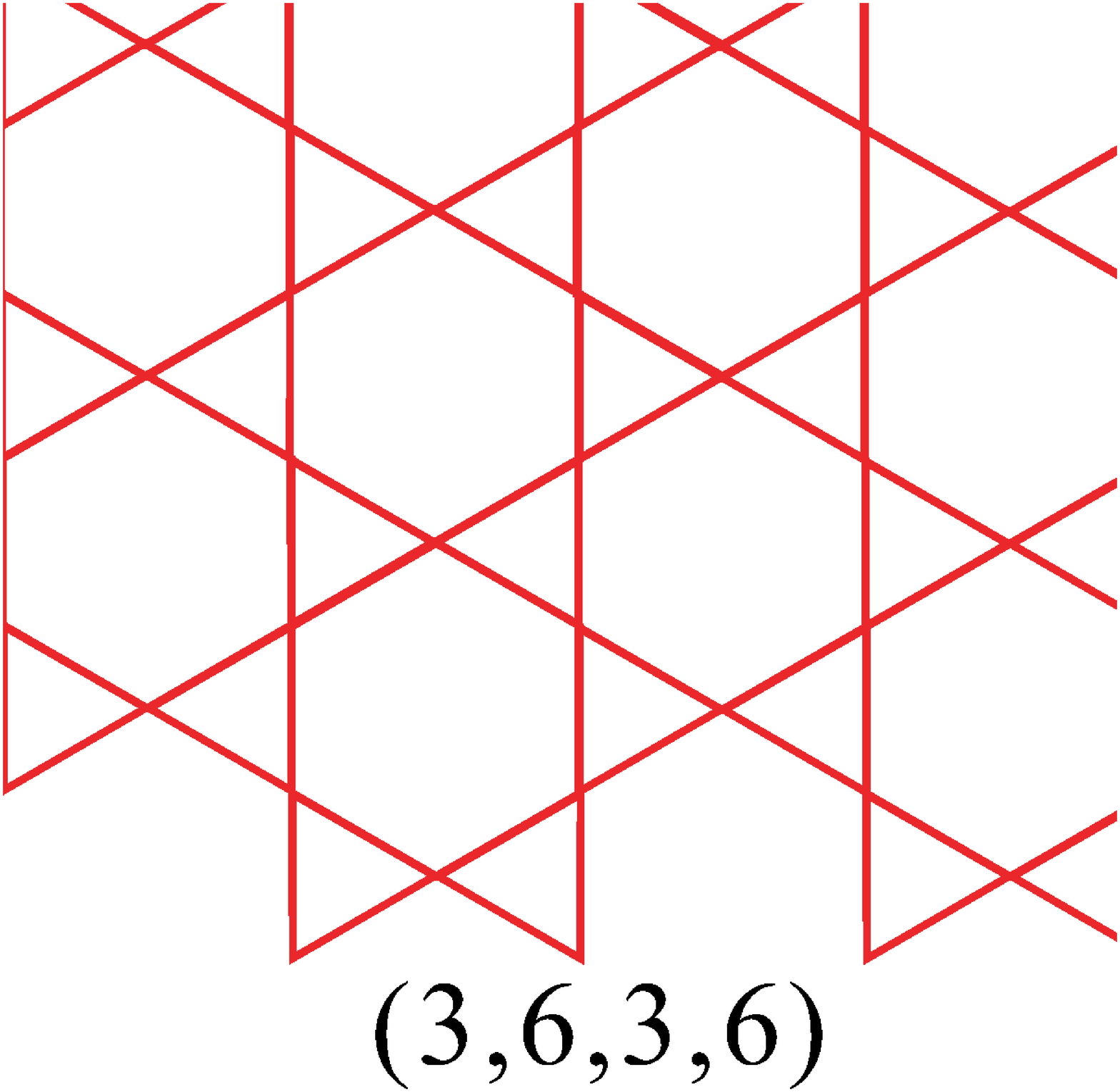}\hspace{0.8cm}
\includegraphics[width=2.1cm]{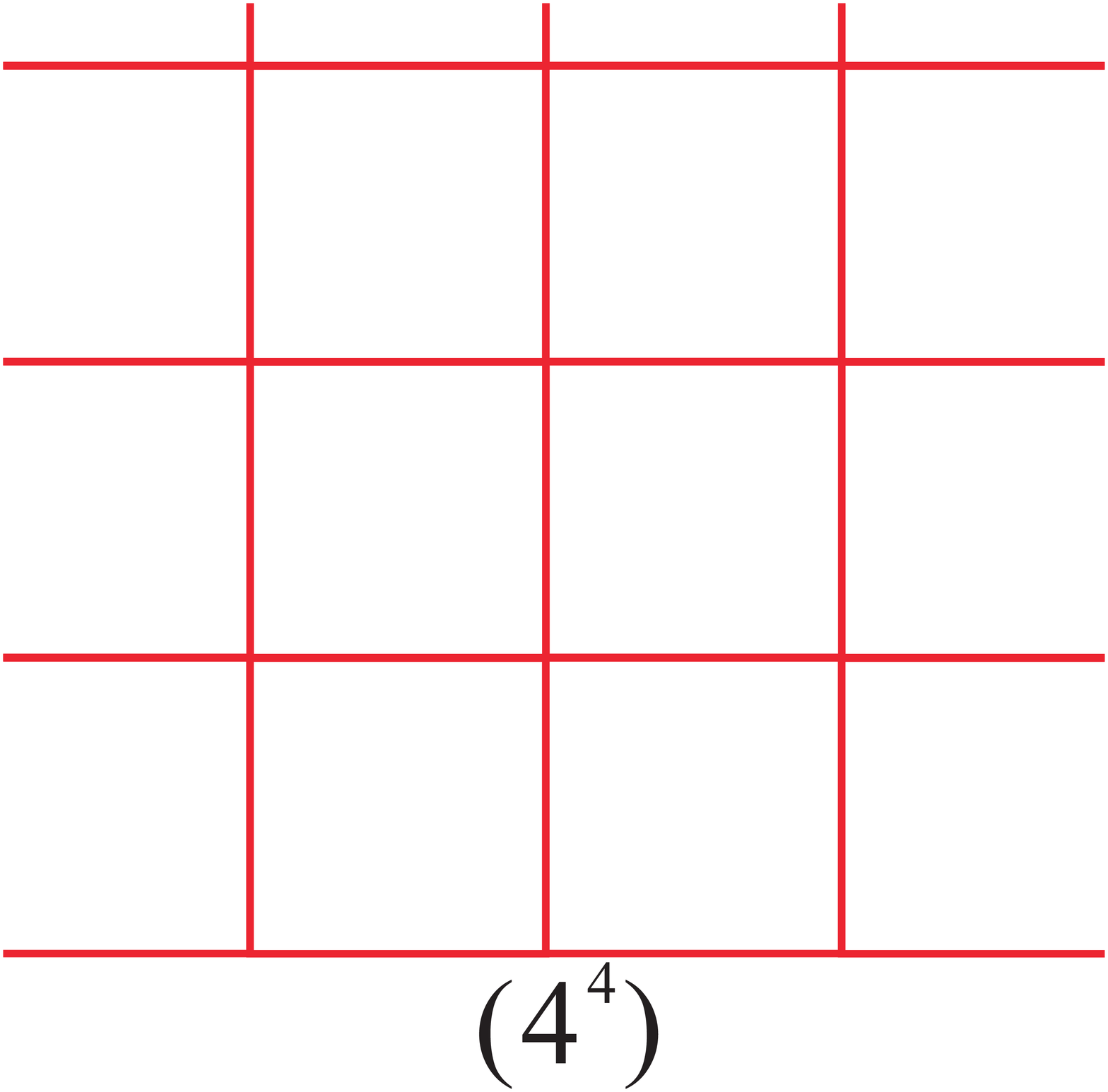}\hspace{0.8cm}\vspace{0.6cm}
\includegraphics[width=2.1cm]{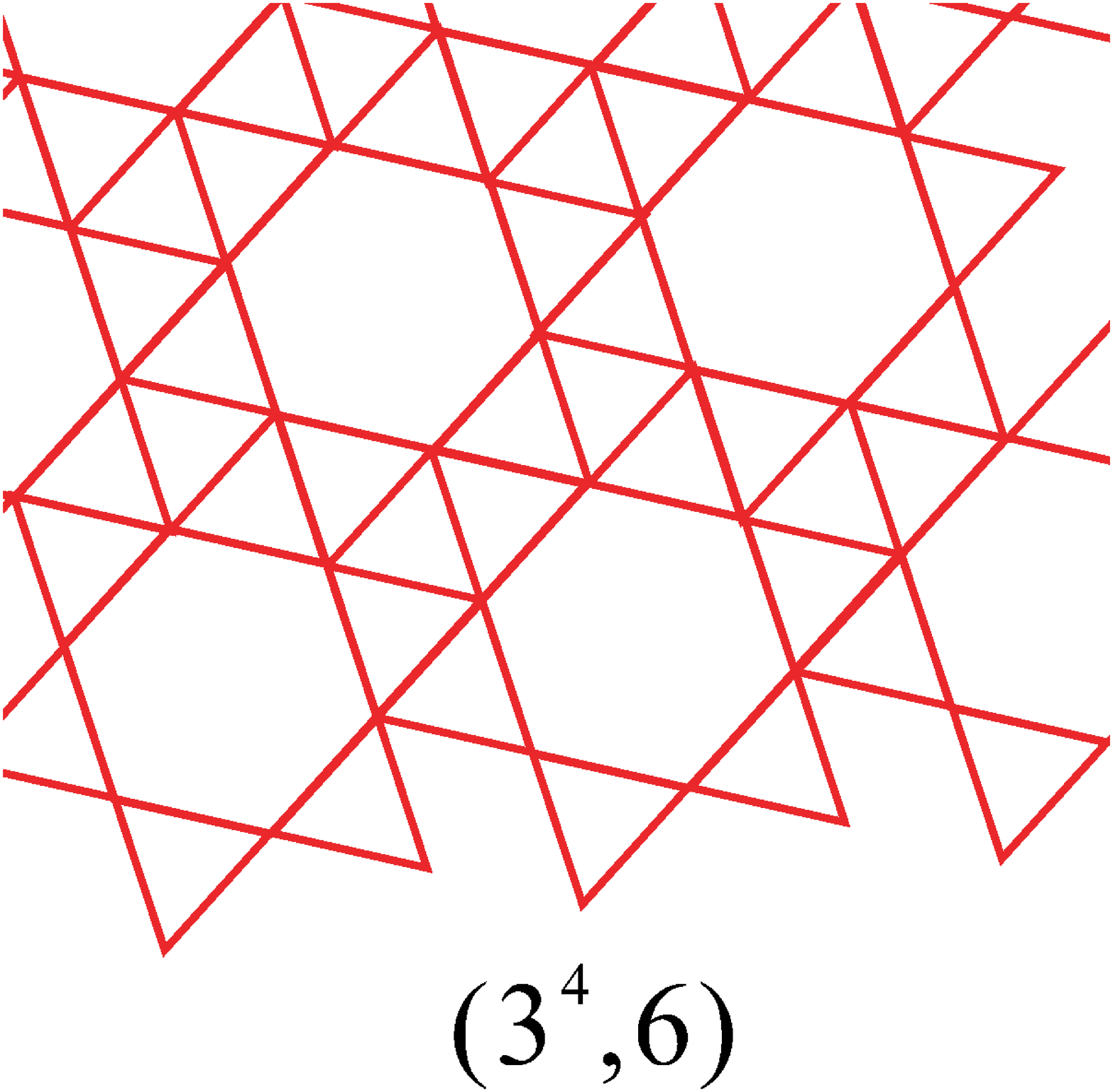}\hspace{0.8cm}
\includegraphics[width=2.1cm]{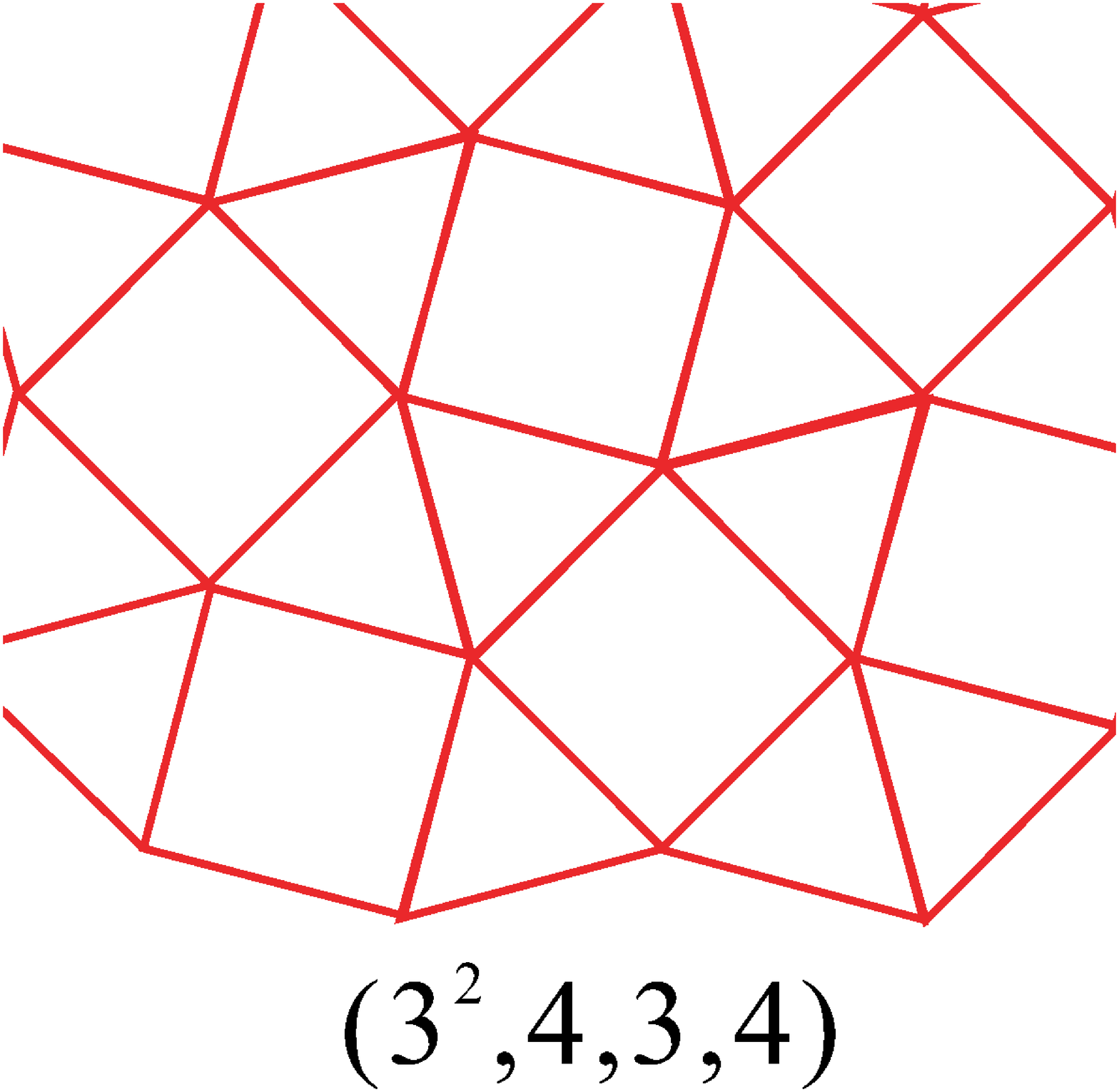}\hspace{0.8cm}
\includegraphics[width=2.1cm]{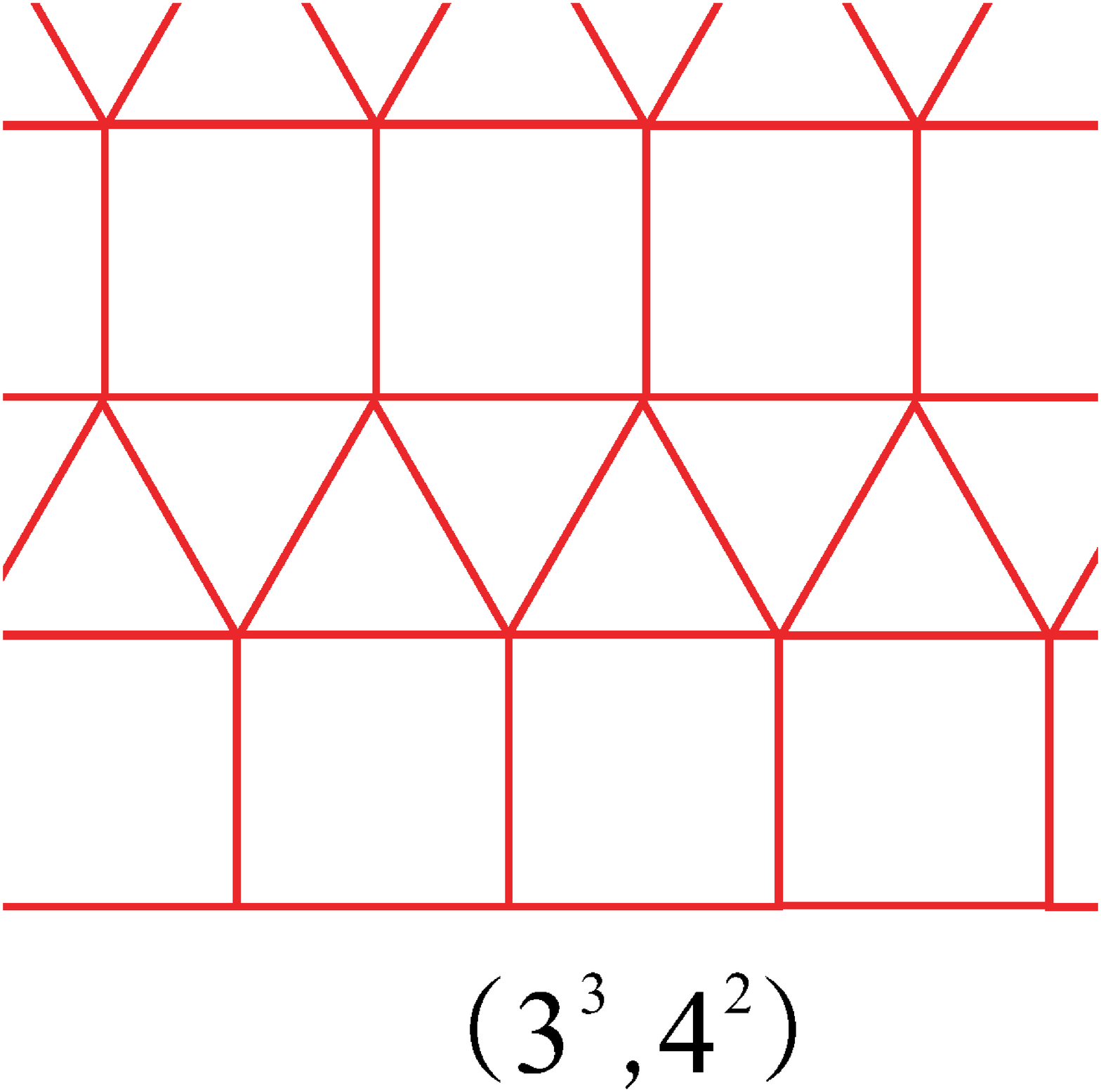}\hspace{0.8cm}
\includegraphics[width=2.1cm]{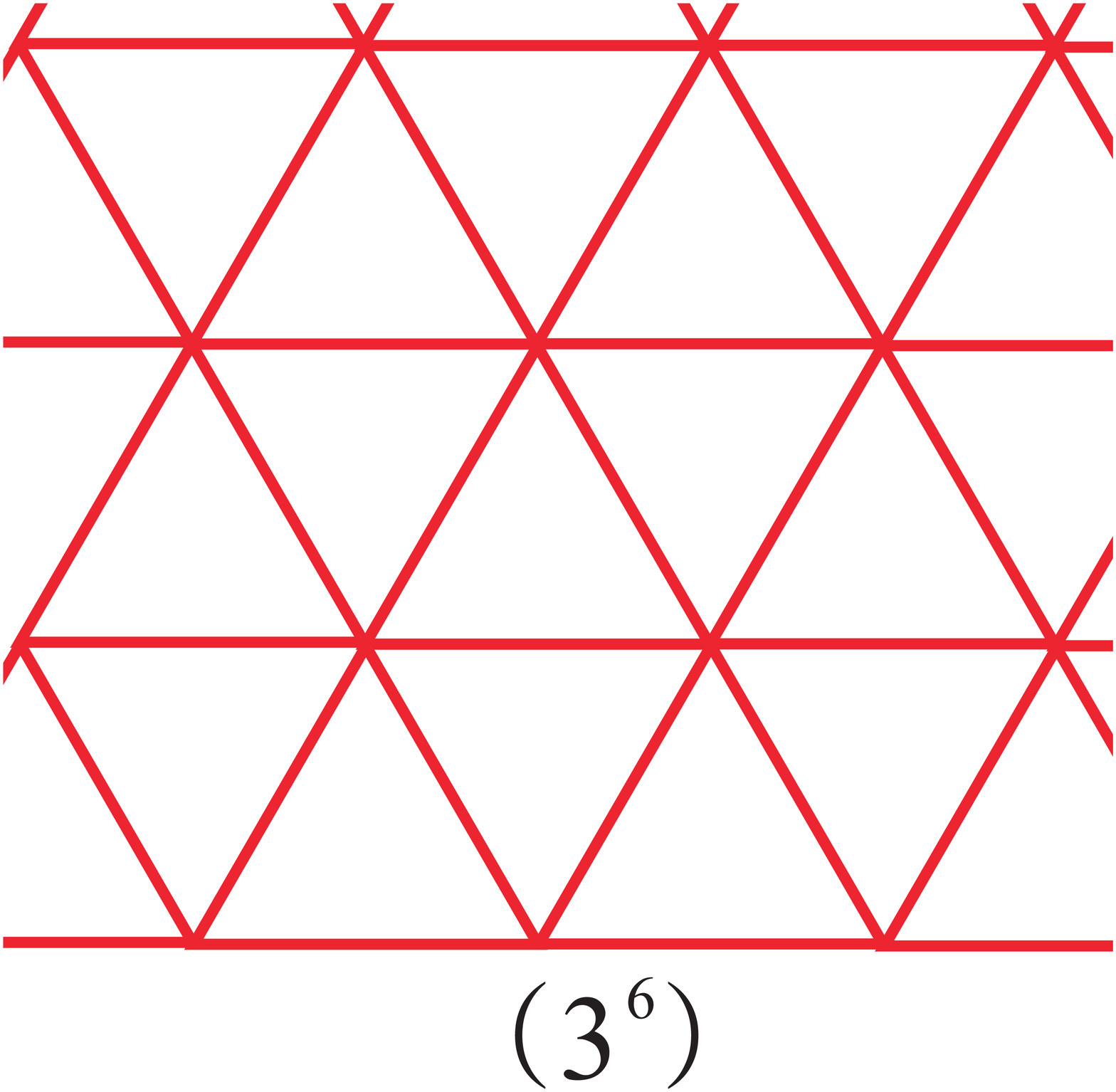}
\caption{The 11 planar Archimedean lattices. The index gives the lattice
name in the terminology explained in the text.}
\label{Archimedean}
\end{figure}

Lattices in which all sites are equivalent are known as Archimedean.
This category includes the honeycomb, square, kagome, and triangular
lattices. There are 11 planar Archimedean lattices, all of which are
shown in Fig.~\ref{Archimedean}. On an Archimedean lattice, the
coordination number is the same for every site. The planar lattice is
equivalent to a tiling of polygons, each site belonging to different
polygons, but with the number and type of polygons to which each site
belongs being the same. If the coordination number of the Archimedean
lattice is $z$, the lattice is said to be ``$n$-colorable'' for any
$n \geq z$; although this creates an AF intersite color condition,
there is no known relation between $n$ and $q_c$.

The Archimedean lattices have systematic names. For any given vertex, the
attached polygons are listed (for example in clockwise order) by their
number of edges. While this process generates multiple names for several
lattices, depending on the starting polygon, the convention is to choose
the lexicographically shortest name by using exponents to abbreviate two
or more consecutive entries. Thus the square lattice is also known as
$(4,4,4,4)$, or $(4^4)$, and this notation is used in Fig.~\ref{Archimedean}.

\begin{figure}[t]
\includegraphics[width=2.1cm]{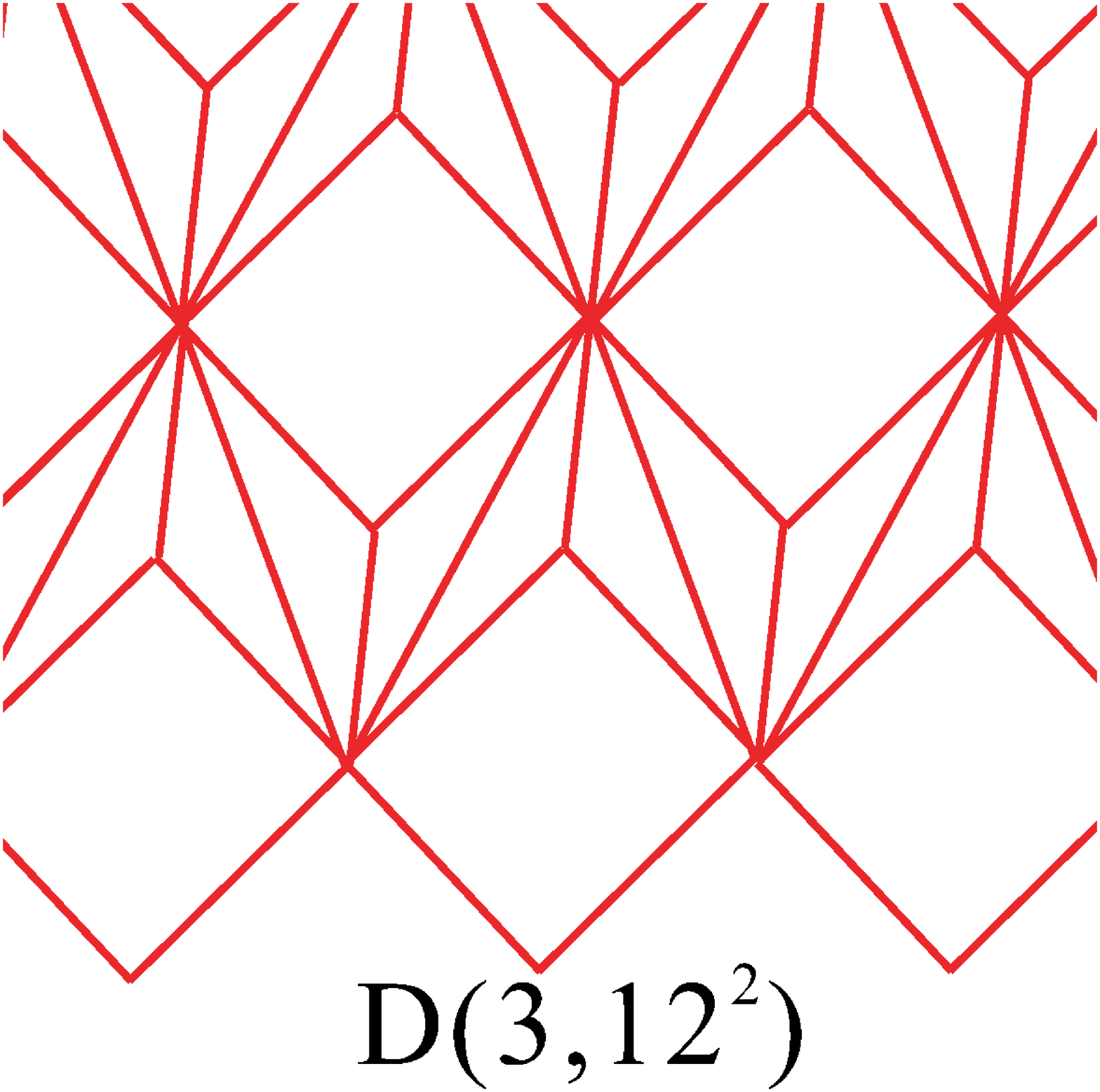}\hspace{0.8cm}\vspace{0.6cm}
\includegraphics[width=2.1cm]{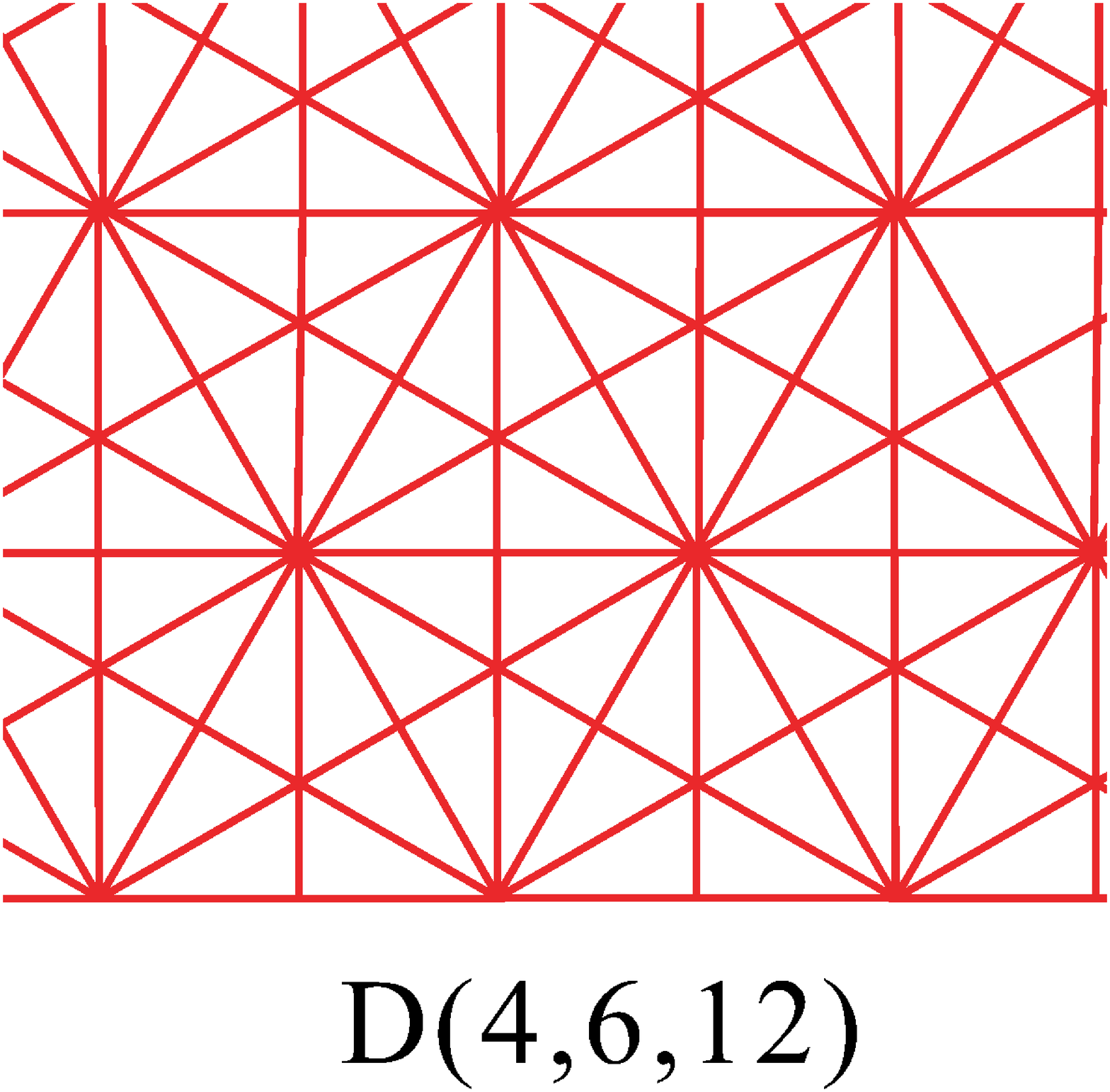}\hspace{0.8cm}
\includegraphics[width=2.1cm]{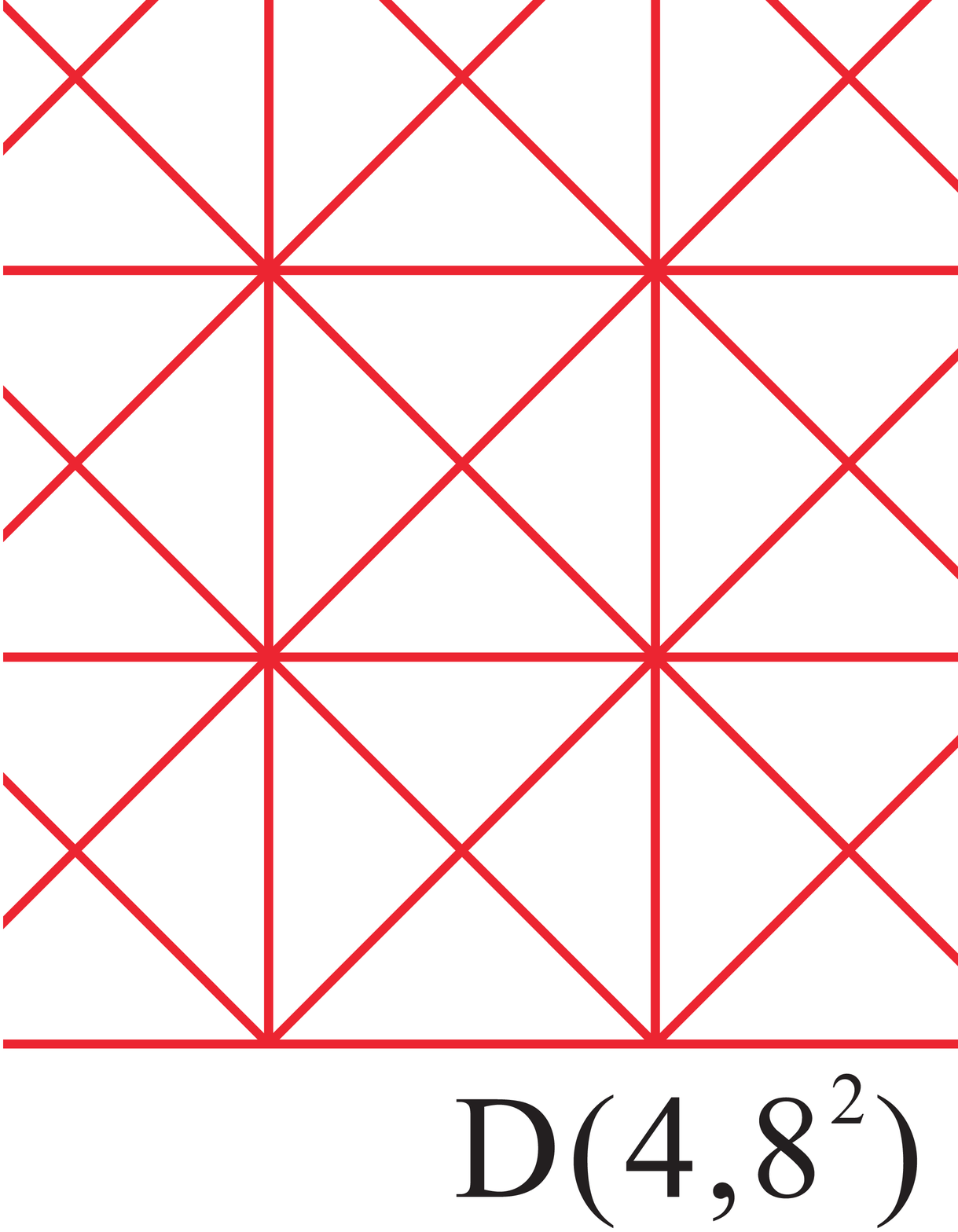} \hspace{0.8cm}
\includegraphics[width=2.1cm]{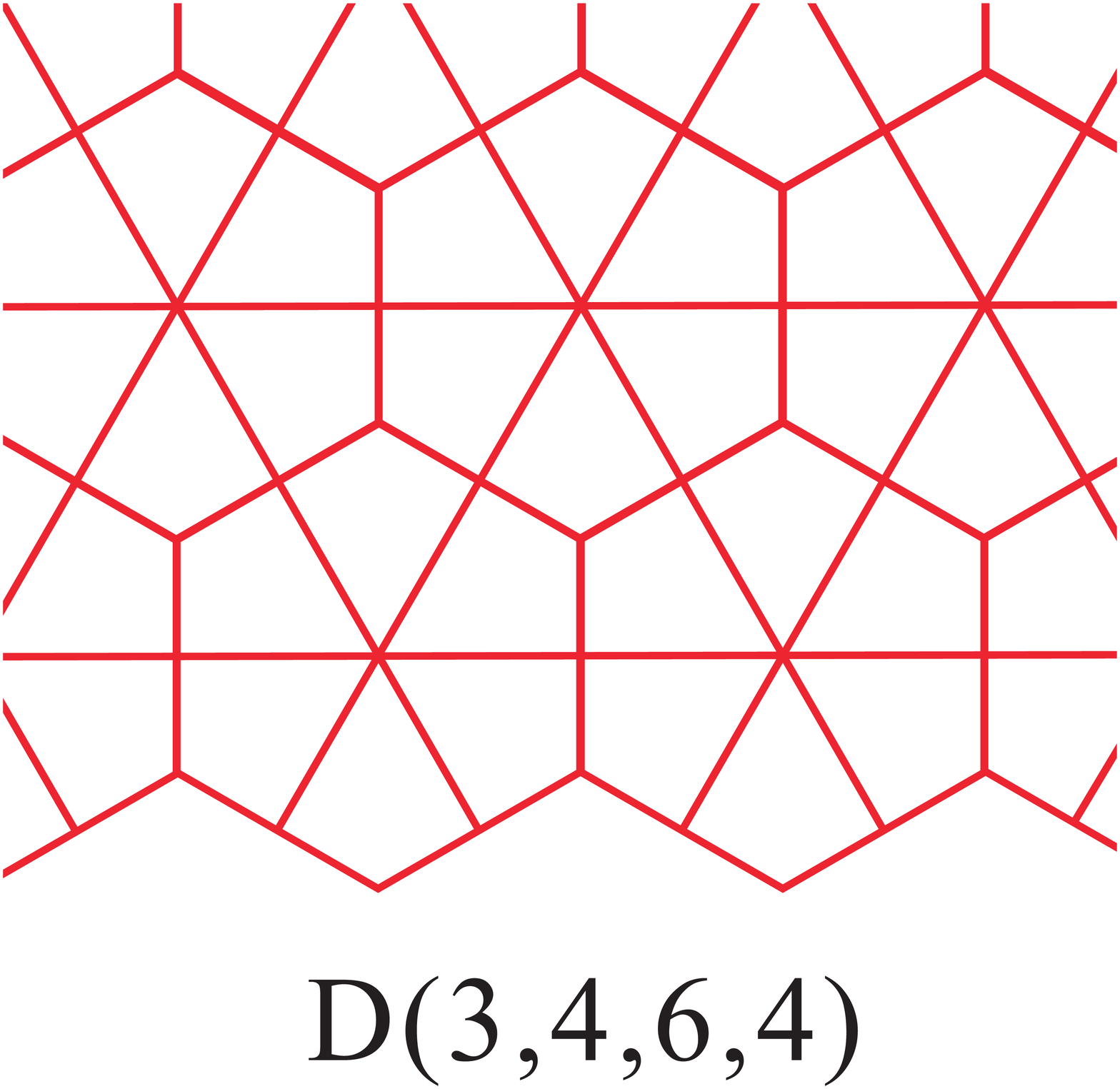}\hspace{0.8cm}\vspace{0.6cm}
\includegraphics[width=2.1cm]{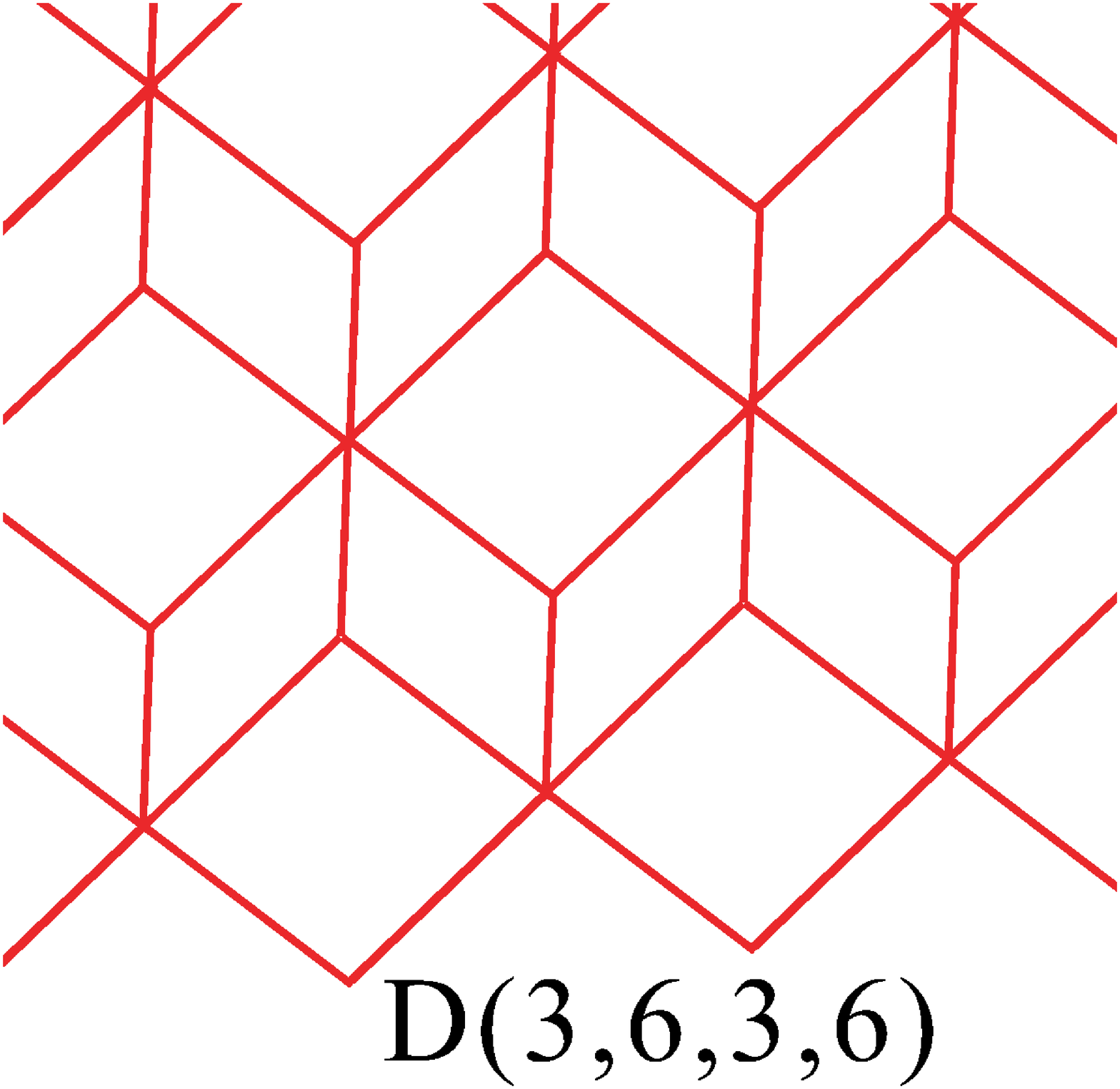}\hspace{0.8cm}
\includegraphics[width=2.1cm]{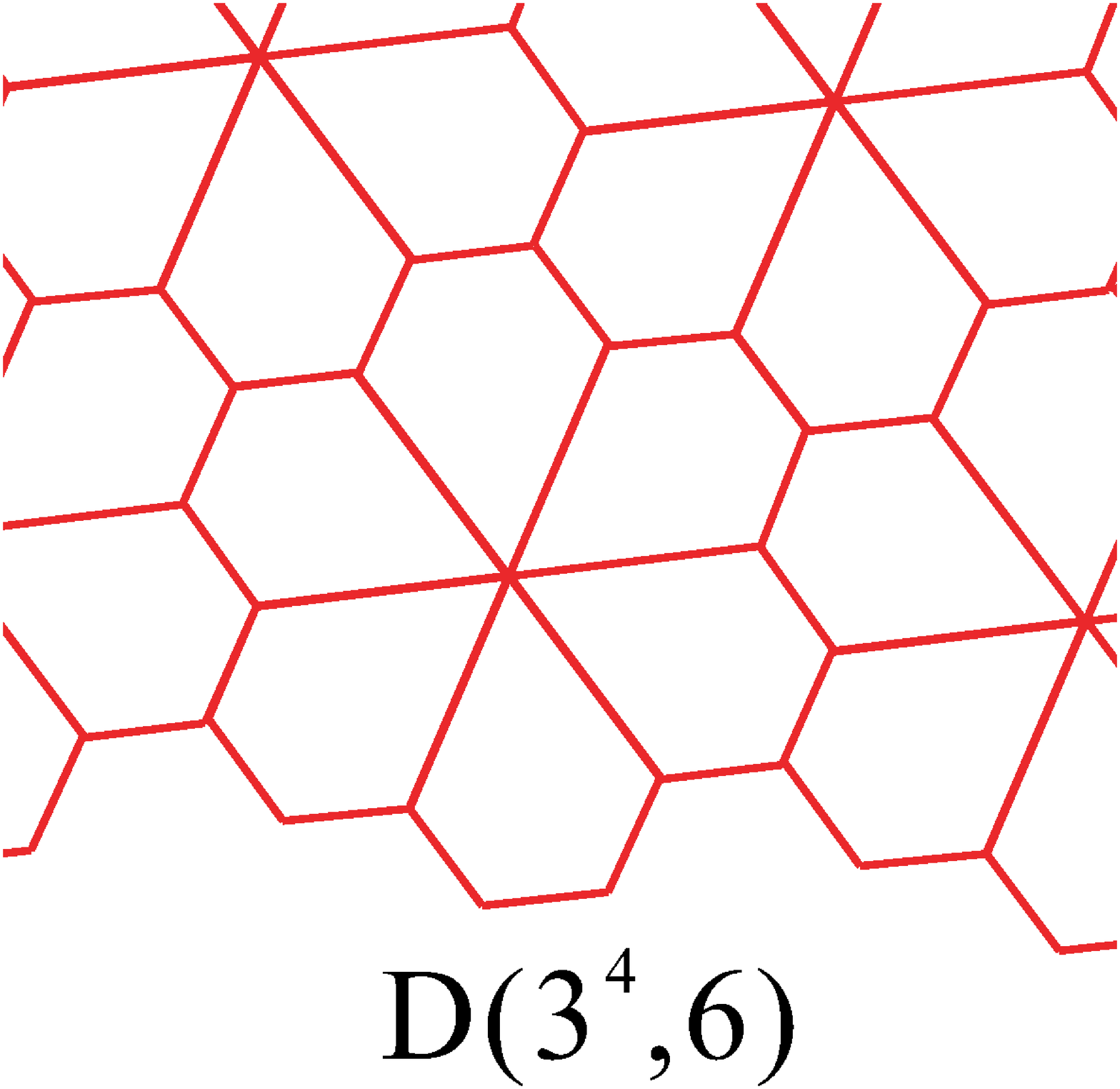}\hspace{0.8cm}
\includegraphics[width=2.1cm]{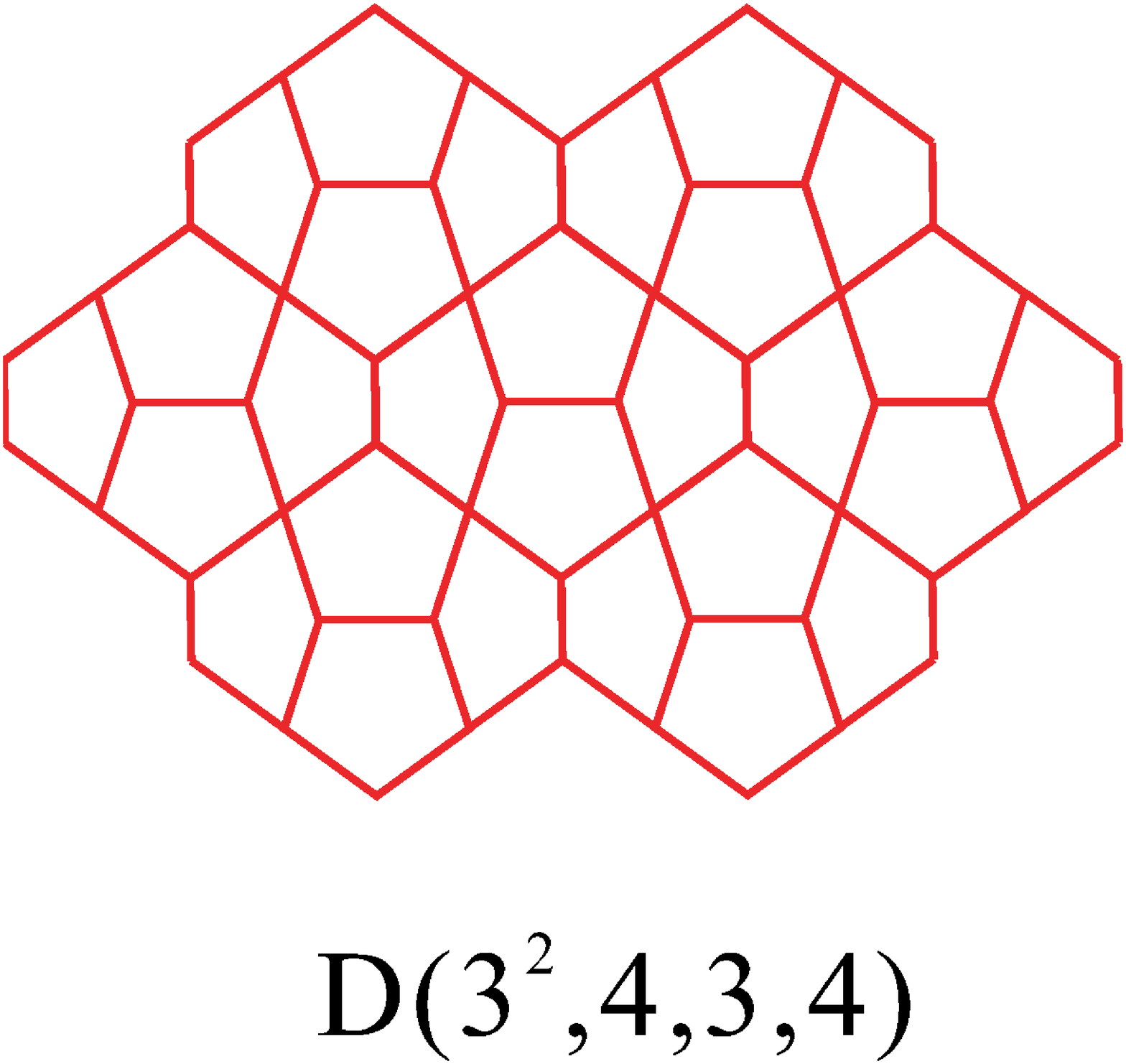}\hspace{0.8cm}\vspace{0.6cm}
\includegraphics[width=2.1cm]{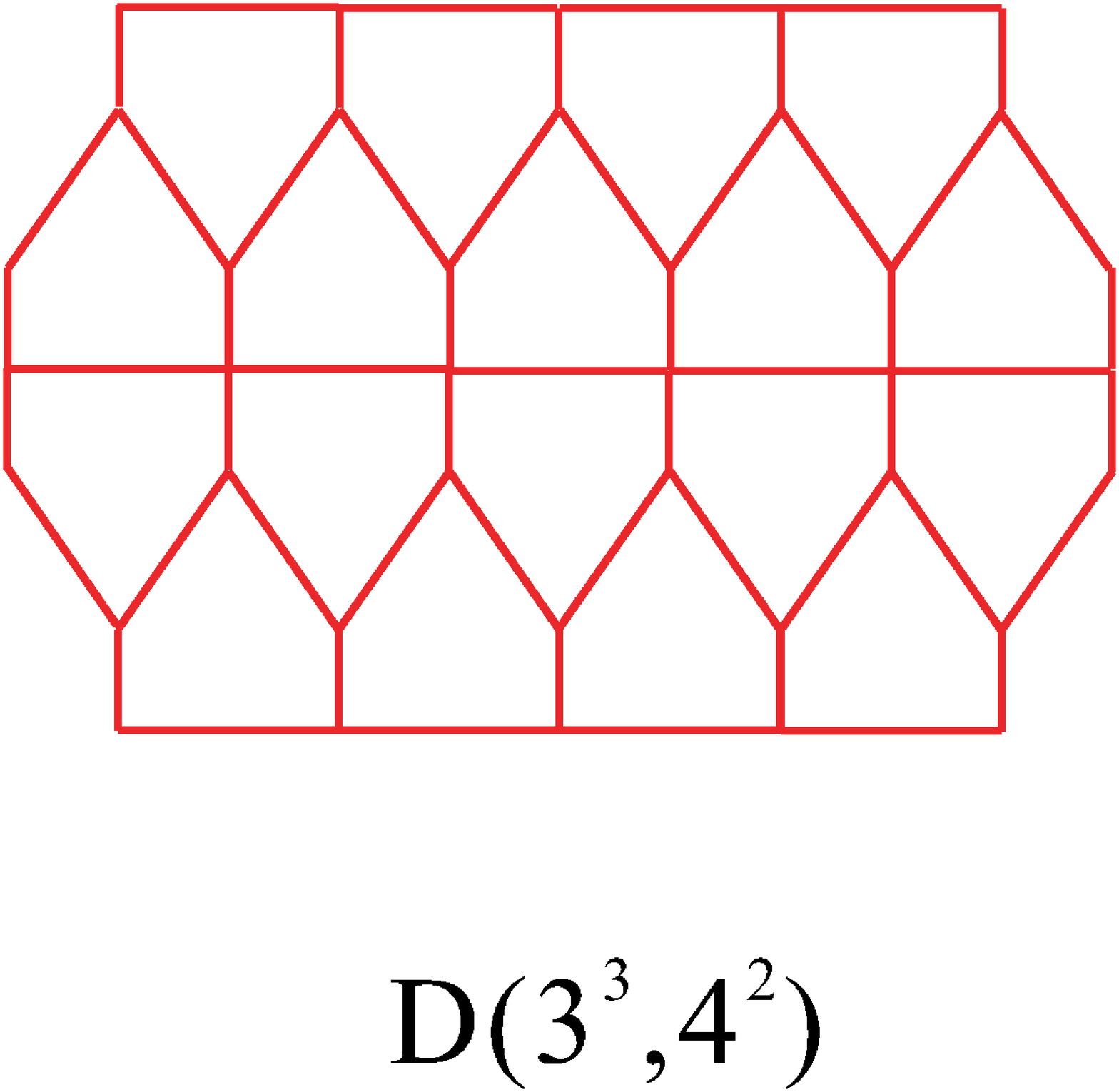}\hspace{0.8cm}
\caption{The set of Laves lattices, irregular planar lattices obtained
as the non-Archimedean duals of the Archimedean lattices. The label gives
the terminology for the dual Archimedean lattice.}
\label{Laves}
\end{figure}

The dual transformation of a planar lattice is defined by adding one site
at the center of each polygon and connecting these new sites to those of all
neighboring polygons. This is a vertex-to-face, face-to-vertex, edge-to-edge
transformation, and is reversible. The square lattice is manifestly self-dual
and the honeycomb and triangular lattices are mutually dual. However, the dual
lattices of the remaining eight Archimedean lattices are not Archimedean;
clearly, the centering sites of the different polygons in these eight
lattices have different connectivity. These are known as the Laves lattices,
and they are shown in Fig.~\ref{Laves}. The Laves lattices with integer
average coordination number ${\bar z}$ play an important role in our
considerations, and here we will study in detail the diced lattice
($D(3,6,3,6)$, ${\bar z} = 4$), the Union-Jack lattice ($D(4,8^2)$,
${\bar z}$ = 6), and the centered diced lattice ($D(4,6,12)$, ${\bar z} = 8$).
According to the four-color theorem \cite{four_coloring1,four_coloring2}, a
planar lattice may only be bipartite (such as the diced lattice), tripartite
(such as the Union-Jack lattice), or quadripartite. The lattices we
investigate in this paper are either bipartite or tripartite.

A bipartite lattice contains only two sublattices of unconnected sites,
and can be generated in one of two ways. One type is a lattice formed only
by polygons with an even number of edges [square, honeycomb, $(4,8^2)$,
$(4,6,12)$, also the diced lattice ($D(3,6,3,6)$)]. On these lattices,
each site A is connected only to sites of type B, and vice versa, and
each polygon is composed alternately of A and B sites. The second type is
the decorated lattice, formed by adding a site to each edge of a starting
lattice. The original lattice sites and the decorating sites belong to
different sublattices, and by taking a partial trace over the Potts variables
on the decorating sites, the $q$-state AF Potts model on a decorated lattice
can always be mapped onto a ferromagnetic Potts model with the same $q$ on
the original lattice \cite{decorated_AF_F}. The $q = 2$ AF Potts (Ising) model
on the bipartite lattice is always ordered at low temperature and disordered
at high temperatures, with a finite-temperature phase transition. The $q = 3$
AF Potts model on a bipartite lattice is more complicated, and its ground
state can be disordered, critical, or ordered. Typical examples of these
cases are respectively the honeycomb, square, and diced lattices. There is
in general no finite-temperature phase transition for the $q = 4$ AF Potts
model on bipartite lattices.

Most of the planar lattices in Figs.~\ref{Archimedean} and \ref{Laves} are
tripartite. A tripartite lattice must contain some polygons with odd edge
numbers (such as triangles or pentagons). The sublattices of a tripartite
lattice may be determined uniquely if the lattice is formed purely by
triangles [triangular, Union-Jack ($D(4,8^2)$), centered-diced ($D(4,6,12)$)].
The $q = 3$ Potts models on these lattices have complete AF long-range order
in the ground state, with one of the three states on each sublattice. This
order can be melted by thermal fluctuations, leading to a finite-temperature
phase transition; if the lattice contains two inequivalent sublattices with
unequal coordination numbers, two finite-temperature phase transitions are
possible (Sec.~VI). The sublattices for most tripartite lattices [kagome
(3,6,3,6), square-kagome (4,$8^2$), (3,4,3,6), the dilute centered-diced
lattices introduced in Sec.~VII] are not unique, and $q = 2$ and 3 AF Potts
models may again have ordered, critical, or disordered ground states on these
lattices. For the $q = 2$ AF Potts model, any order in the ground state will
be partial, because the model is frustrated on a tripartite lattice.

\section{Tensor-Based Numerical Methods}

The development of numerical methods for condensed matter and lattice systems
based on tensor-network representations \cite{Verstraete2008_AP57-143-224} is
motivated by developments in quantum information theory, and a great deal of
progress has taken place in the last five years. Tensor-based numerical
methods have already been used to study spin
\cite{Murg2005_PRL95-057206,jiang_2008_PRL,Xie2009_PRL103-160601,
Zhao2010_PRB81-174411,huihai_2012_prb,Zhiyuan_PRX_2014},
bosonic \cite{Jordan2009_PRB79-174515}, and fermionic models
\cite{Barthel2009_PRA80-042333,Kraus2010_PRA81-052338},
and to deal with quantum critical systems \cite{Vidal2007_PRL99-220405} and
topological quantum phase transitions \cite{Gu2008_PRB78-205116}. They have
been combined with Monte Carlo techniques \cite{Sandvik2007_PRL99-220602}
to take advantage of the best features of both methods and they have been
extended to deal with classical systems such as classical XY models
\cite{Jifeng}.

When dealing with models in classical statistical mechanics, the quantity
expressed as the contraction of a tensor network is the partition function.
In one dimension, this quantity is a product of matrices, which is easy to
evaluate. In higher dimensions, the appropriate representation is by a
network of tensors whose rank matches the coordination number of the
lattice; in this situation, the dimension of the tensors obtained after
each contraction step increases if the same amount of information is to
be stored, and so a truncation is required to keep the contraction under
control. A large number of methods has been developed to perform this
truncation, including the tensor renormalization group (TRG)
\cite{Levin2007_PRL99-120601}, second renormalization group
\cite{Xie2009_PRL103-160601,Zhao2010_PRB81-174411}, infinite time-evolving
block decimation (iTEBD) \cite{Vidal2007_PRL98-070201,Orus2008_PRB78-155117},
corner transfer matrix \cite{Orus2009_PRB80-094403,Or'us2012_PRB85205117},
plaquette renormalization group \cite{Wang2011_PRE83-056703}, and a
renormalization-group method based on higher-order singular value
decomposition (HOSVD) \cite{Xie2012_PRB86-045139}. Here we summarize three
of these methods for pedagogical purposes, and during our analysis of AF
$q$-state Potts models we considered a number of approaches in the process
of optimizing our calculations, but at the end all of the numerical results
presented in Secs.~IV-VII were obtained using iTEBD.

\begin{figure}[t]
\includegraphics[width=8.1cm]{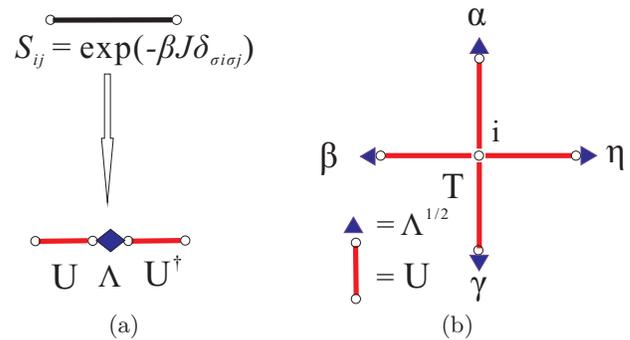}
{\centerline {(a) \qquad\qquad\qquad\qquad\qquad\qquad (b) \qquad }}
\caption{(Color online) Schematic representation of the expression of the
partition function of the $q$-state Potts model as a tensor network. (a) Eigenvalue
decomposition of the Boltzmann factor $S_{ij}$ for a bond. (b) Construction
of the local tensor $T$ by contraction of four $U$ matrices and
renormalization by the square root of the eigenvalues $\Lambda $.}
\label{local_tensor}
\end{figure}

\subsection{Partition Function and Thermodynamics}

It is always possible to find a tensor-network representation for the partition
function of a classical model \cite{Zhao2010_PRB81-174411,Jifeng}. In the
example of the $q$-state Potts model on a square lattice, one may define the
Boltzmann factor associated with each bond $\langle ij \rangle$ as
\begin{equation}
S_{ij} = \exp(-\beta J\delta_{\sigma_{i}\sigma_{j}}),
\end{equation}
where $\sigma_{i}$ denotes the Potts variable on site $i$. As represented
schematically in Fig.~\ref{local_tensor}, an eigenvalue decomposition for
$S$ yields
\begin{equation}
S_{ij} = \sum\limits_{\alpha} U_{i\alpha} \Lambda_{\alpha} U_{j\alpha},
\end{equation}
where $U_{i\alpha}$ is a unitary matrix and $\alpha = 0, 1, \dots, q-1$ because
$S_{ij}$ is a $q \times q$ matrix for the $q$-state Potts model. Now the
partition function can be expressed as
\begin{equation}
Z = \sum\limits_{\{\sigma \}} \prod\limits_{\langle ij\rangle } S_{ij} \;\; =\;\;
\sum\limits_{\{\alpha \}} T_{\alpha \beta \gamma \eta } T_{\alpha \epsilon \zeta \theta }
\dots,
\label{partition_tensor_network}
\end{equation}
where
\begin{equation}
T_{\alpha \beta \gamma \eta } = \sum\limits_{i} U_{i\alpha} U_{i\beta} U_{i\gamma} U_{i\eta}
(\Lambda_{\alpha} \Lambda_{\beta} \Lambda_{\gamma} \Lambda_{\eta})^{1/2},
\end{equation}
i.e.~as a network of tensors $T$ constructed from the bond eigenvalues and
eigenvectors. The rank of $T$ is determined from the number of bonds per site
of the tensor lattice, which is often the coordination number $z$ of Sec.~II.
From above, the bond dimension of each index is $q$. There are many ways to contract this
tensor network and in this section we review the TRG/SRG and iTEBD methods,
which represent respectively the two primary classes of technique, namely
variational, renormalization-group approaches that converge to infinite
size and power or projection approaches that are already (by translational
invariance) in this limit.

\subsection{TRG/SRG}

The tensor renormalization group (TRG) \cite{Levin2007_PRL99-120601} is a
real-space coarse-graining method proposed by Levin and Nave in 2007. After
each coarse-graining step, both the topology of the lattice and the rank and
dimension of the tensor remain the same, but the size of the lattice is
only half (in general) of its original size. The method proceeds by first
decomposing the tensor and then recombining new tensors, but the details
depend on the lattice topology and are best illustrated by example.

\begin{figure}[t]
\includegraphics[width=8.5cm]{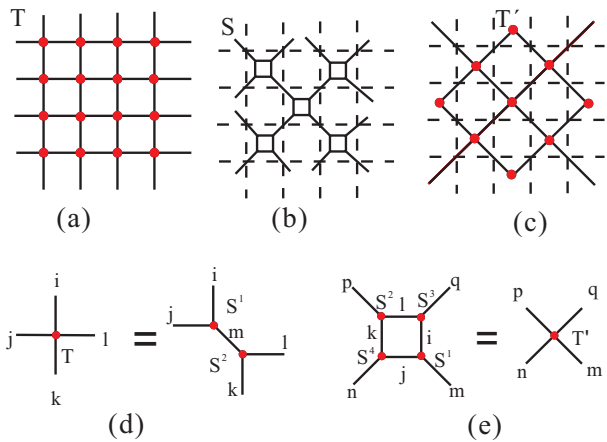}
\caption{(Color online) Two TRG steps on the square lattice. (a) The original tensor
network on the square lattice representing the partition function as in
Eq.~(\ref{partition_tensor_network}). (b) After a SVD of the local tensor
following alternating ``stretching'' directions on the two different
sublattices, the square-lattice tensor network is transformed to one on
the Archimedean ($4$, $8^2$) lattice (Fig.~\ref{Archimedean}). (c) Groups
of four rank-3 tensors are then contracted to form one new rank-4 tensor
on a lattice with half the number of sites of the original square lattice.
(d) SVD of a local tensor, defined by Eq.~(\ref{sqtrg1}). (e) Contraction
of four rank-3 tensors into a rank-4 tensor, as defined by Eq.~(\ref{sqtrg2}).}
\label{sqtrg}
\end{figure}

On the square lattice (Fig.~\ref{sqtrg}), each iteration requires two steps.
First the rank-4 site tensor is decomposed into two rank-3 auxiliary tensors,
with a choice of indices following alternating ``stretching'' directions on
the two different sublattices. Specifically, by combining two indices the
rank-4 tensor becomes a matrix (rank-2 tensor) whose SVD yields a set of
singular values, which are absorbed into the two unitary bond matrices. By
expanding the combined index one obtains two rank-3 tensors,
\begin{eqnarray}
T_{ijkl} & = & \sum_{m} U_{ij,m} \lambda_m V_{kl,m},
\nonumber \\
S^1_{ijm} & = & U_{ij,m} \sqrt{\lambda_m},
\label{sqtrg1}
\\
S^2_{klm} & = & V_{kl,m} \sqrt{\lambda_m}, \nonumber
\end{eqnarray}
where $U$ and $V$ are unitary matrices, and $\lambda$ is a diagonal
singular-value matrix. The partition function is represented as a tensor
network defined on the Archimedean $(4,8^2)$ lattice (Fig.~\ref{Archimedean}).
If the dimension of the bond index for the tensor $T$ is $d$, the dimension
of index $m$ is $d^2$ [Eq.~(\ref{sqtrg1})]. This bond dimension grows during
the renormalization process and when $d = D$, the maximum bond dimension we
can retain due to the limits set by our computational resources, a truncation
is required to prevent divergence on repeated iteration. Here the natural
approach is to cut the dimension of $m$ according to the relative sizes of
the singular values and to keep the $D$ largest ones. The second step of the
iteration is to contract the four rank-3 tensors on the $(4,8^2)$ squares
into a new rank-4 tensor,
\begin{equation}
T'_{mnpq} = \sum_{ijkl}S^1_{ijm} S^4_{jkn} S^2_{klp} S^3_{liq},
\label{sqtrg2}
\end{equation}
as a result of which both the topology of the tensor network and the
dimension of the local tensor are unchanged. Thus each iteration step
forms a new square lattice whose tensor-network representation contains
only half as many sites. If the iteration is repeated $n$ times, the size
of the tensor network shrinks to $2^{-n}$ of the original, giving easy access
to the thermodynamic limit by renormalization methods.

However, in the TRG approach the tensor is truncated according to its
singular values, which is in essence a local approximation. In fact the
same pair of sites is connected by (many) other paths in the lattice and
a more consistent approach is to consider the effect of this ``environment''
in order to perform the truncation globally, which is the concept of the
second renormalization group (SRG) method \cite{Xie2009_PRL103-160601}.

The partition function can be expressed as
\begin{equation}
Z = \rm{Tr}[T_{i} T_{i}^{e}]
\end{equation}
where $T_{i}^{e}$ is the environment contribution, meaning that from all
lattice sites other than $i$. It is not possible to deduce this environment
tensor rigorously (as otherwise one would have a rigorous expression of the
partition function, which is not available for most models), but its effect
can be included optimally by truncating the local tensor $T_i$ in order to
minimize the truncation error of $Z$. Specifically, a SVD of the environment
tensor yields
\begin{equation}
 T^e_{ij,kl} = \sum_{n} U^e_{ij,n} \Lambda^e_{n} V^e_{kl,n}
 \end{equation}
 and thus the partition function becomes
 \begin{eqnarray}
 Z & = & \rm{Tr} [T U^e \Lambda^e V^e] = \rm{Tr} [V^e TU^e \Lambda^e] \nonumber \\
 & = & \rm{Tr}[(\Lambda^e)^{1/2} V^e T U^e (\Lambda^e)^{1/2}].
\end{eqnarray}
 If one defines
\begin{equation}
\tilde{T}_{n_1n_2} = (\Lambda^e)^{1/2}_{n_1} V^e_{kl,n_1} T_{ij,kl} U^e_{ij,n_2}
(\Lambda^e)^{1/2}_{n_2},
\label{etildem}
\end{equation}
then the partition function is
\begin{equation}
 Z = \rm{Tr} \tilde{T}
\end{equation}
and the minimization of
its error is the same as minimizing that of $\tilde{T}$. By a further SVD,
\begin{equation}
\tilde{T} = \tilde{U} \tilde{\Lambda} \tilde{V}
\end{equation}
and the truncation may be performed according to $\tilde{\Lambda}$. By
substituting the truncated $\tilde{T}$ back into Eq.~(\ref{etildem}) one
obtains
\begin{equation}
T = V^e (\Lambda^e)^{-1/2} \tilde{T} (\Lambda^e)^{-1/2} U^e
\end{equation}
and thus the two new rank-3 tensors appearing at the first TRG iteration step
are given in a fully consistent approach by
\begin{eqnarray}
S^1 & = & V^e (\Lambda^e)^{-1/2} \tilde{U} (\tilde{\Lambda})^{1/2}, \\
S^2 & = & (\tilde{\Lambda})^{1/2} \tilde{V} (\Lambda^e)^{-1/2} U^e.
\end{eqnarray}

Once the environment tensor has been obtained, one may then deduce and
truncate the local tensor $T_i$, then follow the steps of TRG to update the
tensors in the renormalized lattice and thus complete a full cycle of SRG
iteration. Repeating this procedure leads finally to the partition function
in the thermodynamic limit, from which full thermodynamic information may
be obtained (Secs.~IV and V). The SRG method was found to improve the
precision of the free energy for the two dimensional Ising model by $2$
to $5$ orders of magnitude over the TRG result \cite{Xie2009_PRL103-160601}.
Further details of the SRG technique may be found in
Refs.~\cite{Xie2009_PRL103-160601} and \cite{Zhao2010_PRB81-174411}.

\begin{figure}[t]
\includegraphics[width=7.5cm]{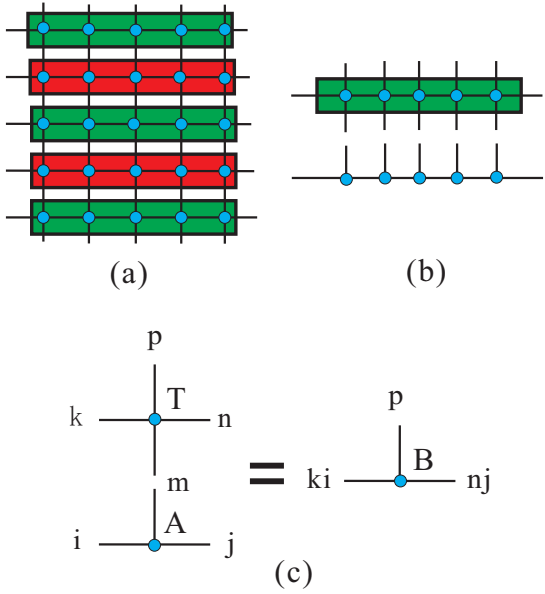}
\caption{(Color online) Schematic representation of iTEBD. (a) Tensor network
on the square lattice expressed as an infinite product of transfer matrices
in the vertical direction, with each block a transfer matrix. (b) Definition
of the transfer matrix and corresponding matrix-product state. (c) Local
action of a transfer matrix on a matrix-product state as shown in
Eq.~(\ref{local_TEBD})}
\label{tebd1}
\end{figure}

\subsection{iTEBD}

A tensor network may be regarded as an infinite product of operators, or
transfer matrices. Thus to contract the tensor network, one need only know
the dominant eigenvector of the transfer matrix, and thus the power method
can be used in the same way as in matrix algebra. This concept is the same
as using a projection method to obtain the ground state of a quantum system.
Let the local tensor (generalized transfer matrix) $T_{pkmn}$ be applied to
the random but translationally invariant matrix-product state (MPS)
$A^{m}_{ij}$, as represented in Fig.~\ref{tebd1}(a), then one obtains
a new MPS [Fig.~\ref{tebd1}(c)]
\begin{equation}
\sum\limits_{m} T_{pkmn} A_{ij}^{m} = B_{(ki),(nj)}^{p} =
B_{i^{\prime }j^{\prime }}^{m^{^{\prime }}}.
\label{local_TEBD}
\end{equation}
The dimension of the local matrix, $B$, for the new MPS is $qD$, where as in
the TRG/SRG case (Sec.~IIIB), $D$ is the maximum bond dimension that can be
retained for the MPS and a truncation is required to keep the process under
control. A unitary transformation of the new MPS places it in the canonical
form \cite{canonical_form}, which for an MPS with open boundary conditions is
the form satisfying the conditions

\begin{enumerate}
\item[1)] $\sum\limits_{^{m _{i}}} A^{m _{i}} A^{m _{i}\dagger } = I \;\; \forall
\;\; 1 \le i \le L$;

\item[2)] $\sum\limits_{^{m _{i}}} A^{m _{i}\dagger } \Lambda^{i-1}A^{m _{i}} =
\Lambda ^{i} \;\; \forall \;\; 1 \le i \le L$;

\item[3)] $\Lambda^{0} = \Lambda^{L} = 1$, with all other $\Lambda^{i}$ being
$D_i \times D_i$ diagonal matrices, which are positive, full-rank, and have
$Tr{\Lambda^{i}} = 1$.
\end{enumerate}

Here $A^{m _{i}}$ is the local matrix on site $i$. The dimensions of the first
and last matrices are respectively $1 \times D_1 $ and $ D_L \times 1 $. If the
index $m$ of the local matrix $A^{m}_{ij}$ is taken as the index of the local
basis for a quantum system, then the MPS represents the quantum state of a
one-dimensional system,
\begin{equation}
|\varphi \rangle = \sum\limits_{\{m\}} \rm{Tr} (\prod\limits_{i}A^{m_{i}}) |m_{1},m_{2},
\dots m_{L} \rangle.
\end{equation}
It can be proved that if the one-dimensional chain is cut between sites $i$
and $i+1$, the eigenvalue of the corresponding reduced density matrix is
$\Lambda^{i+1}$.

The values of $\Lambda$ in the canonical form specify the truncation of the
local matrix, which means retaining only the index corresponding to the $D$
largest $\Lambda$ matrices. This process is repeated, meaning repeated
application of the operator $T_{pkmn}$, until the MPS has converged. The
converged MPS is the approximate dominant eigenvector of the transfer
matrix. The tensor network may thus be written as the contraction of an
infinite product of these matrices and in this case the thermodynamic
quantities can be obtained directly by diagonalizing the local matrix
\cite{Vidal2007_PRL98-070201, Orus2008_PRB78-155117}.

For all of the models we study in this work (Secs.~IV-VII), the final tensor
network for the partition function is defined on the square lattice. Although
every tensor network is uniform, meaning the local tensor is the same on each
site, we use a two-sublattice MPS in all our calculations on this lattice in
order to capture any possible spontaneous breaking of symmetry.

\section{Entropy-Driven Phase Transitions}

\subsection{Diced Lattice with $q = 3$}

\begin{figure}[t]
\includegraphics[width=2.6cm]{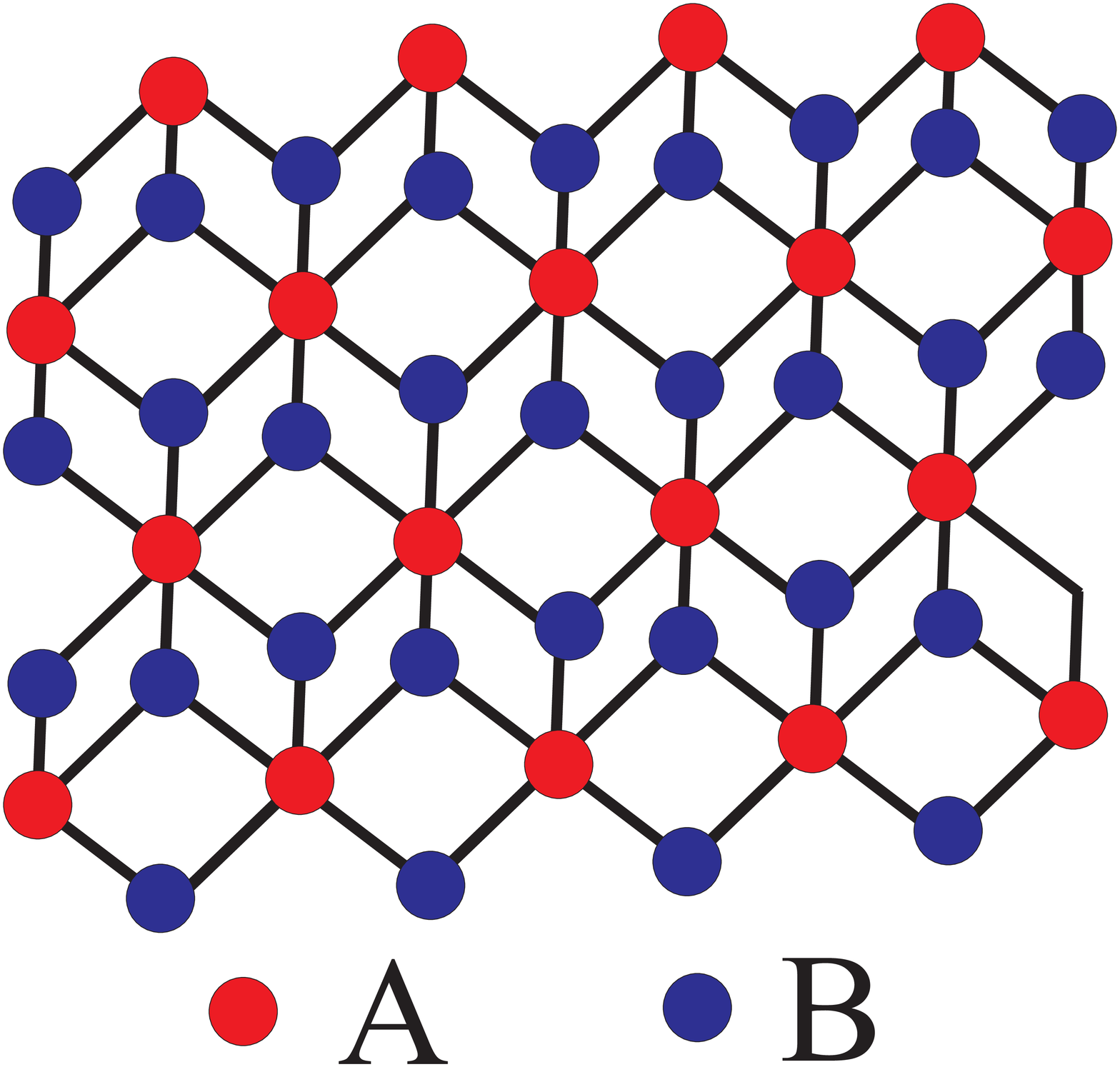}\hspace{0.2cm}
\includegraphics[width=2.6cm]{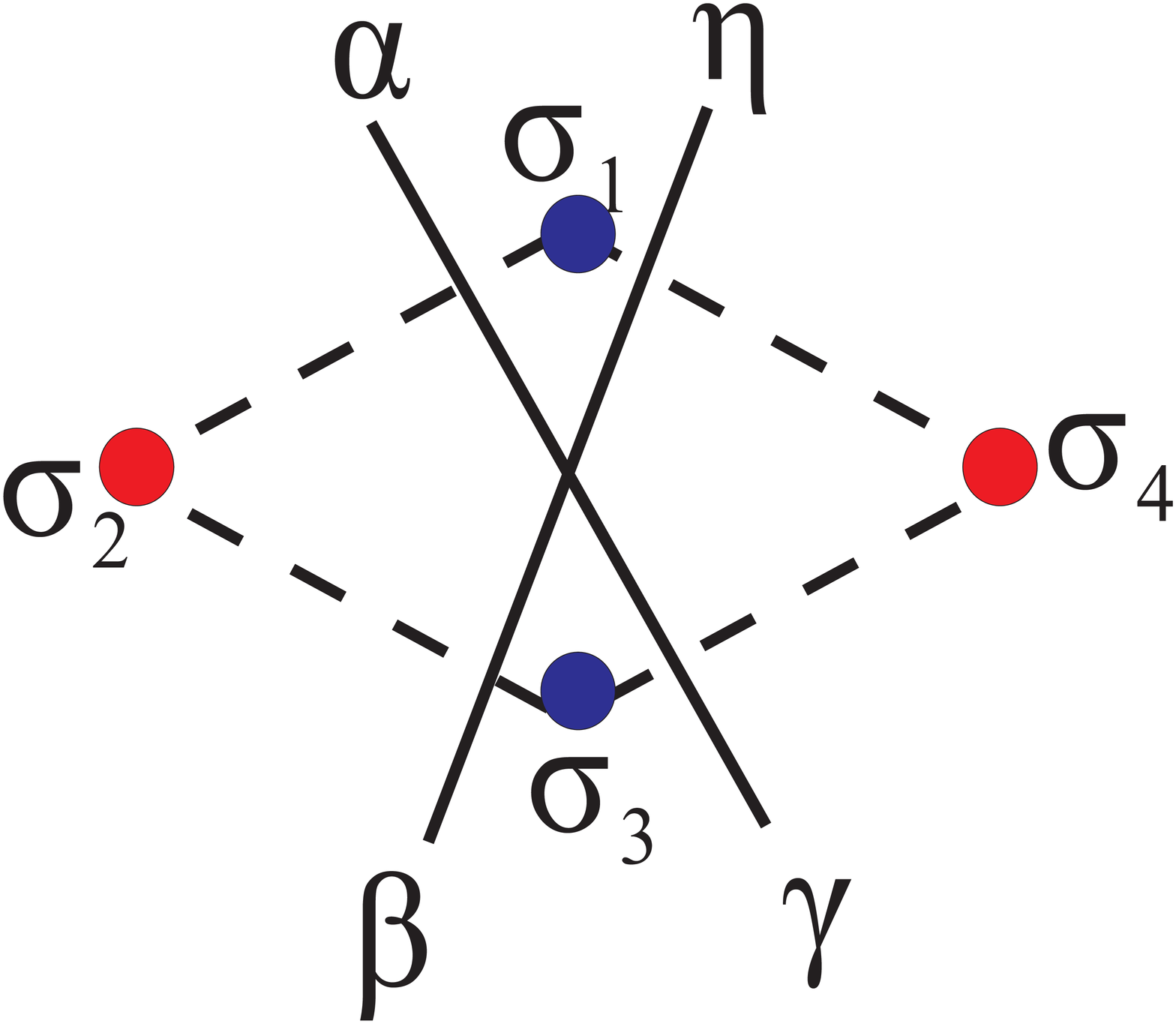}\hspace{0.2cm}
\includegraphics[width=2.6cm]{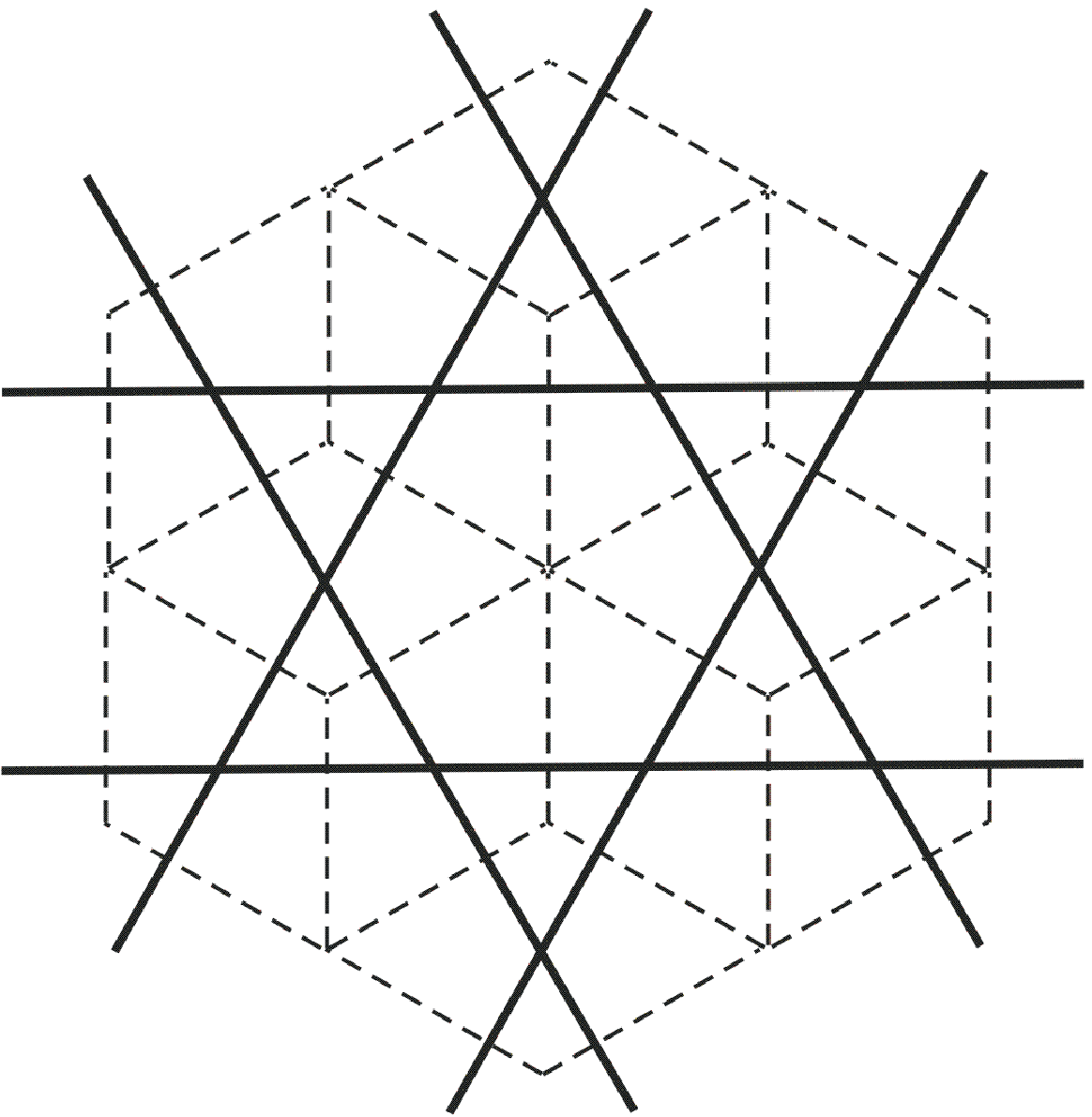}
{\centerline {(a) \quad\qquad\qquad\qquad (b)
\qquad\qquad\qquad\quad (c)}}
\caption{(Color online) (a) Diced lattice. A sites (red) have $z_A = 6$,
while B sites (blue) have $z_B = 3$. (b) Definition of tensors for each
unit cell of the diced lattice. (c) By introducing one tensor in each rhombus
of the original diced lattice, as represented in panel (b), the partition
function on the diced lattice (dashed lines) may be expressed as the
contraction of a network of tensors defined on the sites of the kagome
lattice (solid lines). }
\label{diced}
\end{figure}

The AF $q = 3$ Potts model on the diced lattice provides an excellent
example of an entropy-driven phase transition to a state of partial order,
by which is meant order on a subset of the lattice sites. Thus we begin
the presentation of both the physical ideas and our numerical results by
considering this case. The diced lattice [Fig.~\ref{diced}(a)] is dual to
the kagome lattice, and is composed of a triangular lattice of sites of
one sublattice (A) decorated by centering sites (centered in each triangle)
of the other sublattice (B). On this bipartite lattice, sites A are
sixfold-coordinated by sites B ($z_A = 6$) but sites B are only
threefold-coordinated by sites A ($z_B = 3$), whence the average
coordination number is ${\bar z} = 4$ and there are twice as many B
sites as A sites ($N_A = N_B/2 = N/3$).

With AF interactions, neighboring sites favor different Potts states
$\sigma_i$ (\ref{epm}). A three-state model on a bipartite lattice has
redundant degrees of freedom with which to ensure that every bond is
satisfied and the ground state will be highly degenerate. The two most
obvious possibilities for partially ordered configurations minimizing
the bond energy are as follows. One is that the A sites [red in
Fig.~\ref{diced}(a)] order, choosing for example $\sigma_i (i \in A)
 = 0$, leaving the B-sites (blue) to choose $\sigma_i(i \in B) = 1$ or
$\sigma_i(i \in B) = 2$ at random. The other is that the B-sites order
with the same $\sigma_i$ and the A sites are random. In both cases,
ordering occurs only on a subset of the lattice sites, but every bond
in the system can achieve its lowest energy, which is $0$. We comment that
the combined set of all these ordered configurations does not exhaust
the total possible ground-state configurations. However, these two types
of partially ordered state contribute to a very large residual entropy
in the ground state. At this point, simple physical intuition suggests
that, on lowering the temperature, the A sublattice will order, not because
these are the highly coordinated sites but because the number of states
with the B sublattice disordered is much greater and therefore the entropy
is maximized.

Of course this is the correct answer, and both the qualitative and
quantitative details are well known in the literature. It was proven by
Kotecky {\it et al.} \cite{Kotecky2008_PRL101-030601} that there is a
finite-temperature phase transition in this model, and by calculating the
sublattice magnetization using the Wang-Swendsen-Kotecky cluster algorithm
these authors confirmed the existence of long-range partial order on the A
sublattice at low temperatures.

\begin{figure}[t]
\includegraphics[width=8.0cm]{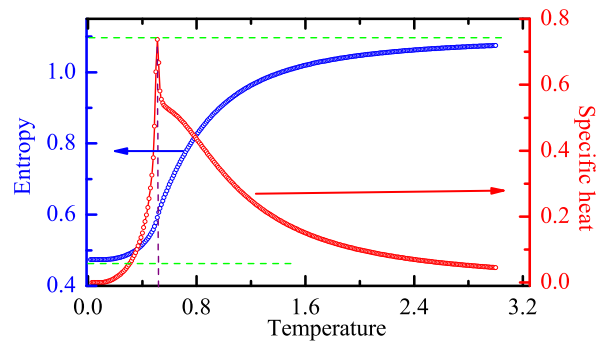}
\caption{(Color online) Entropy and specific heat for the $q = 3$
antiferromagnetic Potts model on the diced lattice. The results are
obtained by iTEBD with $D = 40$. The entropy is
shown in blue, with the green dashed line denoting its low- and
high-temperature limits. The specific heat, shown in red, has a divergence
at the phase transition. The purple dashed line denotes the critical point
obtained by Monte Carlo simulations \cite{Kotecky2008_PRL101-030601}.}
\label{dcq=3-CS}
\end{figure}

This well-understood model for partial order provides an excellent example
to benchmark our methods. Figure \ref{diced}(b) illustrates the definition
of the local tensor for this model. We first define the variable dual to
$\sigma_i$ in each rhombus as
\begin{eqnarray}
\alpha & = & \sigma_{2} - \sigma_{1} \pmod q, \nonumber \\
\beta & = & \sigma_{3} - \sigma_{2} \pmod q, \nonumber \\
\gamma & = & \sigma_{3} - \sigma_{4} \pmod q, \nonumber \\
\eta & = & \sigma_{4} - \sigma_{1} \pmod q,
\label{tensor_dual}
\end{eqnarray}
noting that these four dual variables are not independent, but are related
by the constraint
\begin{equation}
\alpha + \beta - \gamma -\eta = 0 \pmod q.
\end{equation}
The local tensor on the dual lattice is
\begin{equation}
T_{\alpha \beta \gamma \eta} = \exp [-\frac{\beta}{2} (\delta_{\alpha ,0} +
\delta_{\beta,0} + \delta_{\gamma,0} + \delta_{\eta ,0})]
\end{equation}
and defines a tensor network on the kagome lattice. This can be reconnected
to a square-lattice tensor network by SVD \cite{Zhao2010_PRB81-174411} and
the iTEBD method is used to contract the network.

The quantities required for a basic characterization of the thermodynamic
response of a system are the free energy
\begin{equation}
F = - k_{\rm B} T \ln Z,
\end{equation}
the entropy
\begin{equation}
S(T) = - \frac{\partial F }{\partial T}
\end{equation}
 and the specific heat
\begin{equation}
C(T) = - T \frac{\partial^2 F }{\partial T^2}.
\end{equation}
We have calculated these quantities,
either from $Z$ by RG methods (TRG/SRG, Sec.~IIIA) or directly by projection
methods (iTEBD, Sec.~IIIB); Fig.~\ref{dcq=3-CS} shows our results for the
entropy and the specific heat, which were published previously in
Ref.~\cite{Chen2011_Phys.Rev.Lett.107-165701}. The strong divergence
of the specific-heat curve indicates the occurrence of a second-order phase
transition. By analyzing the thermodynamic quantites alone, we obtained a
transition temperature $T_c/J = 0.508(1)$; however, a detailed consideration
of the structure of the local tensor can be used to obtain a very much more
accurate estimate of $T_c$ \cite{rwxcnx}. The Monte Carlo result is $T_c/J
 = 0.507510(8)$ \cite{Kotecky2008_PRL101-030601}, a value lying within the
error bar of the thermodynamic tensor-network result and therefore validating
the method.

The entropy provides some straightforward insight into the nature of the
low-temperature phase. If the minority (A-sublattice) sites are ordered
but the majority (B-sublattice) sites are disordered with a choice of the
two remaining Potts states, the total number of states in the ground manifold
is $2^{N_{B}}$, where $N_{B} = 2N/3$ for a system of $N$ sites. The entropy per
site would therefore be $S_{\rm d}^A (0) = (2/3) \ln 2 = 0.462098$. In contrast,
if the B-sublattice is ordered, the entropy per site is only $S_{\rm d}^B(0) =
(1/3) \ln 2 = S_{\rm d}^A(0)/2$. We indeed conclude that a state of A-sublattice
order will be selected. The zero-temperature limit of the entropy we calculate
is $S_{{\rm d}, q=3} (0) = 0.473839$, which is slightly larger than the ideal
value $S_{\rm d}^A(0)$, indicating an additional minor contribution from
further spin configurations in the ground manifold where the A sites
continue to fluctuate. The ``ideal'' low- and high-temperature limits,
$S_{\rm d}^A (0) = (2/3) \ln 2 $ and $S_{\rm d} (\infty) = \ln 3$ are shown
by the green dashed lines in Fig.~\ref{dcq=3-CS}.

\begin{figure}[t]
\includegraphics[width=2.6cm]{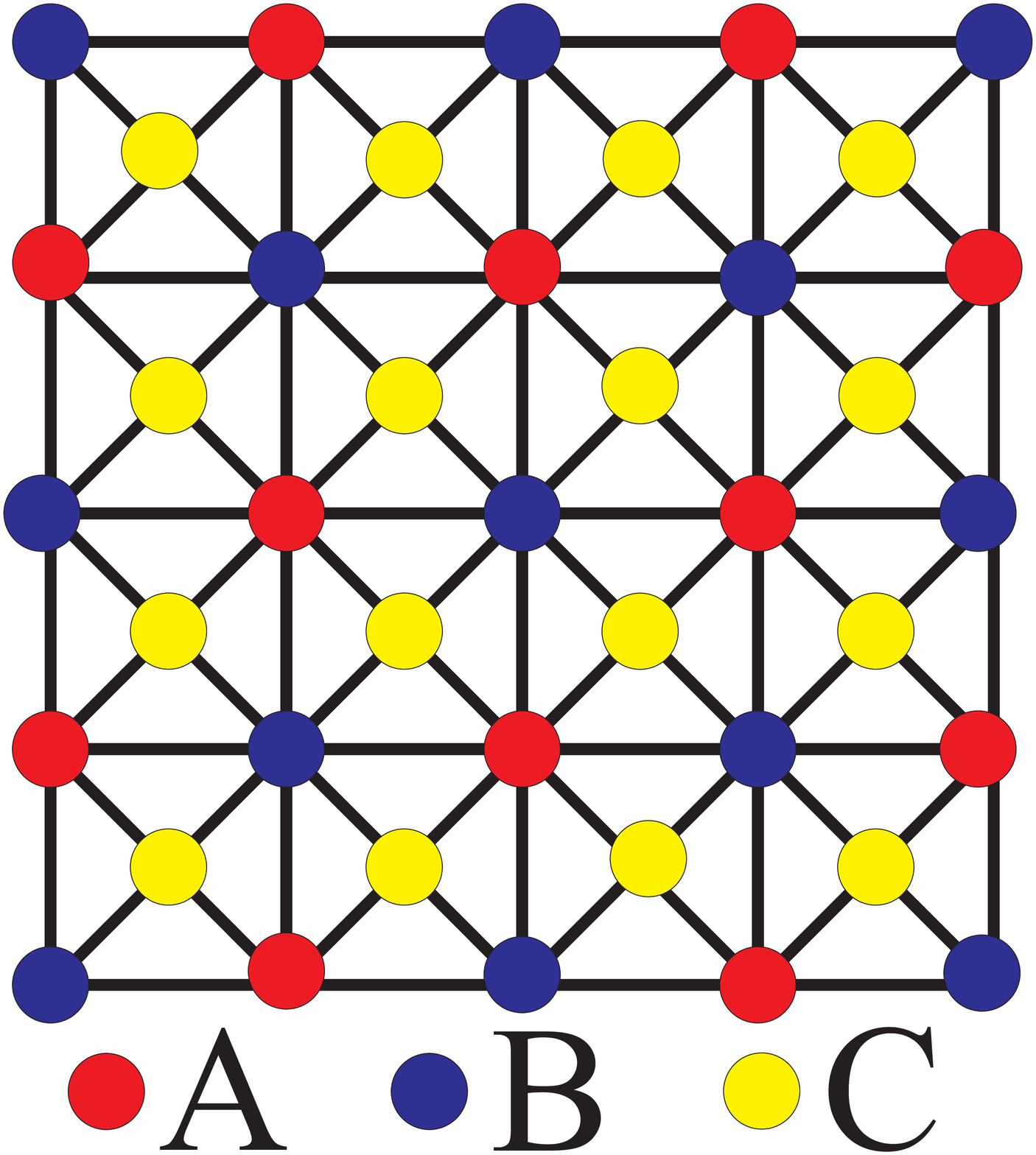}\hspace{0.2cm}
\includegraphics[width=2.6cm]{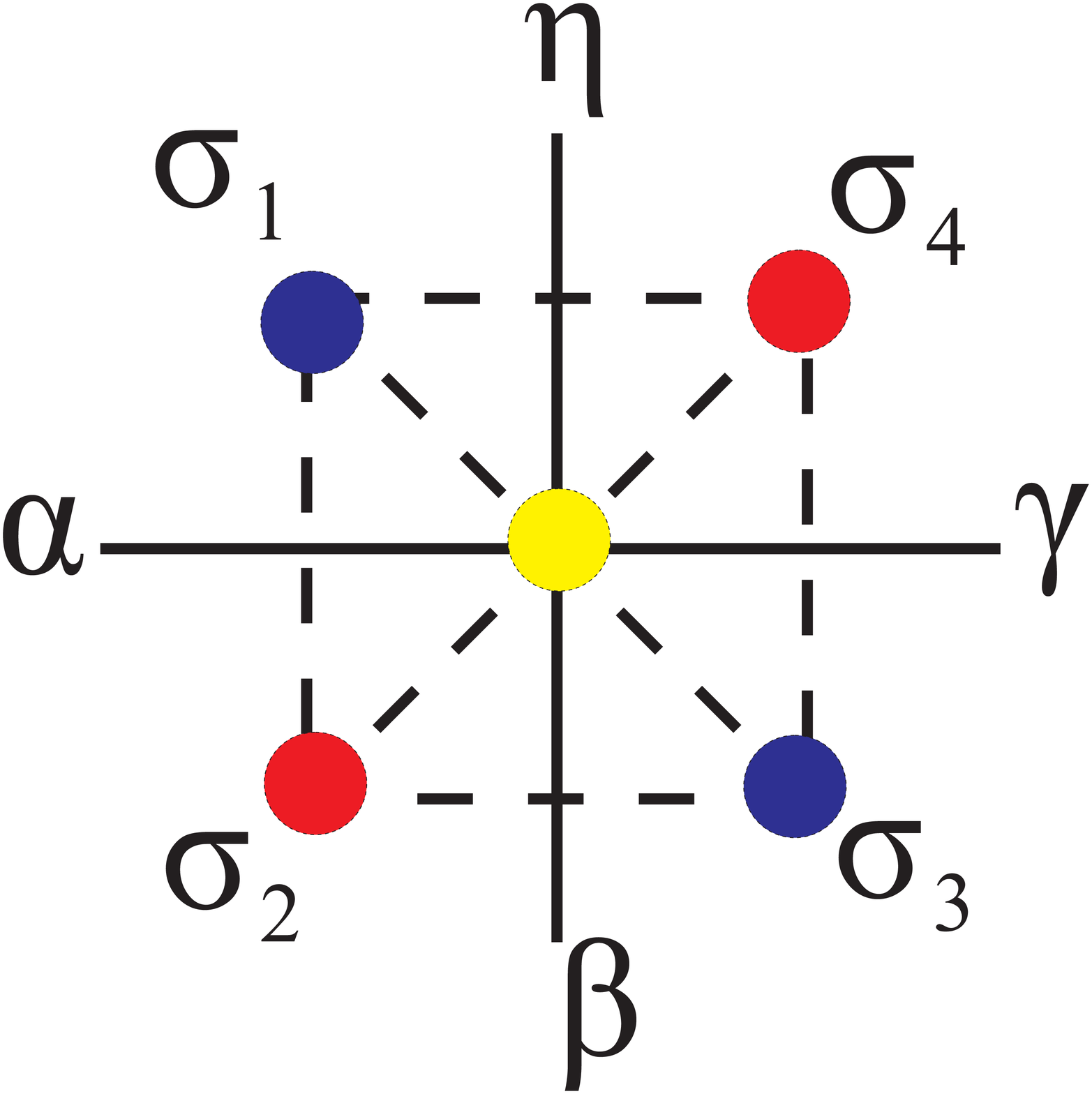}\hspace{0.2cm}
\includegraphics[width=2.6cm]{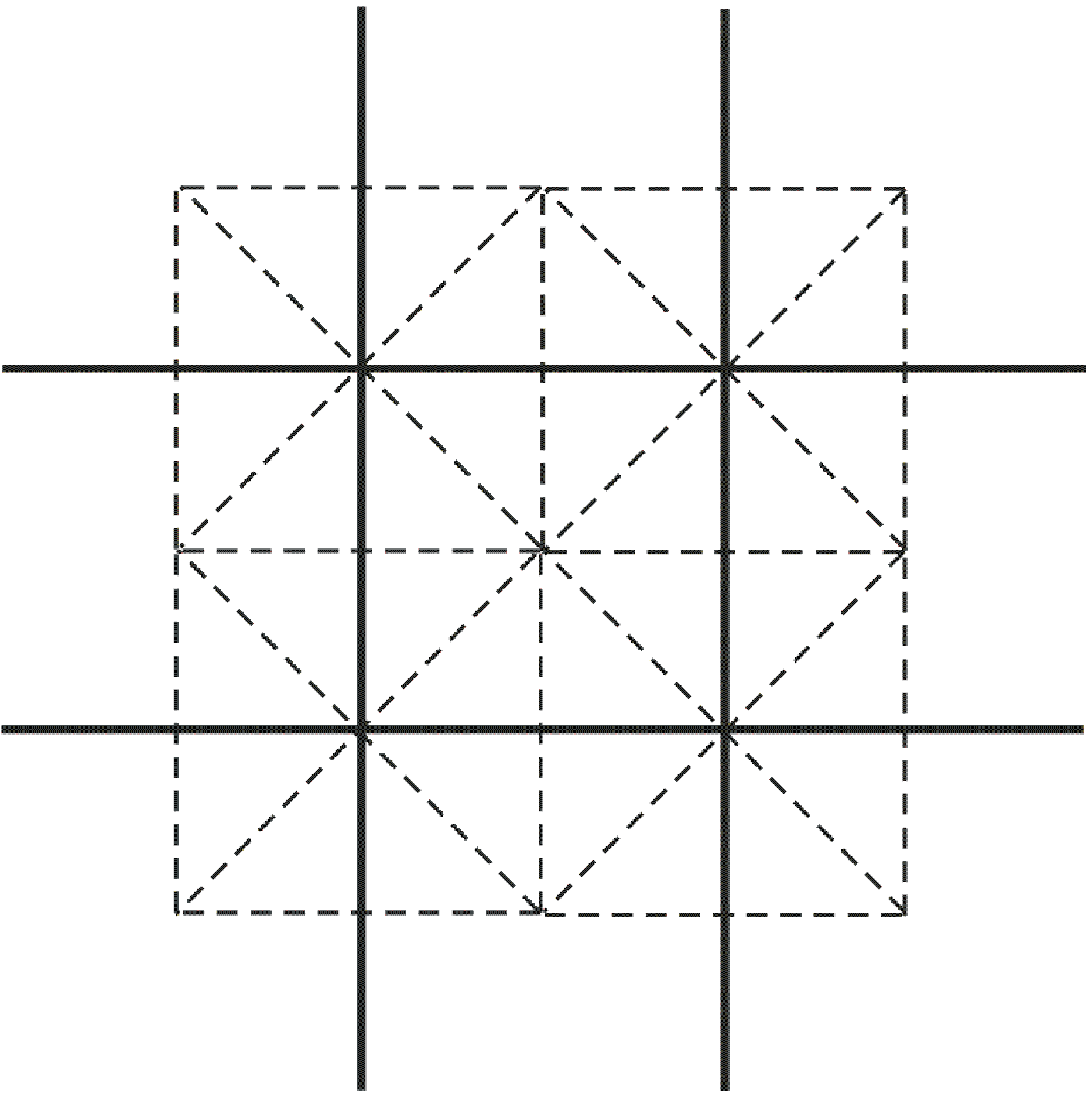}
{\centerline {(a) \quad\qquad\qquad\qquad (b)
\quad\qquad\qquad\qquad (c)}}
\caption{(Color online) (a) Union-Jack lattice. Sites in sublattices A
(red circles) and B (blue) have coordination numbers $z_A = z_B = 8$, while
those in sublattice C (yellow) have $z_C = 4$. (b) Definition of tensors for
each unit cell of the Union-Jack lattice; the center site (yellow) is denoted
$\sigma_5$ in the text. (c) By introducing one tensor in each unit cell of
the original Union-Jack lattice, as represented in panel (b), the partition
function of the Union-Jack lattice (dashed lines) is expressed after summation
over the center sites [(b) and Eq.~(\ref{T_Union_Jack})] as the contraction of
tensors defined on the sites of the square lattice (solid lines).}
\label{uj-lattice}
\end{figure}

\subsection{Union-Jack Lattice with $q = 4$}

A considerably more challenging case of partial ordering is found in
the Union-Jack lattice. This is a square lattice with additional center
sites in each square, shown in Fig.~\ref{uj-lattice}(a). Sites in the
two sublattices of the square lattice are each eightfold-coordinated, $z_A
 = z_B = 8$, while those on the centers have $z_C = 4$; because there are
twice as many C sites as A or B sites, $N_A = N_B = N_C/2 = N/4$, the system
has an integral average coordination, ${\bar z} = 6$. One may therefore expect
some comparison with the triangular lattice, where $z = 6$ and $q_c = 4$,
making (Sec.~II) the 4-state Potts model on the triangular lattice critical
at $T = 0$.

To consider the possibility of partially ordered states minimizing the
bond energy, we begin with one square unit cell. After assigning a Potts
state ${\sigma_i}$ to the center site, there are three other states for the
four corner sites, and thus at least one of the diagonal pairs must be in
the same state. This motivates the possibility of long-range order on just
one of the A or B sublattices, which could also be anticipated from the
previous subsection.

To determine the local tensor in a tensor-network formulation, we first
define the variable dual to $\sigma_i$ in the same way for the diced lattice
in Eq.~(\ref{tensor_dual}). The most straightforward way to proceed is to
trace out the Potts variable $\sigma_5$ in the middle of the square by
introducing a temporary variable
\begin{equation}
\theta = \sigma_{1} - \sigma_{5} \pmod q,
\end{equation}
in terms of which the local tensor is
\begin{eqnarray}
T_{\alpha \beta \gamma \eta } & = & e^{-\frac{\beta }{2}(\delta_{\alpha,0} + \delta_{\beta,0}
 + \delta_{\gamma,0} + \delta_{\eta,0})} \nonumber \\
& & \times \sum_{\theta }e^{-\beta (\delta_{\theta,0} + \delta_{\theta + \alpha,0}
 + \delta _{\theta + \alpha +\beta,0} + \delta_{\theta + \eta,0})}.
\label{T_Union_Jack}
\end{eqnarray}
The resulting square-lattice tensor network is then handled optimally by the
iTEBD method.

\begin{figure}[t]
\includegraphics[width=8.0cm]{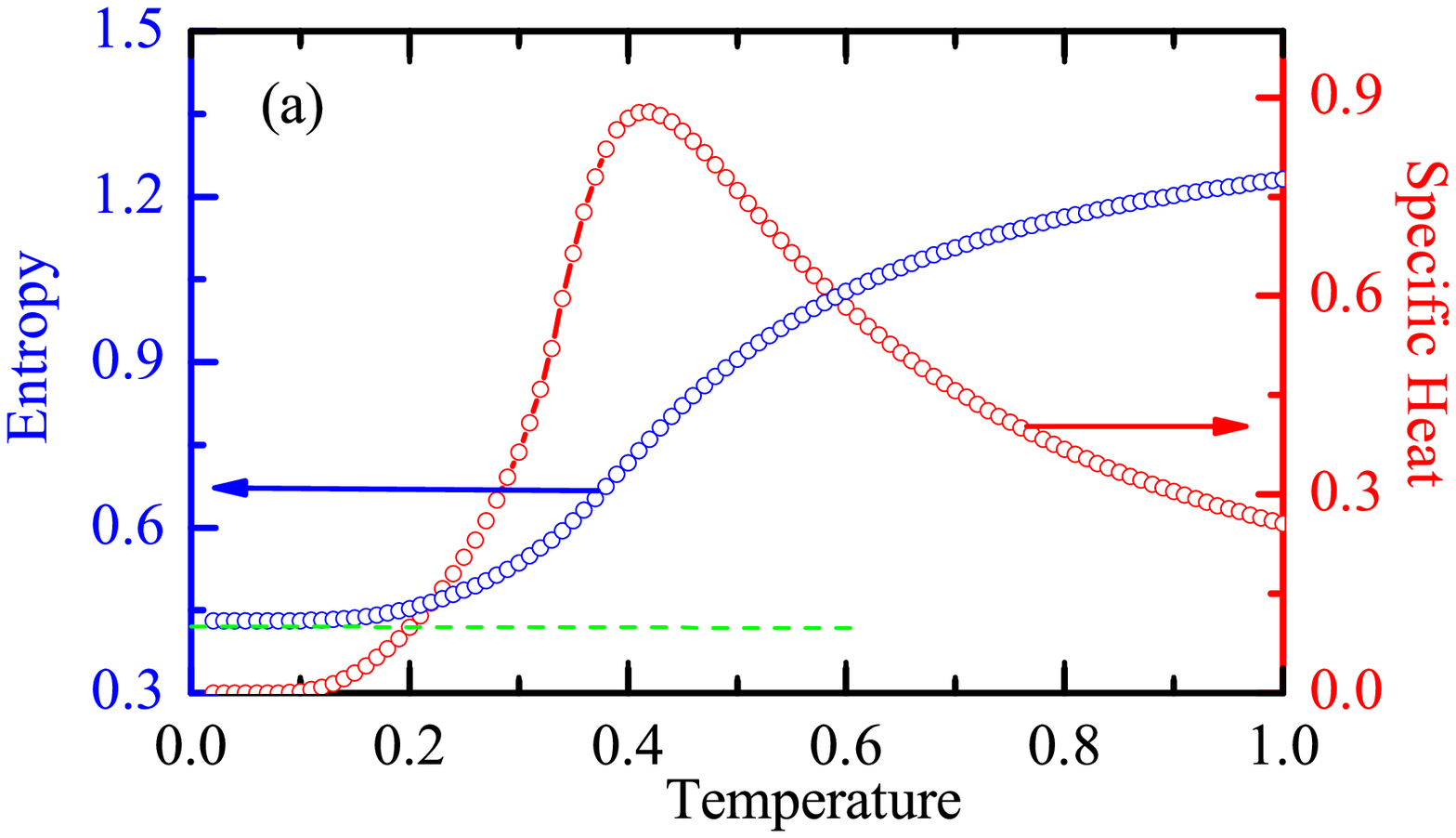}
\includegraphics[width=8.0cm]{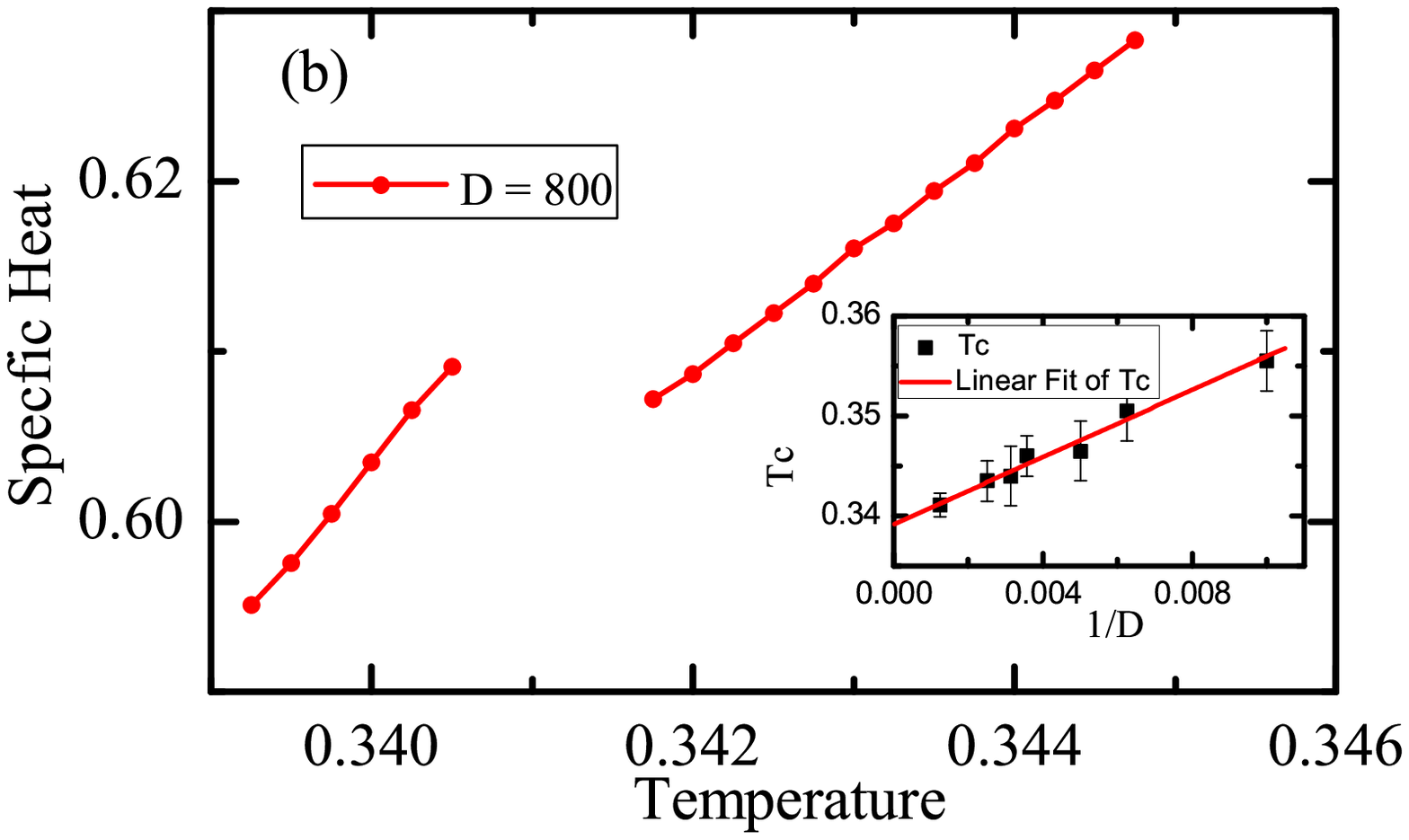}
\caption{(Color online) (a) Entropy (blue) and specific heat (red) of the
AF $q = 4$ Potts model on the Union-Jack lattice. The dashed green line is
the entropy derived from the $q = 3$ Potts model on the decorated square
lattice, which is relevant (see text) if either the A or the B sublattice is
ordered. (b) Detail of the specific heat, showing a subtle discontinuity
at the transition to partial order. Inset: scaling of the specific-heat
discontinuity as a function of tensor dimension $D$.}
\label{ujsc}
\end{figure}

The presence of partial order is indicated by a phase transition. The entropy
and especially the specific-heat curves illustrated in Fig.~\ref{ujsc}(a)
show no apparent discontinuities, and could on cursory inspection be taken
as a sign that the model is at best critical, with $q_c = 4$. However, a
sufficiently detailed investigation of the specific heat, shown in
Fig.~\ref{ujsc}(b), reveals that it is in fact discontinous, and this
was one of the key results of Ref.~\cite{Chen2011_Phys.Rev.Lett.107-165701}.
Very sophisticated calculations were required, by two different tensor-based
approaches (Sec.~III) and using a systematic increase in the tensor dimension,
which in Fig.~\ref{ujsc}(b) we also denote by $D$, to extract of the behavior
of this feature. We were able to conclude that the discontinuity does remain
finite on extrapolating $D \rightarrow \infty$, and that the partial ordering
transition occurs at a temperature $T_c/J = 0.339(1)$. The immediate question
is why this transition should be so weak while that in the diced lattice is
so strong. The immediate answer is that this should be evidence of a strong
competition between candidate partially ordered states, for example those
with only A-sublattice order and those with only B-sublattice order, and
that this competition almost prevents the system from ordering at all.

To examine the partially ordered ground state in more detail, we begin by
considering the entropy. For simultaneous order on both A and B sublattices,
for example $\sigma_i = 0$ on A and $\sigma_i = 1$ on B, sites on sublattice
C may choose $\sigma_i = 2$ or $3$ at random to satisfy every bond. Because
there are $2^{N/2}$ such states, one would expect to find $S_{AB}(0) = (1/2)
\ln 2 = 0.346573$. Our numerical result for the zero-temperature entropy is
very much larger, $S_{\rm UJ} (0) = 0.43097359$, indicating that configurations
with both A- and B-sublattice order do not play a significant role in the
ground state. If only A sublattice sites are ordered in one state, then the
B and C sites form a decorated square lattice with three remaining degrees
of freedom. These sites are clearly highly energetically correlated, but if
this hypothesis for the ground state is relevant then their behavior should
be given by that of the AF $q = 3$ Potts model on the decorated square
lattice. If the zero-temperature entropy of this model is expressed as
$S_{{\rm DS}, q=3}(0) = \ln \zeta$, the requirement for the ground state to be
dominated by configurations with partial order only on a single sublattice
is that $\zeta^{3N/4} > 2^{N/2}$, or $\ln \zeta > (2/3) \ln 2 = 0.462098$.
Our tensor-network calculation \cite{Chen2011_Phys.Rev.Lett.107-165701}
gives the result $S_{{\rm DS},q=3}(0) = 0.56106936$ and thus the condition is
clearly satisfied, meaning that partial order appears only on one of the
sublattices (A or B). Continuing with the approximation of a decorated
square lattice, one would expect to find that $S_{\rm UJ} (0) = 3
S_{{\rm DS}, q=3}(0) /4 = 0.420802$. The deviation between this value and the
exact numerical result above quantifies the contributions to the ground
manifold of configurations where neither sublattice A nor B is ordered.

The qualitative knowledge that a highly degenerate ordered state exists for
the $q = 4$ Potts model on the Union-Jack lattice has immediate connections
to a number of other problems in statistical physics. Because the fundamental
unit of the Union-Jack lattice is a triangle, there is a mapping between
the 4-state Potts model on this lattice and the 3-bond coloring problem
on its dual lattice \cite{triangular_qc}, which is the 4-8 lattice [marked
as (4,$8^2$) in Fig.~\ref{Archimedean}]. If the four states $\sigma_i = 0,
1, 2, 3$ are represented by the vertices of a tetrahedron, then three
different colors are required to mark the inequivalent pairs of edges.
At zero temperature, every triangle of the Union-Jack lattice must take
one of the configurations of the faces of this tetrahedron, with no two
bonds of the same color touching. After the dual transformation, the bonds
sharing the same vertex on the 4-8 lattice are always of different colors,
and the manifold of solutions to the 3-bond coloring problem is established.
The total number of configurations for the ground state on the 4-8 lattice,
$W_{4-8}^{N_{4-8}}$, has been calculated exactly by mapping the bond coloring
problem further to a solved model on the square lattice
\cite{Fjaerestad2010_J.Stat.Mech.2010-P01004}. The result is $W_{4-8} =
1.24048$, and because $N_{4-8} = 2 N_{\rm UJ}$, one may deduce that $S_{\rm UJ}
(0) = 2 \ln W_{4-8} = 0.430997$, which coincides to two parts in $10^{-5}$
with the result we obtain numerically.

Further, the bond-coloring problem on the 4-8 lattice is equivalent
to the fully-packed loop (FPL) model on the same lattice. FPL models
consider all configurations of non-crossing closed loops that may be drawn
along the edges of the lattice, with every vertex visited by one loop. A
loop covering on the 4-8 lattice may be derived from a 3-bond (red, blue,
green) coloring by drawing loops on those edges which are red or blue, but
not on green edges. Thus every vertex will be visited by a loop, no loops
may touch, and because each vertex has two red or blue edges then all loops
are closed. The correspondence on the 4-8 lattice between fully-packed-loop
and 3-bond-coloring models is well established
\cite{Fjaerestad2010_J.Stat.Mech.2010-P01004}. The partition function
of a FPL model is
\begin{equation}
Z = \sum_{G} n^{N_{L}},
\label{ezfpl}
\end{equation}
where the fugacity $n$ is the weight of every loop, $N_{L}$ is the number of
loops, and the sum is over all configurations $G$ of loops. Because the edges
of each loop may be 'red-blue-red-blue' or 'blue-red-blue-red,' the fugacity
is $n = 2$. The $n = 2$ FPL model is known \cite{Jacobsen_loop} to
be critical on both the square and the honeycomb lattices, but not on the
4-8 lattice, which is completely consistent with the existence of partial
order on the 4-8 lattice.

Finally, if a vertex is placed at the midpoint of every edge of the 4-8
lattice, and all vertices on neighboring bonds are connected, one obtains
the square-kagome lattice \cite{triangular_qc}, a non-Laves lattice
composed of triangles, squares, and octagons. This is a bond-to-site
transformation, and so the 3-bond coloring model on the 4-8 lattice is
equivalent to the 3-vertex coloring model on the square-kagome lattice.
Thus at zero temperature the AF $q = 3$ Potts model on the square-kagome
lattice is equivalent to the AF $q = 4$ Potts model on the Union-Jack
lattice.

\subsection{Centered Diced Lattice with $q = 4$ and 5}

\begin{figure}[t]
\includegraphics[width=2.6cm]{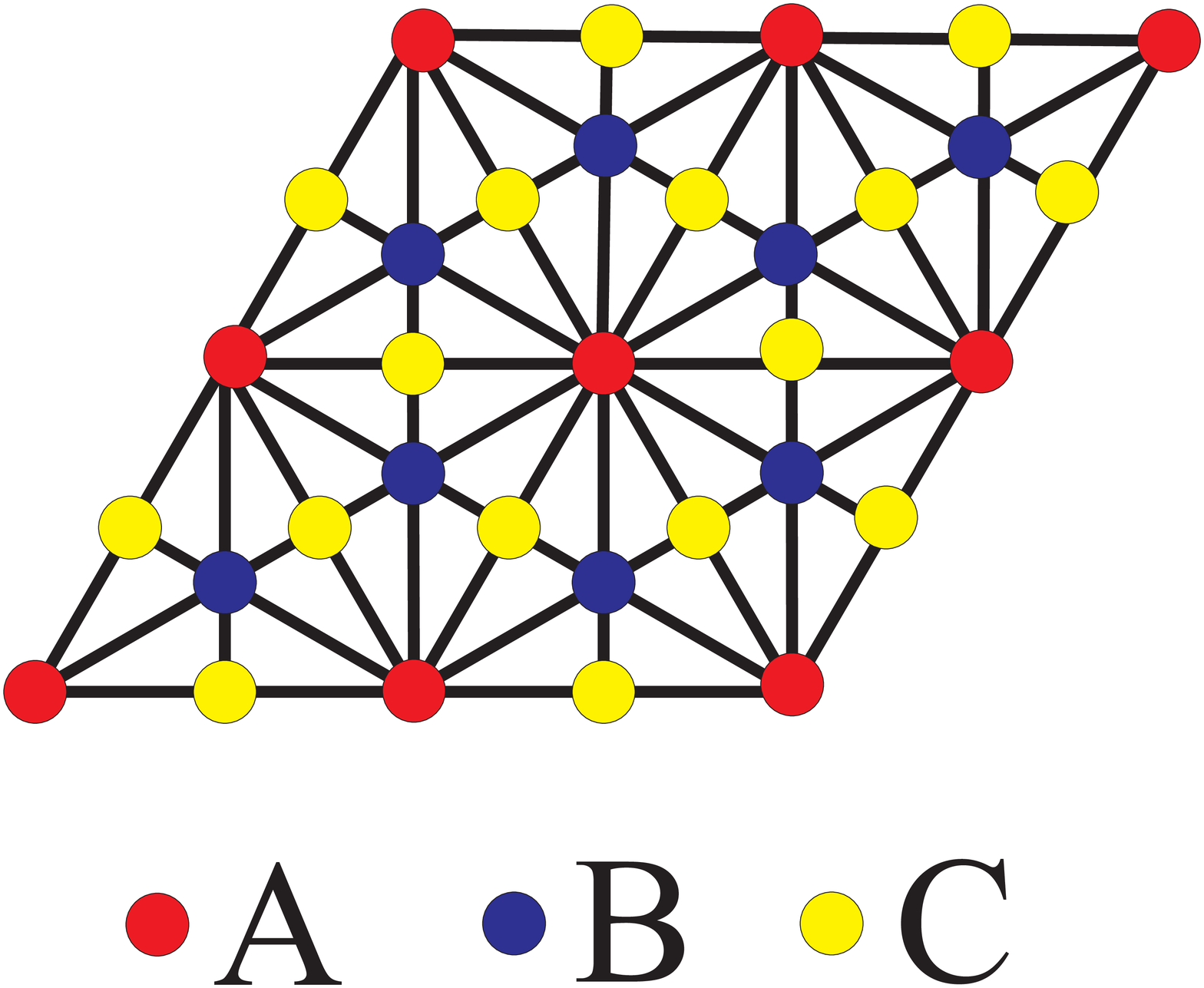}\hspace{0.2cm}
\includegraphics[width=2.6cm]{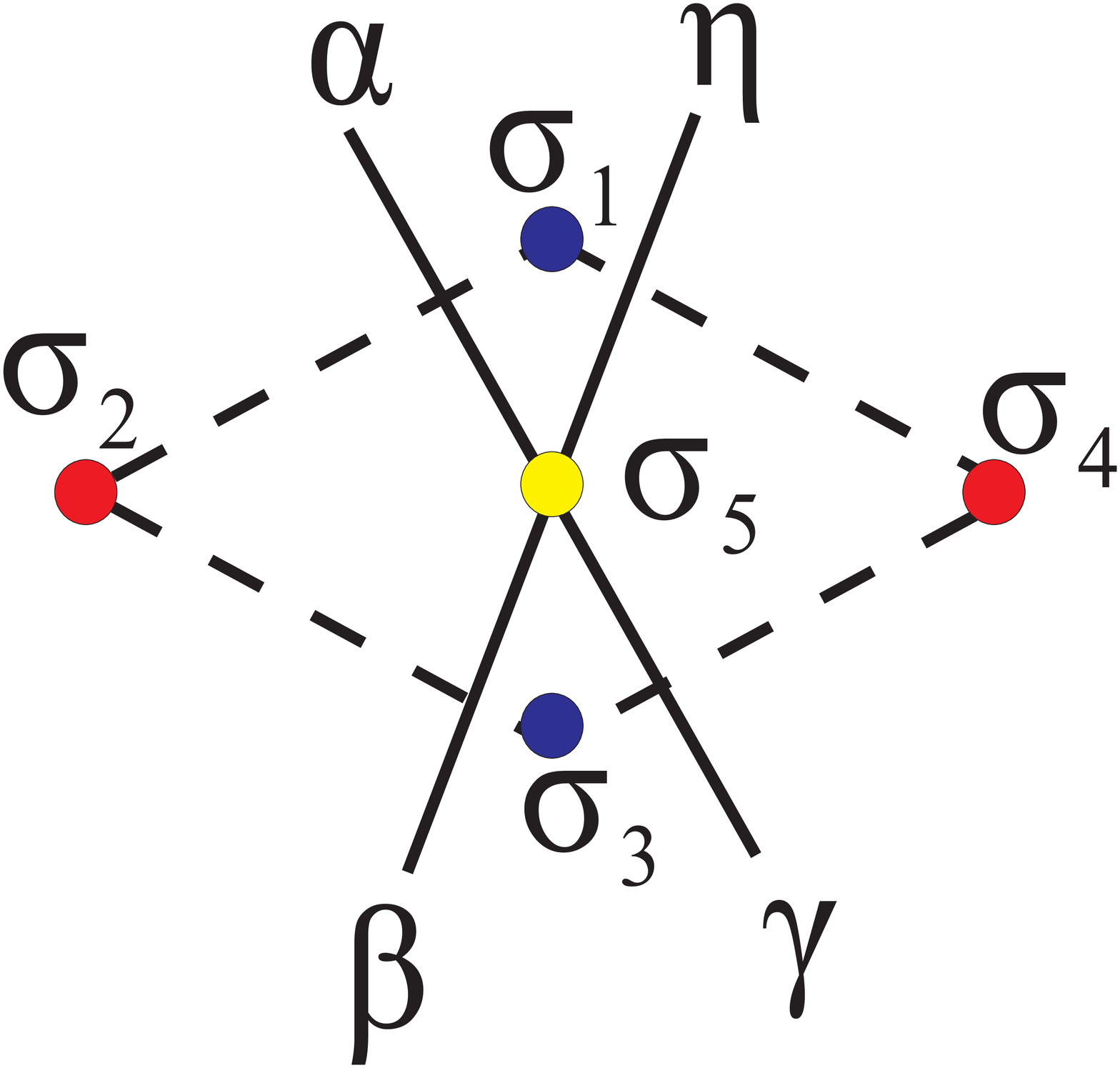}\hspace{0.2cm}
\includegraphics[width=2.6cm]{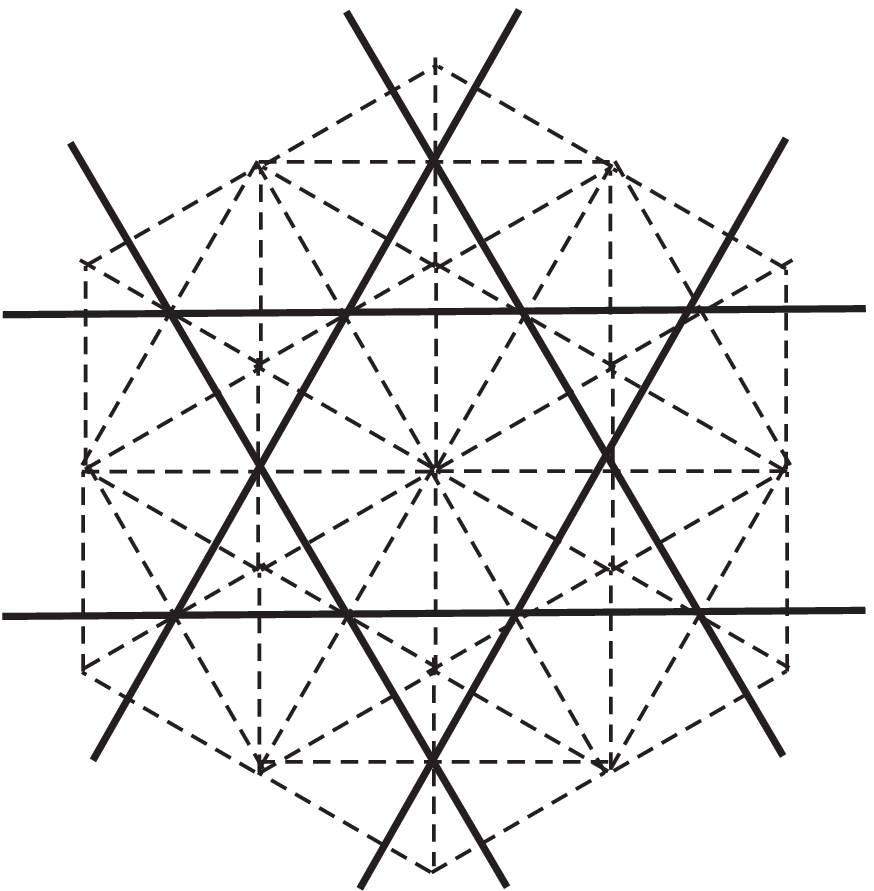}
{\centerline {(a) \quad\qquad\qquad\qquad (b)
\quad\qquad\qquad\qquad (c)}}
\caption{(Color online) (a) Centered diced lattice: $z_A = 12$ (red),
$z_B = 6$ (blue), and $z_C = 4$ (yellow). (b) Definition of tensors for
each unit cell of the centered diced lattice. (c) By introducing one tensor
in each unit cell of the original centered diced lattice, as represented in
panel (b), the partition function on the centered diced lattice (dashed lines)
may be expressed as the contraction of tensors defined on each site of the
kagome lattice (solid lines).}
\label{cd-lattice}
\end{figure}

If an extra site is added to the center of each rhombus in the diced lattice
and is connected to all its neighbors, one obtains the centered-diced
lattice, also known as the bisected-hexagonal or $D(4,6,12)$ lattice
[Fig.~\ref{cd-lattice}(a)]. Like the Union-Jack lattice, the centered
diced lattice is tripartite, is composed entirely of triangles and has
${\bar z} = 6$, with sublattice site coordinations $z_A = 12$, $z_B = 6$,
and $z_C = 4$ and site numbers $N_A = N_B/2 = N_C/3 = N/6$.

From the intuition developed for irregular lattices in the preceding
subsections, one expects that a $q = 4$ Potts model on this lattice will
show an ordering transition to a state of partial order on the
highly-coordinated A sublattice. The lack of competition between
different ordering configurations suggests that the transition should
be rather robust, more similar to that in the diced lattice than to the
Union-Jack case. Indeed we presented these considerations as predictions
in Ref.~\cite{Chen2011_Phys.Rev.Lett.107-165701} and we provide the
complete quantitative details here.

By working with a centered four-site unit cell, the local tensor for the
centered diced lattice is same on the Union-Jack lattice (\ref{T_Union_Jack}),
with the difference appearing only in the topology of the tensor network. By
an SVD transformation of the same type as that made in our treatment of the
diced lattice \cite{Zhao2010_PRB81-174411}, we obtain a square-lattice tensor
network and compute the thermodynamic quantities by the iTEBD method.

\begin{figure}[t]
\includegraphics[width=7.5cm]{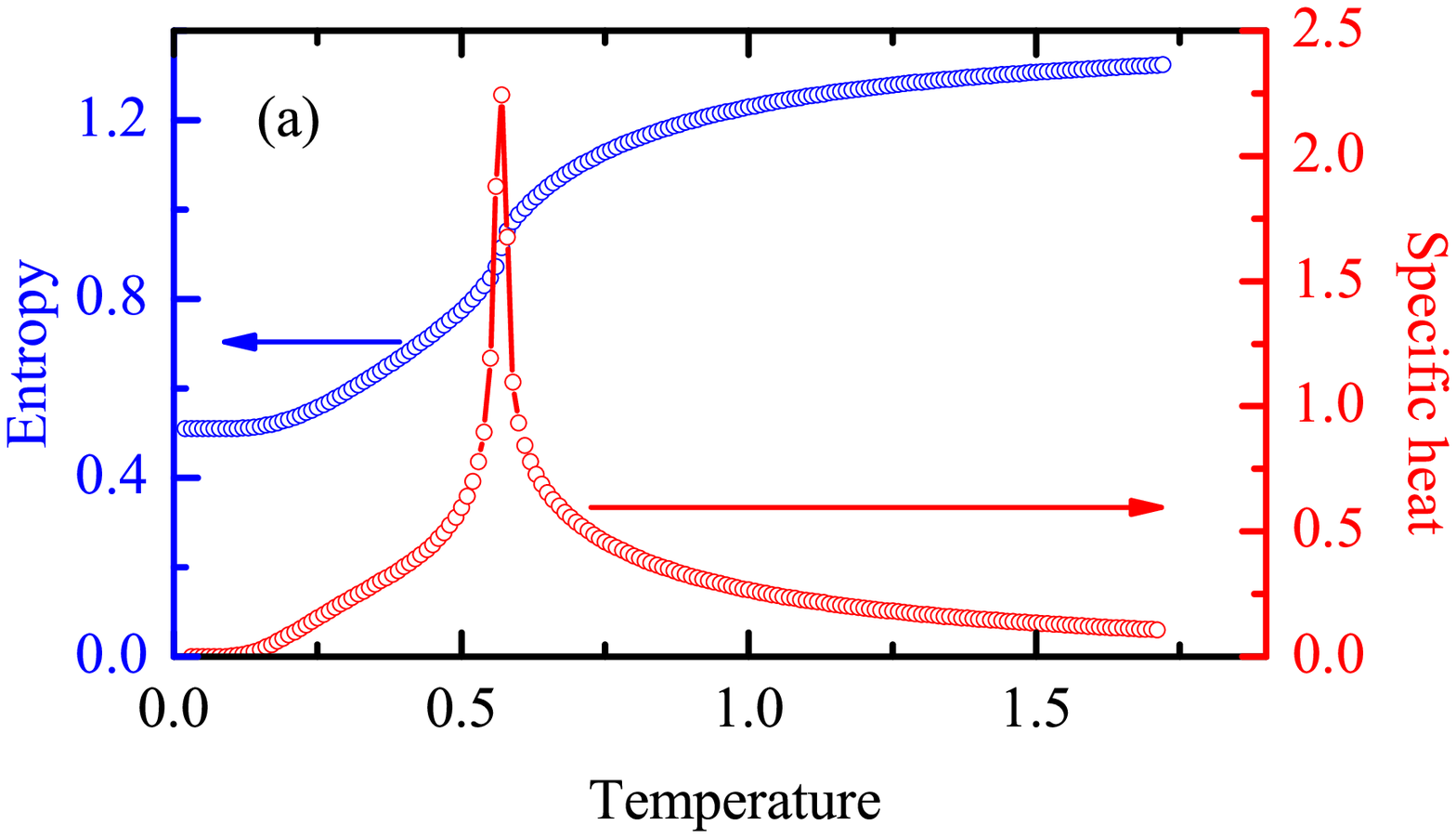}
\includegraphics[width=7.5cm]{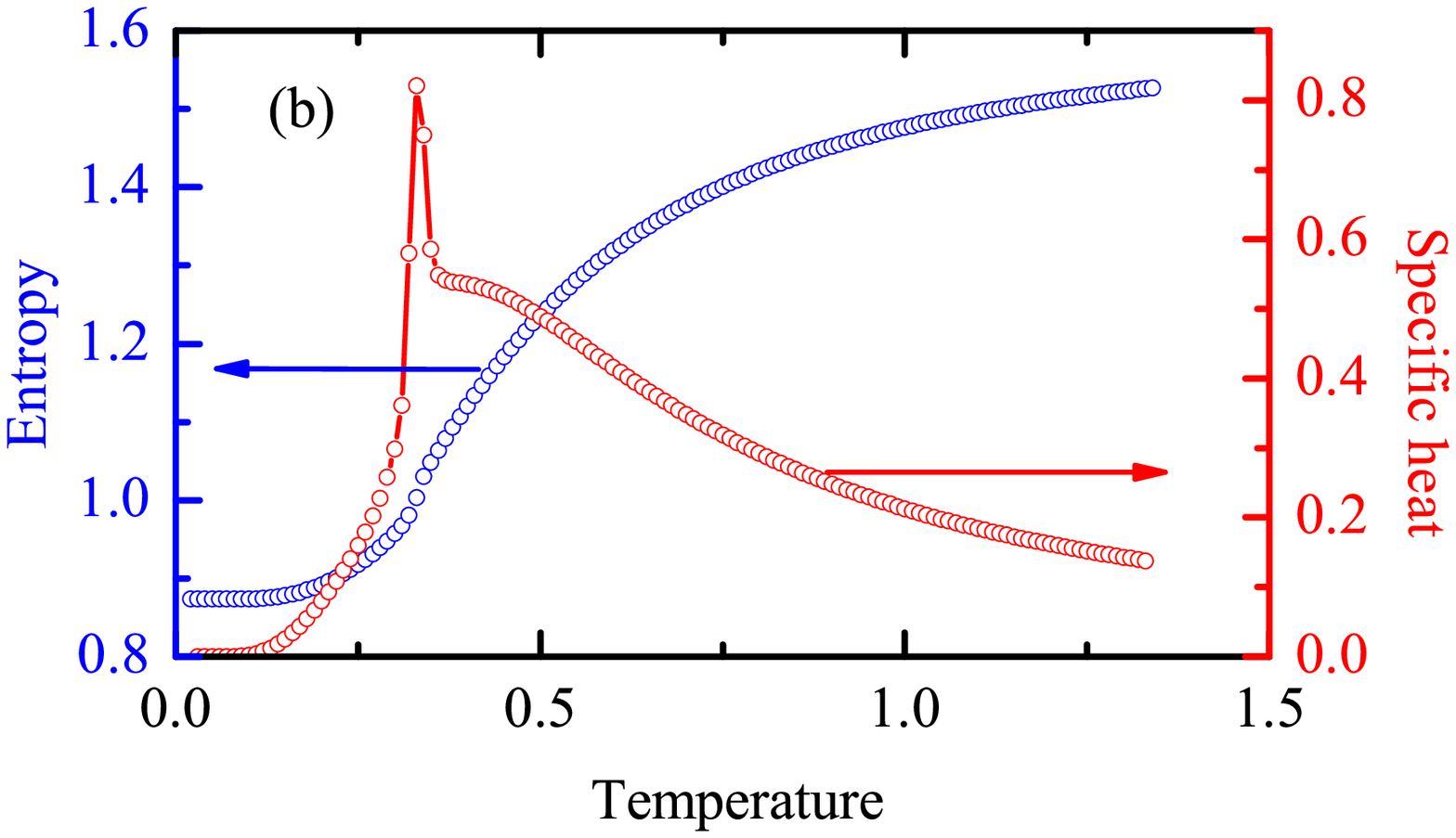}
\caption{(Color online) Entropy and specific-heat of the isotropic AF Potts
model on the centered diced lattice. The results are obtained by iTEBD with
$D = 40$. (a) $q = 4$. (b) $q = 5$. }
\label{cdesh}
\end{figure}

The entropy and specific heat of the $q = 4$ case are shown in
Fig.~\ref{cdesh}(a). Indeed the specific heat demonstrates the presence
of a very robust transition at $T_c/J = 0.56(1)$. Here and henceforth we
quote transition temperatures with an accuracy of $0.01 J$, because our
primary focus is the qualitative presence and nature of the transition,
but we stress that our tensor-based methods allow the value of $T_c$ to
be computed to very high accuracy if required \cite{rwxcnx}. As above,
the nature of the partially ordered state on the centered diced lattice
may also be inferred from the low-temperature limit of the entropy. If
indeed only the A sublattice is ordered, the sites of sublattices B and
C form a $q = 3$ decorated honeycomb lattice and the zero-temperature
entropies would be given by $S_{{\rm cd}, q = 4} (0) = 5 S_{{\rm dh}, q = 3} (0)
/ 6$. Our calculations give $S_{{\rm cd},q = 4} (0) = 0.510380$ and
$5 S_{{\rm cd}, q = 3} (0) / 6 = 0.510128$, demonstrating that partial
order only on the A sublattice is in fact realized very accurately
for the $q = 4$ model.

As in the case of the Union-Jack lattice, the $q = 4$ Potts model on the centered
diced lattice is also related to a number of other statistical problems.
At zero temperature, it may be mapped to the 3-bond coloring problem on the
Archimedean 4-6-12 lattice (Fig.~\ref{Archimedean}), and to an $n = 2$ FPL
model on this lattice. By similar manipulations it may also be mapped to an
$n = 3$ close-packed-loop (CPL) model on the kagome lattice, to a six-vertex
model on the kagome lattice, and to a ferromagnetic $q = 9$ Potts model at a
temperature $e^{\beta J} = 4$ on both the honeycomb and triangular lattices
\cite{Deng2011_Phys.Rev.Lett.107-150601}. All of these problems are known to
be non-critical, in agreement with our conclusion that the system has quite
robust, if partial, long-range order.

We conclude this section by extending our considerations to the $q = 5$
Potts model on the centered diced lattice, whose entropy and specific
heat are illustrated in Fig.~\ref{cdesh}(b). A finite-temperature phase
transition occurs at $T_c/J = 0.33(1)$, and its signal remains robust even
though its critical temperature is significantly lower than the $q = 4$
case. As above, the nature of the partial order may be verified by comparing
the zero-temperature entropy of the model with that of the decorated honeycomb
lattice formed by the sites of sublattices B and C, which have $q = 4$
remaining Potts states, in the event of full A-site order. We obtain
$S_{{\rm cd}, q = 5} (0) = 0.873635$ and $5 S_{{\rm dh}, q = 4} (0) / 6 = 0.867564$,
indicating a minor but discernible entropic contribution from additional
configurations minimizing the bond energy without A-site order.

To our knowledge, the centered diced lattice is the only planar lattice yet
known to have long-range order when $q = 5$, and therefore it possesses the
largest $q_c$ known in two dimensions. Our initial study of irregular lattices
\cite{Chen2011_Phys.Rev.Lett.107-165701} was followed up by a further analysis
of the centered diced geometry \cite{Deng2011_Phys.Rev.Lett.107-150601}, which
predicted that $q_c = 5.397(5)$.

\section{Order parameter}

The most accurate way to characterize the partially ordered state is to
determine the order parameter, which is the sublattice magnetization $M$.
Vanishing of the order parameter at the phase transition also offers
an alternative to the specific heat for determining the presence of a
transition and the exact value of $T_c$. The fact that $M = - \partial
\ln Z (H) / \partial H$ is only a first derivative of the free energy, while
the specific heat is a second derivative, makes it possible to determine
the location of the transition from $M$ using significantly larger values
of the tensor bond dimension $D$.

To calculate the sublattice magnetization most efficiently, we add a very
small field $H$ to one sublattice (which we are able to choose from the
results of Sec.~IV), as shown in Eq.~(\ref{epm}). We compute the quantity
\begin{equation}
M = \frac{1}{N_{\cal L}} \langle \sum\limits_{i \in {\cal L}} \delta_{\sigma_i,0}
\rangle - \frac{1}{q},
\label{mag_d}
\end{equation}
which is the probability of finding the $N_{\cal L}$ sites of sublattice
${\cal L}$ in the state $\sigma_i = 0$ selected by the field term. The
average value $1/q$ is subtracted to ensure that the order parameter is
zero when the system is in its high-temperature disordered phase.

\begin{figure}[t]
\includegraphics[width=7.5cm]{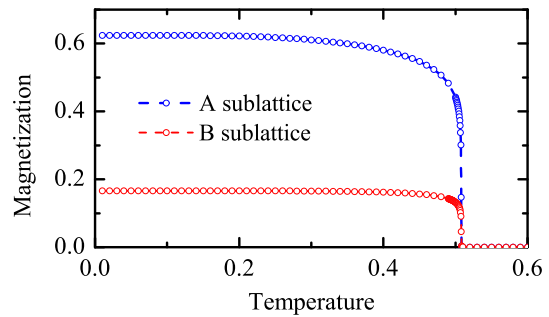}
\caption{(Color online) Magnetization of the A sublattice for the AF $q = 3$
Potts model on the diced lattice. The results are obtained by iTEBD with
$D = 40$. Shown also is the probability of sites on the B sublattice being
in one Potts state $\sigma_i$ different from that of the A sites.}
\label{dq=3-M}
\end{figure}

\subsection{Diced Lattice with $q = 3$}

Figure \ref{dq=3-M} shows the magnetization of the AF $q = 3$ Potts model
on the diced lattice. We show results not only for the probability of
finding $\sigma_i = 0$ on the A sublattice but also for the probability of
finding $\sigma_i = 1$ or 2 on the B sublattice. Both curves exhibit the
typical behavior of a second-order phase transition, with the order parameter
(determined from the A sublattice) going continuously to zero at $T_c$. The
low-temperature limiting value for the magnetization of the A sublattice
is $M_{\rm d} (0) = 0.62426$, somewhat lower than the perfect-order result
$M_{\rm d}^0 (0) = 2/3 = 1 - 1/3$, from which we deduce that the ground state
of the diced lattice retains fluctuations suppressing the partially ordered
state by approximately 6\%.

The B-sublattice quantity is neither a magnetization nor an order parameter,
but its value is $0.166202$, which is very close to the value $1/6 = 1/2 -
1/3$ obtained in the event of a completely random distribution of the B
sites between states $\sigma_i = 1$ and 2. Indeed, if 6\% of the A-sublattice
sites are not ordered with $\sigma_i = 0$, the density of B-sublattice sites
with no $\sigma_i = 0$ neighbor is of order $0.06^3$; because these contain
only limited information, we do not present calculations for the non-ordered
sublattices in the other systems discussed in this section.

Thus the entropy and the sublattice magnetization both demonstrate that the
low-temperature phase is a state of partial order in which A-sublattice sites
order in one Potts state while B-sublattice sites are disordered, choosing
the remaining two Potts states at random and contributing to the very high
ground-state entropy. The sublattice magnetization provides clearer insight
into the deduction made from the entropy in Sec.~IV that there exist
ground-state configurations where the A sublattice is not fully ordered,
and these configurations cause the departures from the ideal values observed
in our exact numerical results.

\begin{figure}[t]
\includegraphics[width=8.0 cm]{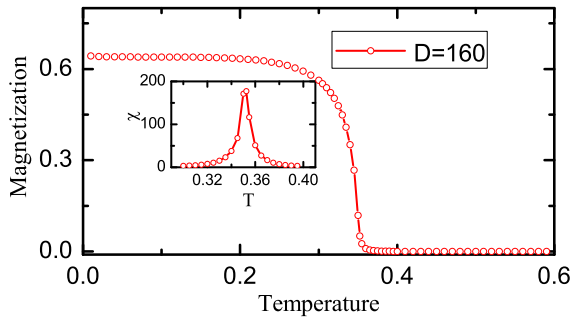}
\caption{(Color online) Magnetization of the A sublattice for the AF $q = 4$
Potts model on the Union-Jack lattice. Inset: corresponding susceptibility
$\chi = \partial M / \partial H$.}
\label{ujmk}
\end{figure}

\subsection{Union-Jack Lattice with $q = 4$}

We computed the magnetization and susceptibility of the $q = 4$ Potts model on
the Union-Jack lattice in Ref.~\onlinecite{Chen2011_Phys.Rev.Lett.107-165701},
and show the results in Fig.~\ref{ujmk} to retain the completeness of this paper.
Despite the fact that the phase transition is very difficult
to identify in the specific heat (Fig.~\ref{ujsc}), it is clearly visible
in the magnetization as a rapid but continuous drop of the sublattice
magnetization. While the tensor-network calculation of $M(T)$ can indeed
be performed with significantly higher values of $D$ than for $C(T)$, it
is largely the nature of $M$ as a true order parameter that makes it a
superior indicator of the phase transition. A further robust indicator
of the transition is the susceptibility, defined as $\chi = \partial M /
\partial H$, which diverges on approaching the transition. However, we
comment that the temperature grid used in the preparation of Fig.~\ref{ujmk}
does not allow a determination of the critical temperature $T_c/J = 0.339(1)$
more accurate than the result of the detailed analysis performed in Sec.~IV.
The low-temperature limiting value we compute for the order parameter is
$M_{\rm UJ} (0) = 0.6428$, which is some 14\% less than the ideal value $3/4
 = 1 - 1/4$. The sublattice magnetization provides a clear indication
of the discrepancy between the exact result and the state of perfect
A-sublattice order. While we are unaware of an analytical relationship
between the magnetization discrepancy and the entropy discrepancies
calculated in Sec.~IVB, our calculations allow a quantitative determination
of this connection for all lattices, and similar results will appear again
in Sec.~VII.

\begin{figure}[t]
\includegraphics[width=8.0 cm]{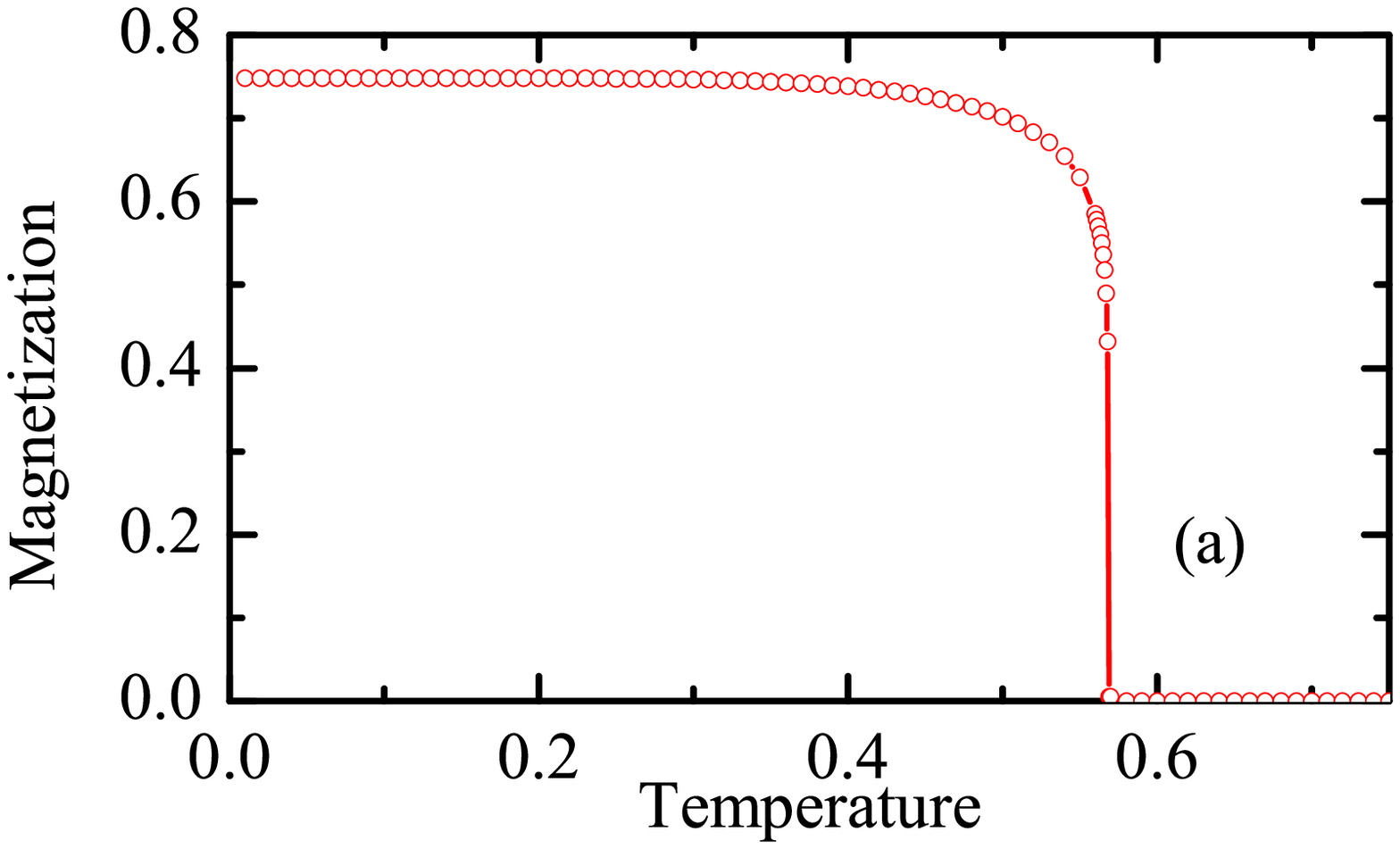}
\includegraphics[width=8.0 cm]{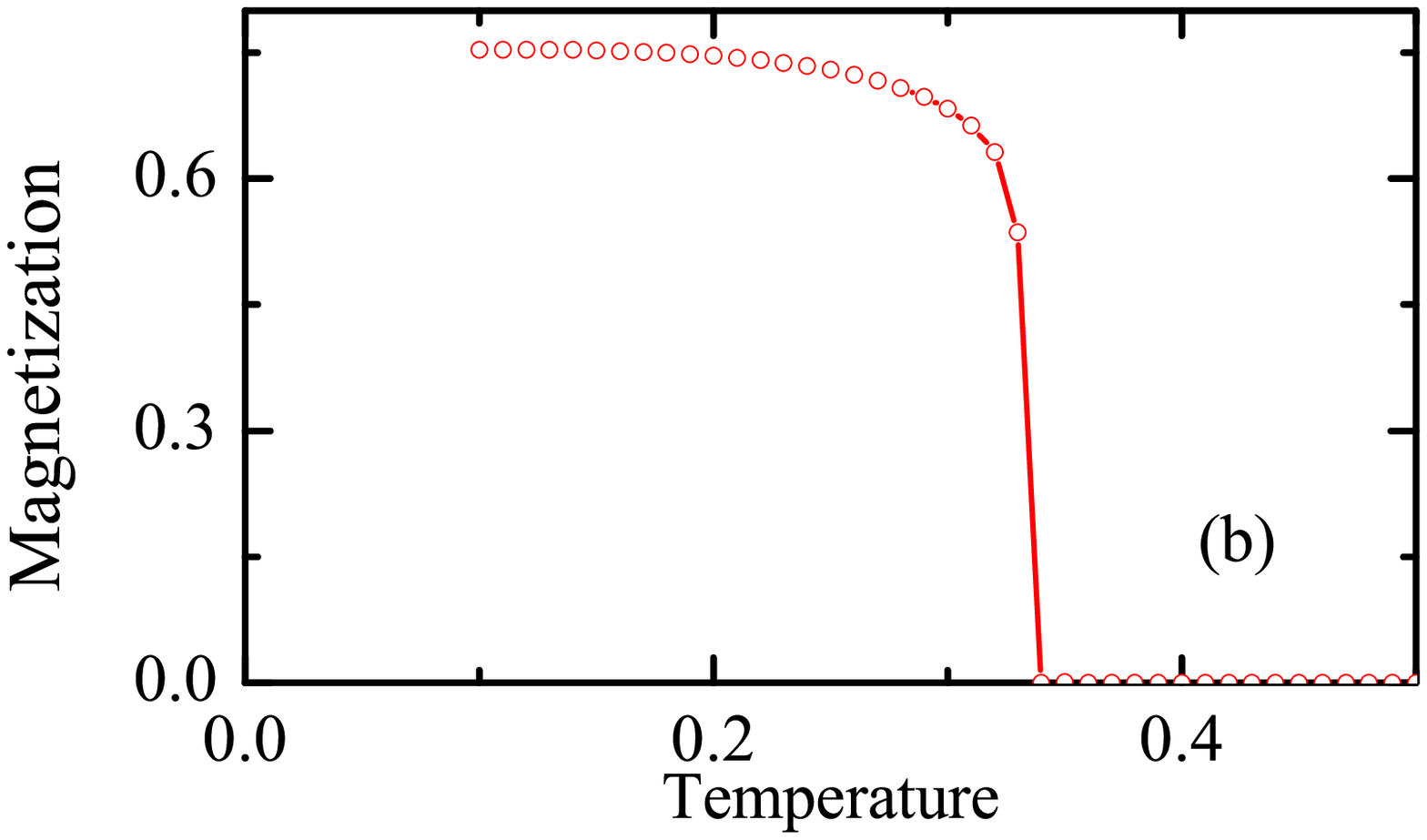}
\caption{(Color online) Magnetization of the A sublattice for the AF Potts
model on the centered diced lattice. The results are obtained by iTEBD with
$D = 30$. (a) $q = 4$. (b) $q = 5$. }
\label{cdm}
\end{figure}

\subsection{Centered Diced Lattice with $q = 4$ and 5}

We conclude this section by computing the magnetization of the A sublattice
for the centered diced lattice, which is presented in Fig.~\ref{cdm}(a) for
$q = 4$ and in Fig.~\ref{cdm}(b) for $q = 5$. Both curves show clear,
second-order phase transitions occurring respectively at critical
temperatures $T_c/J = 0.56(1)$ and 0.33(1). The low-temperature limit of
the magnetization for the $q = 4$ case is $M_{{\rm cd}, q = 4} (0) = 0.74845$,
which is very close to the ideal value $0.75 = 1 - 1/4$, as expected for a
system with the very robust one-sublattice partial order suggested by our
entropy calculations for the decorated honeycomb lattice (Sec.~IVC). For the
$q = 5$ model, we find that $M_{{\rm cd}, q = 5} (0) = 0.7540$ in comparison with
an ideal value of $0.80 = 1 - 1/5$, illustrating clearly a 5.5\% departure
from perfect order arising as a consequence of the very high degeneracy of
ground-state configurations in this high-$q$ case.

\section{Intermediate Partial Order and Multiple Phase Transitions}

In the preceding sections we have considered only models with the
same interaction $J$ between Potts variables on every pair of sites
[Eq.~(\ref{epm})], i.e. despite the inequivalent sites, the bonds have
equivalent strengths. In this section, we relax this constraint to
illustrate the phenomenon of multiple phase transitions within a single
Potts model. These can occur on a number of different lattices and for
specific $q$ values whose common feature is that states of partial order
appear at intermediate temperatures, between complete order at low
temperature and complete disorder at high temperature.

\subsection{Union-Jack Lattice with $q = 2$}

The AF $q = 2$ Potts model is equivalent to the Ising model, which has been
solved exactly on the Union-Jack lattice \cite{Vaks1966_JETPLetters22-820}.
We consider the system with inequivalent interactions, taking those between
A and B sites to have a strength $J_{AB}$ and those of C sites to both A and
B sites to have a strength $J_C$. If the sign of $J_C$ is exchanged, and at
the same time change the definition of the Potts variable $\sigma_i$ on the
C sublattice is changed to $-\sigma_i$, the model is unchanged. Thus the
phase diagram is symmetrical about $J_C = 0$ and for simplicity we consider
only ferromagnetic values of $J_C$ ($J_C < 0$). When $J_{AB}$ is ferromagnetic,
the model has very simple behavior, and in the following discussion we
consider only the case where $J_{AB}$ is AF ($J_{AB} > 0$).

\begin{figure}[t]
\includegraphics[width=6.0cm]{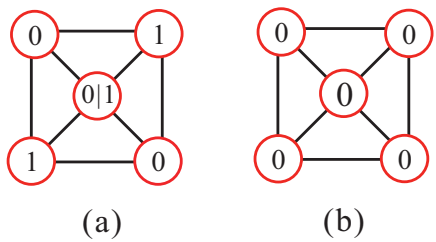}
\caption{Ground-state configurations for the $q = 2$ Potts model on a single
unit cell of the Union-Jack lattice. (a) $J_{AB} \gg |J_C|$. (b) $|J_C| \gg
J_{AB}$. }
\label{Gs_q_2}
\end{figure}

If one considers a single square unit cell, in the limit of large $J_{AB}$,
sublattices A and B will adopt an AF ordered configuration and sublattice
C will be chosen randomly, as represented in Fig.~\ref{Gs_q_2}(a), giving a
ground-state energy per unit cell of $-2 J_{AB}$. In the opposite limit
of large $|J_C|$, the minimum energy is obtained when all the sites in the
lattice are ferromagnetically ordered, as shown in Fig.~\ref{Gs_q_2}(b),
giving a ground-state energy per cell of $-4|J_{C}| + 2 J_{AB}$. By equating
the two limits, one might expect a transition to occur when $J_{AB} = |J_C|$.

\begin{figure}[t]
\includegraphics[width=7.5cm]{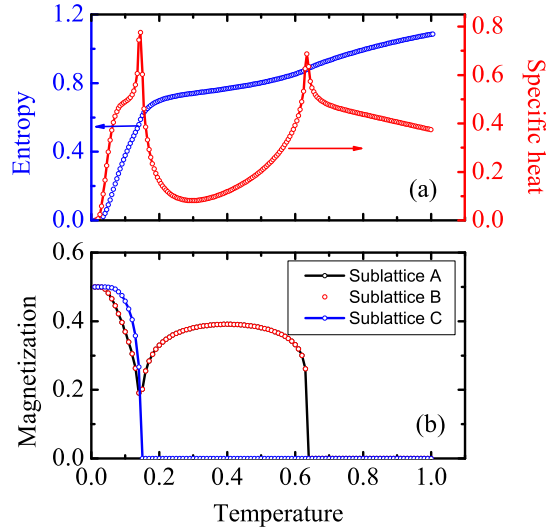}
\caption{(Color online) Thermodynamic properties of the $q = 2$ Potts model
on the Union-Jack lattice, calculated with coupling constants $J_{AB} = 1.0$
and $J_{C} = - 1.05$. The results are obtained by iTEBD with $D = 30$. (a)
Entropy and specific heat. (b) Magnetizations of the A, B, and C sublattices.}
\label{ujq=2}
\end{figure}

The local tensor for this model is defined as in Eq.~(\ref{T_Union_Jack}).
If we choose a parameter ratio far from either limit of the previous
paragraph, $J_{AB} = 1.0$, $J_C = - 1.05$, our results for the entropy,
specific heat, and sublattice magnetizations show complex behavior
(Fig.~\ref{ujq=2}). From the specific heat it appears that two phase
transitions occur, with critical temperatures of $T_{c1} = 0.145(5)$ and
$T_{c2} = 0.635(5)$. However, the exact result \cite{Vaks1966_JETPLetters22-820}
contains not two but three finite-temperature phase transitions for this
parameter ratio. On cooling from the high-temperature disordered phase,
there is a transition to an AB-sublattice AF phase with the C sublattice
disordered, then another, very narrow, phase of complete disorder, and at
low temperatures a ferromagnetic phase. The three critical temperatures
are respectively $0.6348196$, $0.1446858$, and $0.1438721$.

To interpret these results we note from above that, if $|J_C| > J_{AB}$
the ground state should be ferromagnetic on all the three sublattices
[Fig.~\ref{Gs_q_2}(b)]. However, this is a state with zero entropy, whereas
the AF configuration of the A and B sublattices [Fig.~\ref{Gs_q_2}(a)] has
a higher energy (only marginally higher for $J_C/J_{AB} = - 1.05$) but a
massive degeneracy of $2^{N/2}$ in an $N$-site lattice. Thus a moderate
temperature may stabilize this type of partially ordered configuration, with
complete C-sublattice disorder, before order is lost on all three sublattices
at higher temperatures. The most striking aspect of the phase diagram is that
all transitions are continuous (second-order), but the AB-sublattice AF
configuration is so different from the low-temperature ferromagnetic
configuration that the order parameter must vanish completely between
the two phases. The width of this regime of ``fully frustrated disorder''
is, however, so narrow that it can only be resolved in our numerical
calculations by specific targeting \cite{rwxcnx}. We note in addition
that the energy balance allowing this entropy-driven reordering to occur
is also rather delicate, arising only for coupling values $1 < |J_C/J_{AB}|
 < 1.09(1)$.

To verify the nature of the ordered and partially ordered phases, we also
calculate the sublattice magnetization and the results are presented in
Fig.~\ref{ujq=2}(b). We remind the reader that the magnetization is
computed [Eq.~(\ref{mag_d})] with an explicit assumption for the Potts
state $\sigma_i$ of each sublattice. With the assumption of low-temperature
ferromagnetism and an intermediate AF state, the numerical results confirm
the above analysis. We observe in the low-temperature state that the
frustrated AB-sublattice order is suppressed by thermal fluctuations
more rapidly than is the satisfied C-sublattice order.

The existence of multiple phase transitions in this model has been
discussed \cite{Fradkin_PRA14} in a general framework of competing
effective interactions between spin pairs arising due to paths with
different numbers of bonds. We stress that, although this is the only
model we consider in this paper with an explicit frustration, this
frustration is resolved in favor of one (fully ordered) configuration in
the ground state, which is nevertheless supplanted at finite temperatures
by an entropically driven, partially ordered state of a very different
local nature, without altering the frustration parameter.

\subsection{Union-Jack Lattice with $q = 3$}

\begin{figure}[t]
\includegraphics[width=6.0cm]{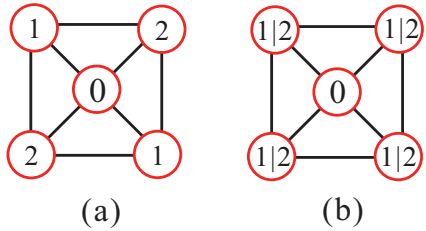}
\caption{Configurations of the AF $q = 3$ Potts model on a single unit cell
of the Union-Jack lattice. (a) Three-sublattice AF-ordered ground-state
configuration. (b) Configuration for the intermediate phase of partial
order arising for $J_C/J_{AB} > 2.2$. }
\label{Gs_q_3}
\end{figure}

We turn to the $q = 3$ Union-Jack lattice and focus on the fully AF regime
($J_{AB} > 0$, $J_C > 0$ ). Because this lattice is tripartite and there is
a unique way of dividing all the sites into three disconnected sublattices,
the ground state of the $q = 3$ AF model is expected to be a traditional
three-sublattice AF ordering of the type shown in Fig.~\ref{Gs_q_3}(a), but
the fact that this is a state of zero entropy suggests the possibility of
more complex physics at finite temperatures.

\begin{figure}[t]
\includegraphics[width=7.5cm]{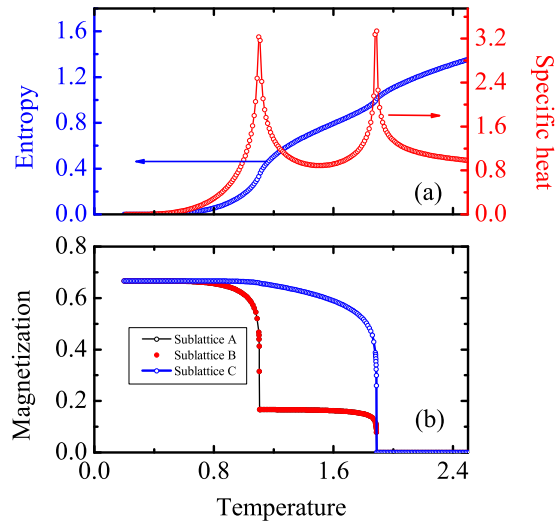}
\caption{(Color online) Thermodynamic properties of the $q = 3$ Potts model
on the Union-Jack lattice, calculated with coupling constants $J_{AB} = 1.0$
and $J_{C} = 5.0$. The results are obtained by iTEBD with $D = 30$. (a)
Entropy and specific heat. (b) Magnetizations of the A, B, and C sublattices.}
\label{ujq=3}
\end{figure}

To investigate this situation, we compute the entropy, specific heat, and
sublattice magnetizations for a range of values of the parameter ratio
$J_C/J_{AB}$. The definition of the local tensor is the same as in the
$q = 2$ case [Eq.~(\ref{T_Union_Jack})]. Figure \ref{ujq=3} shows our
results for $J_{AB} = 1$ and $J_C = 5$, where again two finite-temperature
phase transitions are clearly visible in the specific heat
[Fig.~\ref{ujq=3}(a)], with critical temperatures $T_{c1} = 1.10(1)$ and
$T_{c2} = 1.89(1)$. In the absence of an exact solution for this model, we
deduce the nature of the phases at low and intermediate temperatures by
calculating the sublattice magnetizations for the same parameters, as shown
in Fig.~\ref{ujq=3}(b).

At low temperatures, the system adopts the three-sublattice AF configuration
as expected. However, when the temperature exceeds $T_{c1}$, only the
C-sublattice order is preserved but the A- and B-sublattice order is
destroyed by thermal fluctuations. Sites in these two sublattices do not
become completely random but remain ``polarized'' by their strong AF
interaction with the C sites; thus if the C sublattice has $\sigma_i = 0$,
the A and B sites have a random choice between $\sigma_i = 1$ and 2, a
result reflected in their finite magnetization of approximately $1/6 = 1/2
 - 1/3$ in the regime $T_{c1} < T < T_{c2}$ [Fig.~\ref{Gs_q_3}(b)]. At
temperatures $T > T_{c2}$, entropic demands destroy the C-sublattice order
as well, driving the system into the fully disordered phase.

\begin{figure}[t]
\includegraphics[width=7.5cm]{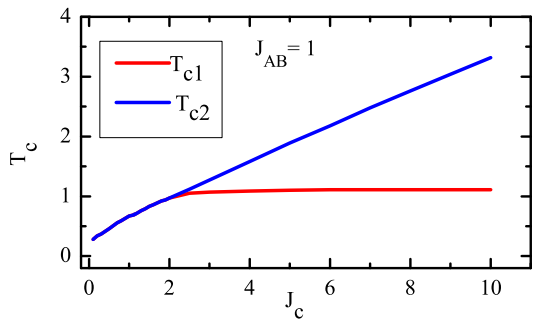}
\caption{(Color online) Phase diagram of the $q = 3$ Potts model on the
Union-Jack lattice. The transition from full to partial order is shown in
red (light grey) and from partial order to disorder in blue (dark grey).
The two transitions merge for $J_C < 2.2$. $J_{AB}$ is fixed to $1$.}
\label{phase_uni_q_3}
\end{figure}

For a complete understanding of these phenomena, in Fig.~\ref{phase_uni_q_3}
we show the phase diagram of the full parameter space. In the regime of large
$J_C$, there are always two phase transitions, of which the lower one ($T_{c1}$)
approaches the value $1.13$ as $J_C \rightarrow \infty$. $T_c/J = 1.13$
is the transition temperature of the Ising model on the square lattice
\cite{square_exact}, and this is the model for the behavior of the
AB-sublattice system when dominant $J_C$ bonds enforce for example $\sigma_i
 = 0$ on all C sites, leaving a $q = 2$ Potts degree of freedom on the A and
B sites. This is fully consistent with the discussion above for the nature of
the first transition, it means that $T_{c1}$ depends only on the coupling
between A and B sites ($J_{AB}$), and it means that the lower transition
is in the universality class of the Ising model.

The upper transition is the loss of C-sublattice order and therefore
$T_{c2}$ scales linearly with $J_C$ when this becomes large. A linear fit
using the data from $J_C = 3$ to $J_C = 10 $ gives the form $T_{c2} = 0.41(2)
J_{C} + 0.292(3)$. This transition has the universality of the ferromagnetic
Ising model. As $J_C/J_{AB}$ becomes smaller, the behavior becomes more
complex and the two transitions merge to a single one at $J_{C} \simeq
2.2$. One may anticipate that for very small values of $J_C$, a further
type of intermediate phase could appear in which the A and B sites retain
AF order but the C sites become random for entropic reasons. However, in
our calculations we find only that $T_c \rightarrow 0$ as $J_C \rightarrow
0$, consistent with the fact that, when $J_C = 0$, one obtains the $q = 3$
Potts model on the square lattice, which is known to be critical at zero
temperature \cite{Salas1998_JSP92-729}.

\subsection{Centered Diced Lattice with $q = 3$}

\begin{figure}[t]
\includegraphics[width=7.5cm]{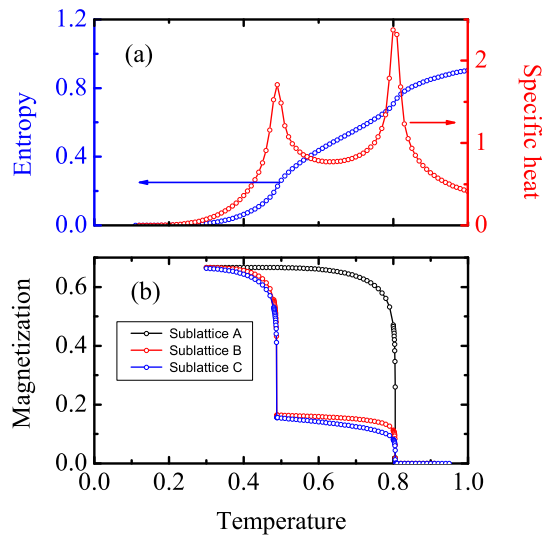}
\caption{(Color online) Thermodynamic properties of the $q = 3$ Potts model
on the isotropic centered diced lattice.  (a) Entropy and specific heat
obtained by iTEBD with $D = 40$. (b) Magnetizations of the A, B, and C
sublattices obtained by iTEBD with $D = 30$.}
\label{cent_q3}
\end{figure}

We complete our analysis of intermediate-temperature partial order by
considering the centered diced lattice with $q = 3$. In common with the
Union-Jack lattice, this geometry is tripartite with only one way of
dividing all the sites into three disconnected sublattices and again
one expects that the ground state should display three-sublattice AF
order for $q = 3$. Restricting our considerations to the isotropic AF
model, meaning $J_{AB} = J_{AC} = J_{BC} = 1$, we calculate the entropy,
specific heat, and sublattice magnetization using the iTEBD method.
From the specific-heat curve in Fig.~\ref{cent_q3}(a), we again find
two finite-temperature phase transitions with $T_{c1} = 0.48(1)$ and
$T_{c2} = 0.79(1)$.

To determine the nature of the intermediate phase in this case, we compute the
sublattice magnetizations shown in Fig.~\ref{cent_q3}. At low temperatures,
the results for all three sublattices converge to their ideal value of $2/3
 = 1 - 1/3$, but in the intervening phase between $T_{c1}$ and $T_{c2}$,
the sites on the B- and C-sublattices are randomized not among all three
Potts states but among only two, giving the value $1/6 = 1/2 - 1/3$.
On the isotropic centered diced lattice, the energy-minimization problem
of removing the order in two of the three sublattices is a subtle one.
As noted in Sec.~IVC and shown in Fig.~\ref{cd-lattice}, the coordination
numbers of sites in the three sublattices are $z_A = 12$, $z_B = 6$, and
$z_C = 4$ but the site numbers are $N_A = N/6$, $N_B = N/3$, and $N_C = N/2$,
from which it is easy to deduce that the bond numbers are $N_{AB} = N_{AC}
 = N_{BC} = N$. Thus for any other coupling ratios (one may imagine a wealth
of different cases depending on the values of $J_{AB}$, $J_{AC}$, and $J_{BC}$),
the bond energy would decide on which sublattice the partial order is
retained. However, in the isotropic case the selection is entropic only,
and thus, as in Sec.~IVC, the partial order remains on the A sublattice,
maximizing the entropic contribution from partial disorder (two of the
three Potts states) on 5/6 of the lattice sites.

We summarize this section by stating that we have explored a number
of models in which partial order emerges as a phase intermediate between
a low-temperature phase of complete order and a high-temperature phase
of complete disorder. In principle one may also expect that this is not a
requirement, in that a sufficiently complex model may contain more than
one type of partially ordered phase, including a partially ordered ground
state, and in this case it would be possible to investigate further types
of sequential phase transition to states of different intermediate partial
order. However, these phases do not emerge from within the confines of the
geometries (Archimedean and Laves lattices only) and coupling constants
(mostly isotropic) we consider. The emergence of intermediate partial order
is neither a consequence of frustration (Sec.~VIA only) nor of anisotropic
couplings (see Sec.~VIC), the difference between the $q = 3$ Union-Jack and
centered diced lattices being a result of their connectivity. Quite generally,
a Potts model possesses a number of symmetries, which depend on both the
geometry of the lattice and on the Potts degeneracy $q$, and these may be
broken partially and sequentially at the different transitions from full
to partial to no order.

\section{Partial Order with Sub-Extensive Residual Entropy}

We conclude our investigation of the different types of partial order in
AF Potts models by considering the form of the configurational entropy.
In all of the models studied in the preceding sections, the partially ordered
ground and intermediate states always possess an extensive degeneracy and
thus a non-zero entropy per site. In this section, we discuss the issue of
partial order in the ground states of systems whose entropy is sub-extensive,
such that the residual (zero-temperature) entropy per site is $0$.

\subsection{Generalized Decorated Square Lattice}

\begin{figure}[t]
\includegraphics[width=8.1cm]{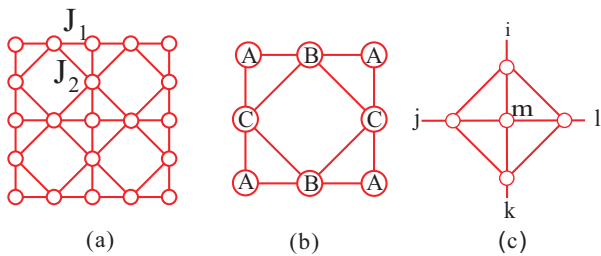}
\caption{(Color online) (a) Generalized decorated square lattice. (b) For the sublattice
partition illustrated, sites in sublattices B and C have coordination numbers
$z_B = z_C = 6$ and those in sublattice A have $z_A = 4$. (c) Definition of
tensors in each unit cell of the lattice.}
\label{cbl}
\end{figure}

We introduced the concept of partial order arising from sub-extensive
residual entropy in a study \cite{gdsl} of the AF Potts model on the
generalized decorated square lattice, particularly the $q = 3$ case,
and summarize the results in this subsection. The Union-Jack lattice
is obtained from the square lattice by adding one site at the center
of each square; if centering sites are placed only on alternate squares,
or equivalently the Union-Jack lattice is alternately diluted, one obtains
the generalized decorated square lattice shown in Fig.~\ref{cbl}(a), with two
different couplings $J_1$ and $J_2$. Like the Union-Jack lattice, this
lattice is tripartite (three-colorable in graph theory), but because
the lattice is formed of both triangular and square polygons, the division
into three sublattices is not unique.

\begin{figure}[t]
\includegraphics[width=7.5cm]{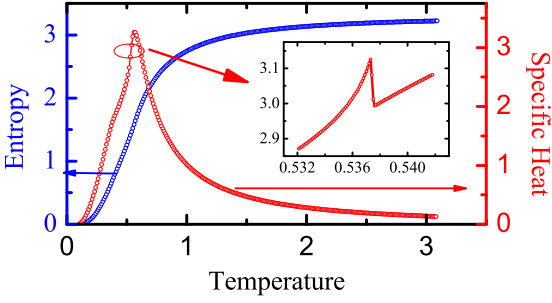}
\caption{(Color online) Specific heat for the $q = 3 $ AF Potts model on the
generalized decorated square lattice with $J_1 = J_2 = 1$. The results are
obtained by iTEBD with $D = 30$. The inset shows the very narrow discontinuity
appearing at $T_c = 0.5373(1)$.}
\label{cbq=3}
\end{figure}

The local tensor for the generalized decorated square lattice is given by
\begin{eqnarray}
T_{ijkl} = & & \sum \limits_{m} \! \exp [ - \beta J_{1} (\delta_{\sigma_i,\sigma_m}
\! \! + \! \delta_{\sigma_j,\sigma_m} \! \! + \! \delta_{\sigma_k,\sigma_m} \! \! + \!
\delta_{\sigma_l,\sigma_m} ) ] \nonumber \\
& & \; \times \; \exp [ - \beta J_{2} ( \delta_{\sigma_i,\sigma_j} \! \! + \!
\delta_{\sigma_j,\sigma_k} \! \! + \! \delta_{\sigma_k,\sigma_l} \! \! + \!
\delta_{\sigma_l,\sigma_i} ) ]
\label{T}
\end{eqnarray}
and represented in Fig.~\ref{cbl}(c). The Potts variables $\sigma_i$ on
each site serve as the indices of the tensors, and the tensors for each
unit cell may be combined to form a square-lattice tensor network. Because
the partition of the lattice into three sublattices is not unique, the
ground state of the AF $q = 3$ Potts model is not expected to be as
straightforward as the (zero-entropy) three-sublattice order of the same
model on the triangular, Union-Jack, or centered diced lattices. Indeed,
the specific heat shown in Fig.~\ref{cbq=3} for the case $J_1 = J_2 = 1$
appears to lack any evidence of a phase transition. However, on careful
inspection it reveals not a divergence but a small discontinuity
qualitatively similar to that of the $q = 4$ Union-Jack lattice (Sec.~IVB),
which marks a phase transition to partial order at $T_c = 0.5373(1)$
(detected more clearly but less accurately through the magnetization
and the susceptibility shown in Ref.~\cite{gdsl}).

To understand the origin of this behavior, we considered \cite{gdsl} the
configurations minimizing the bond energy that make up the ground manifold.
Sites in a single sublattice are expected to order ferromagnetically
because they are separated by pairs of AF bonds. If the B sublattice
in Fig.~\ref{cbl}(b) is ordered with $\sigma_i = 0$, the state on the
intervening lines of A and C sites may be either $121212 \dots 12$ or
$212121 \dots 21$, each with probability 50\%. An analogous state exists
for order only on the C sublattice, but the two are mutually exclusive;
both ordered states break the $\pi/2$ rotation symmetry of the lattice (also
broken on the Union-Jack lattice for $q = 4$, where a similar competition
between ordered states causes the weak transition in $C(T)$). It is the
linear structures of alternating order that hold the key to the properties
of the system. If one calculates the degeneracy of the ground manifold for
a system of size $L \times L$, it is ${\cal S} = 6(2^L - 1)$, a quantity
exponential only in the linear size of the system and not in its volume.
Thus the residual entropy per site is
\begin{equation}
S_{0} = \lim_{L \rightarrow \infty } \frac{\log (6\times 2^{L} - 6)}{L^{2}} = 0,
\end{equation}
vanishing due to the one-dimensional nature of the ground-state degrees
of freedom. However, the selection of a partially ordered ground state
within this model, proceeding in the same way as the extensively degenerate
examples studied in Sec.~IV, indicates that a large but sub-extensive
degeneracy of Potts configurations is sufficient to drive this phenomenon.

\subsection{Dilute Centered Diced Lattices}

We continue by demonstrating that partial order with sub-extensive entropy
can occur more generally than in a single lattice. In the same way that
partial dilution of the centering sites of the Union-Jack lattice leads
to the generalized decorated square lattice, dilution of the centering
sites of the centered diced lattice leads to a further class of irregular
lattices. The centering sites of the centered diced lattice form a kagome
lattice (yellow sites in Fig.~\ref{cd-lattice}), which is a tripartite
geometry offering many ways to divide all the sites into three disconnected
sublattices; the two most common are known as the $k = 0$ and $\sqrt{3}
\times \sqrt{3}$ structures \cite{rc}.

\begin{figure}[t]
\includegraphics[width=4.0cm]{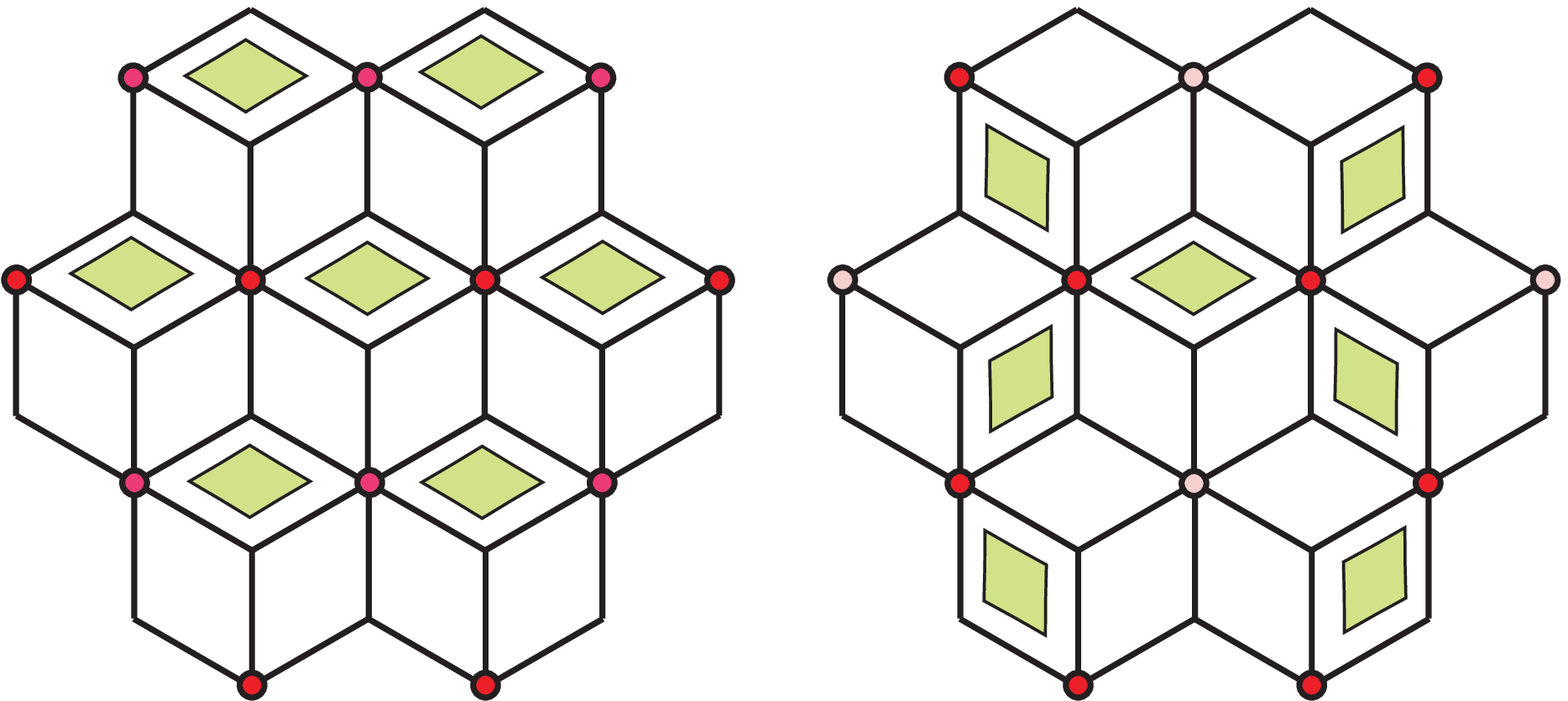} \hspace{0.1cm}
\includegraphics[width=4.0cm]{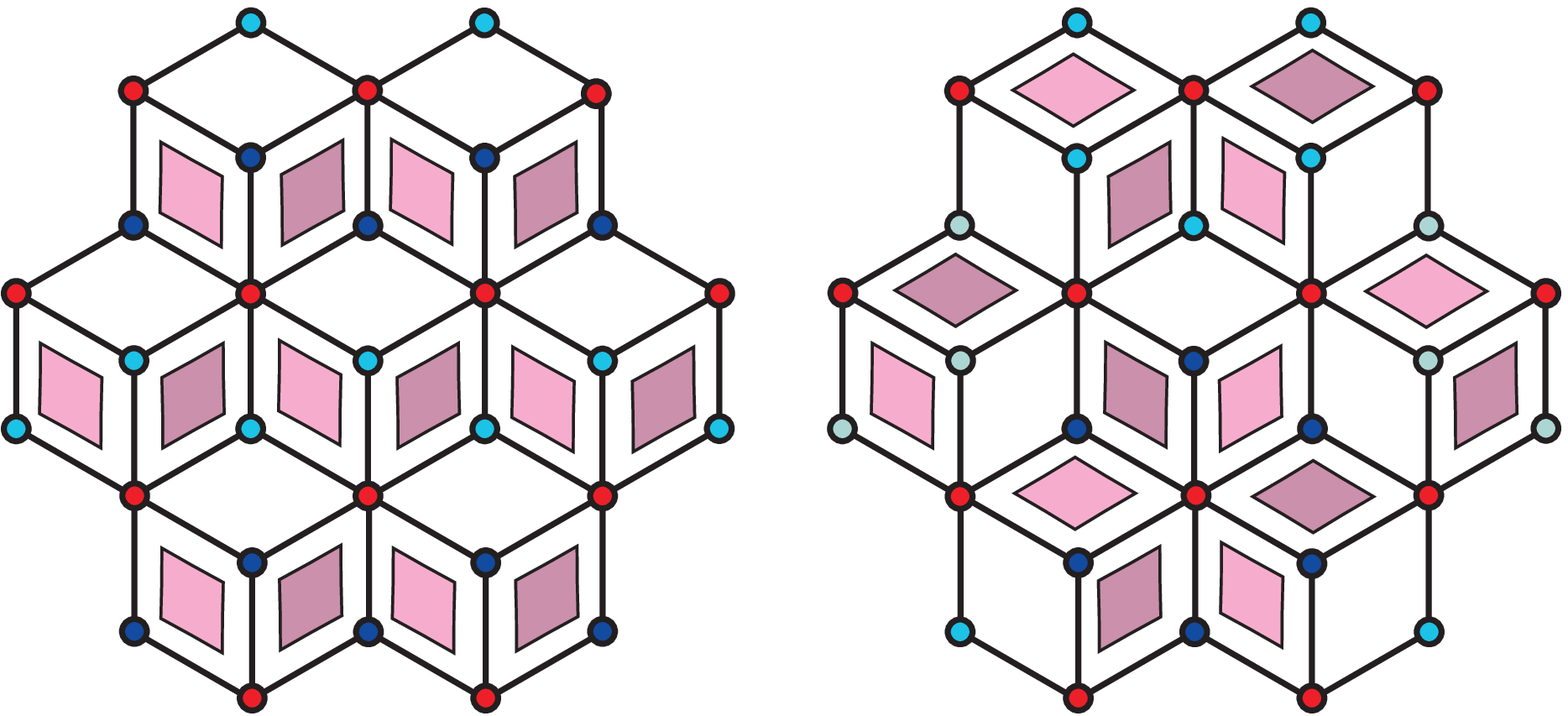}
{\centerline {IA \qquad\qquad\qquad IB \qquad\qquad\qquad IIA
\qquad\qquad\qquad IIB}}
\caption{(Color online) Dilute centered diced lattices. The quadrilaterals
containing centering sites are marked in color. In lattices IA and IB, $1/3$
of the quadrilaterals have center sites, while lattices IIA and IIB have
center sites on $2/3$ of the quadrilaterals. Lattices IA and IIA correspond
to $k = 0$ arrangements on the kagome lattice formed by all of the centering
sites in the centered diced lattice, whereas lattices IB and IIB are
$\sqrt{3} \times \sqrt{3}$ arrangements.}
\label{pc-diced-lattice}
\end{figure}

We consider only commensurate dilutions yielding small unit cells, which
leaves two choices of dilution, namely 1/3 and 2/3. If $2/3$ of the centering
sites are removed, such that only one sublattice of the kagome lattice has a
Potts variable and the other two sublattices are empty, we obtain the lattices
shown in Fig.~\ref{pc-diced-lattice} as IA and IB. If 1/3 of the centering
sites are removed, leaving a regular 2/3 filling, we obtain the lattices
IIA and IIB. We note that lattices IB and IIB contain the additional
complexity of inequivalent A sites (specifically, in lattice IB these
have coordinations 6 and 9, while in IIB they have 9 and 12) and we do not
consider these geometries further; this is equivalent to considering only
the $k = 0$ structures (IA and IIA). If the sublattices A and B of the diced
lattice are labeled as in Sec.~IVC, and the remaining centering sites form
a partial sublattice C, then lattice IA has site numbers $N_A = N/4$, $N_B =
N/2$, and $N_C = N/4$, with respective coordination numbers $z_A = 8$ and
$z_B = z_C = 4$, while lattice IIA has $N_A = N/5$ and $N_B = N_C = 2N/5$
with $z_A = 10$, $z_B = 5$, and $z_C = 4$.

\subsubsection{IIA Lattice}

We begin by considering the IIA lattice with $q = 3$. For every centered
rhombus, every bond can be satisfied if both pairs of diagonal sites have
the same $\sigma_i$. Thus all A sites are mutually ferromagnetic, by
convention with $\sigma_A = 0$, and all B sites between two horizontal
lines of A sites (Fig.~\ref{pc-diced-lattice}, IIA) should also have the
same $\sigma_B \ne 0$, but the values of $\sigma_B$ on the different
(horizontal, zig-zag) lines of B sites are independent, i.e.~$\sigma_B
 = 1$ or 2 at random. The value of $\sigma_C$ is fixed for all C sites once
$\sigma_A$ and the lines of $\sigma_B$ values are known. Thus the ground-state
degeneracy is $2^{L}$, where $L = \sqrt{N_A}$ is the number of lines of A sites.
As a consequence, the residual entropy per site in the thermodynamic limit is
$0$ for the same reason, linear structure formation, as in the $q = 3$ model
on the generalized decorated square lattice (Sec.~VIIA).

\begin{figure}[t]
\includegraphics[width=7.5cm]{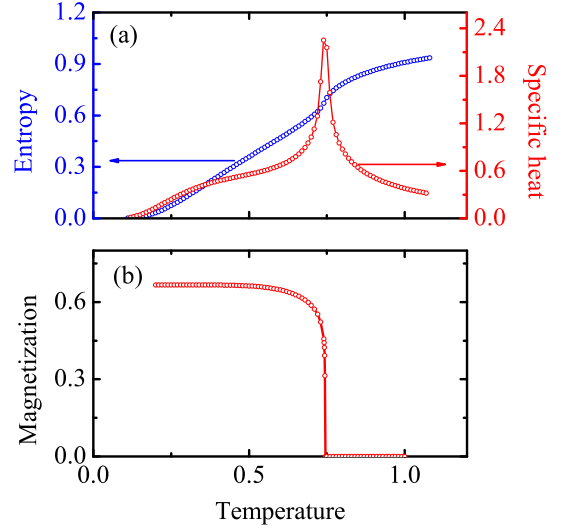}
\caption{(Color online) Thermodynamic properties of the $q = 3$ Potts model
on the IIA lattice of Fig.~\ref{pc-diced-lattice}. The results are obtained
by iTEBD with $D = 40$. (a) Entropy and specific heat. (b) Magnetization of
the A sublattice.}
\label{23_q3}
\end{figure}

Figure \ref{23_q3} shows the entropy, specific heat, and magnetization of
sublattice A for the IIA dilute centered diced lattice. The zero-temperature
entropy per site is $0$, in accord with the analysis above. The peak in the
specific heat indicates that a finite-temperature phase transition occurs at
$T_c = 0.74(1)$. The magnetization of the A sublattice is $2/3 = 1 - 1/3$,
corresponding to perfect order, while the analogous quantities for the B and
C sublattices (not shown) exhibit perfect disorder. Thus the IIA lattice
provides another example of partial order in the ground state selected by
a sub-extensive residual entropy.

\begin{figure}[t]
\includegraphics[width=7.5cm]{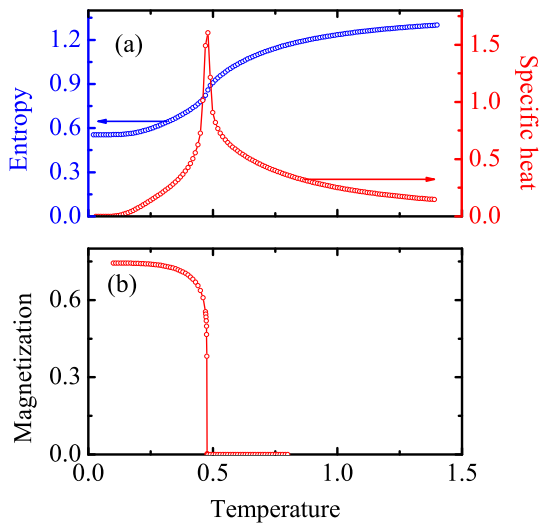}
\caption{(Color online) Thermodynamic properties of the $q = 4$ Potts model
on the IIA lattice of Fig.~\ref{pc-diced-lattice}. The results are obtained
by iTEBD with $D = 40$. (a) Entropy and specific heat. (b) Magnetization of
the A sublattice.}
\label{23_q4}
\end{figure}

By contrast, despite the linear structure of the lattice, the $q = 4$ model
in the same geometry shows very different behavior. In Ref.~\cite{gdsl} we
demonstrated that the AF $q = 4$ Potts model on the generalized decorated
square lattice is critical at zero temperature, its susceptibility approaching
a divergence as $T \rightarrow 0$. However, on the IIA lattice with $q = 4$ we
find a ``conventional'' finite-temperature phase transition at $T_c = 0.48(1)$,
appearing as a clear peak in the specific heat in Fig.~\ref{23_q4}(a) and with
a large residual entropy per site, $S_{{\rm IIA}, q = 4} (0) = 0.5556$.
Straightforward counting arguments for perfect partial order only on the
A, B, or C sublattices reveal respective ground-state degeneracies $g_A
 = 2^{4N/5} (3/2)^{L}$, $g_B = \exp (3 N S_{{\rm ds}, q=3}(0)/5)]$, where
$S_{{\rm ds}, q=3}(0) = 0.561070$ is the residual entropy for the $q = 3$
decorated square lattice (Sec.~IVB), and $g_C = 1.606^{3N/5}$, where the
numerical properties are given by the $q = 3$ diced lattice of the A and
B sites. Clearly a partial order on the A sublattice remains the most
favorable, and indeed our computed entropy is very close to the value
$S_A = (4/5) \ln 2 = 0.5545$ obtained in this case. The zero-temperature
A-sublattice magnetization is $M_{{\rm IIA}, q = 4}^A = 0.7438$
[Fig.~\ref{23_q4}(b)], a result within 0.8\% of the the ideal value
$0.75 = 1 - 1/4$. The discrepancies of both entropy and magnetization
from their ideal values indicate the presence of non-negligible
contributions from different ground-state configurations, but no
changes to the qualitative physics of Sec.~IV.

\subsubsection{IA Lattice}

For completeness, we conclude by considering the IA dilute centered diced
lattice of Fig.~\ref{pc-diced-lattice}. Because this geometry also consists
of linear structures in two dimensions, it is not unreasonable to expect a
further example of partial order with sub-extensive residual entropy. As
noted above, every centered rhombus in an AF Potts model with $q = 3$ has
the same $\sigma_i$ for diagonal pairs of sites, and thus all A sites in
the same horizontal line have the same state, but the A sites on adjacent
lines may take the same or different states $\sigma_i$ with complete
energetic degeneracy. For fully ferromagnetic A sites, for example with
$\sigma_i = 0$, pairs of B sites in every centered rhombus retain two
independent choices, $\sigma_i = 1$ or 2, and the degeneracy is $2^{N/4}$
($N/4$ is the number of centered rhombi). However, if there exists a row
of A sites with $\sigma_i = 1$, then the B sites in this row and both its
neighboring rows become fixed to $\sigma_i = 2$ and the degeneracy falls
to $2^{N/4-3L}$, where $L = \sqrt{N/4}$ is the number of centered rhombi
in a row. Thus configurations in which all A sites have the same value
are dominant in the ground state but one expects a finite residual entropy
of $S_{{\rm IA}, q=3}^A (0) = (1/4) \ln 2 = 0.173286795139986$.

\begin{figure}[t]
\includegraphics[width=7.5cm]{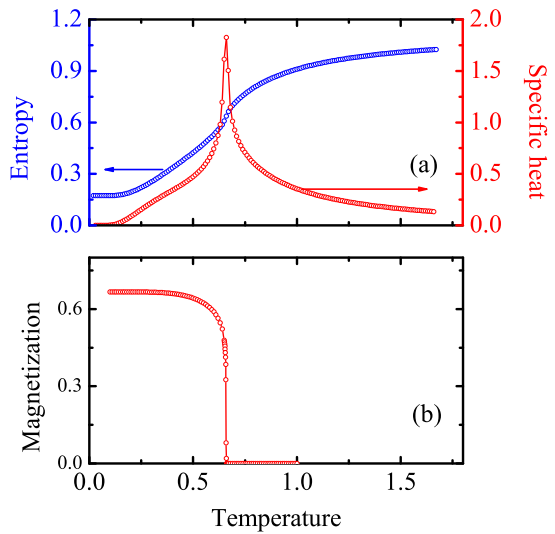}
\caption{(Color online) Thermodynamic properties of the $q = 3$ Potts model
on the IA lattice of Fig.~\ref{pc-diced-lattice}. The results are obtained
by iTEBD with $D = 40$. (a) Entropy and specific heat. (b) Magnetization of
the A sublattice.}
\label{13_q3}
\end{figure}

In Fig.~\ref{13_q3}, we present the entropy, specific heat, and the
A-sublattice magnetization of the AF $q = 3$ Potts model on the IA lattice.
Our numerical result for the zero-temperature entropy, $S_{{\rm IA}, q=3} (0)  =
0.17328679513999$ confirms completely the analytical reasoning. Both the
specific heat and the sublattice magnetization confirm a conventional
entropy-driven phase transition at $T_c = 0.66(1)$ and the magnetization
of the A sublattice in the low-temperature limit is the expected
$2/3 = 1 - 1/3$. We conclude that linear structures may be a necessary
but not a sufficient condition for partial order from sub-extensive
residual entropies in two dimensions, but that the fundamental criterion
for partial order, namely the relationship between ${\bar z}$ and $q$,
remains the dominant determining factor.

We end with the $q = 4$ model on the IA lattice. In this case the
degeneracy is large and the connectivity smaller than the IIA lattice,
so one may even suspect a zero-temperature critical phase with no true
order \cite{gdsl}. Focusing directly on a candidate partially ordered state
with all A sites ferromagnetic ($\sigma_A = 0$), the Potts variables on the
B and C sites in each centered rhombus are independent. If the C sites
have $\sigma = 1$, the two B sites can be $(2,2)$, $(2,3)$, $(3,2)$, or
$(3,3)$, and so the centered rhombus has $ 3 \times 4 = 12$ configurations,
giving a total degeneracy $g_A = 12^{N/4}$. The analogous degeneracies
for full B- and C-sublattice order are $g_B = 2^{N/2}(3/2)^{L}$ and
$g_C = 1.606^{3N/4}$, confirming that if a partially ordered state
exists then it will be on the A sublattice.

\begin{figure}[t]
\includegraphics[width=7.5cm]{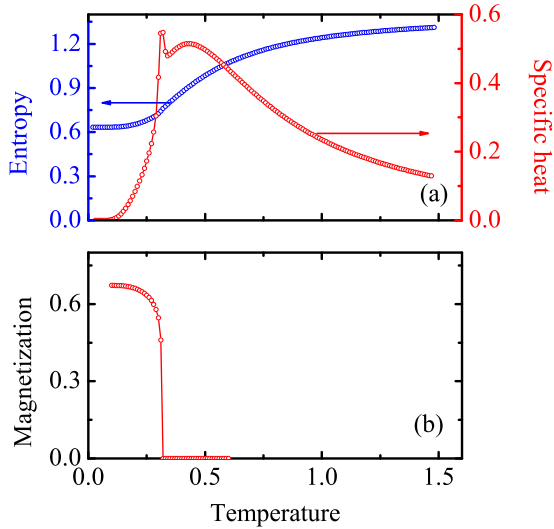}
\caption{(Color online) Thermodynamic properties of the $q = 4$ Potts model
on the IA lattice of Fig.~\ref{pc-diced-lattice}. The results are obtained by
iTEBD with $D = 40$. (a) Entropy and specific heat. (b) Magnetization of the A
sublattice.}
\label{13_q4}
\end{figure}

The entropy, specific heat, and A sublattice magnetization of the $q = 4$
AF Potts model on the IA lattice are shown in Fig.~\ref{13_q4}. A
finite-temperature phase transition does in fact occur, at $T_c = 0.32(1)$.
However, the associated discontinuity is not the dominant feature in the
specific heat, a situation more reminiscent of the $q = 4$ model on the
Union-Jack lattice (Sec.~IVB) and the $q = 3$ model on the generalized
decorated square lattice (Sec.~VIIA), and the transition temperature is
one of the lowest finite values we have encountered. The evidence for a
general lack of robustness to thermal fluctuations in this model is
reinforced by our results for the residual entropy, $S_{{\rm IA}, q=4} (0) =
0.63402$, which lies above the perfect-order value $(1/4) \ln 12 = 0.6212$
by an amount we have seen (Sec.~IV) to be significant. Similarly, although
our results confirm that the partial order is on the A sublattice, the
zero-temperature magnetization $M_{{\rm IA}, q=4} (0) = 0.6725$ lies well below
the ideal value of 0.75.

\section{Summary}

We have performed a detailed analysis of the antiferromagnetic Potts model
in two dimensions, covering a range of lattice geometries and numbers $q$
of degrees of freedom per site. The primary focus of our investigation is
the phenomenon of partial long-range order, which arises in the presence of
high state degeneracies. Quite generally, this partial order sets in at a
``finite-temperature phase transition,'' which is the defining property of
the Potts model in question, but whose properties can differ widely as a
consequence of the interplay between lattice geometry and $q$ value.

An essential ingredient of the partial ordering scenario is the nature
of the lattice. In the absence of frustration, ordering transitions occur
when the site coordination number $z$ constrains the number of degrees of
freedom $q$. For sufficiently large $q$, the AF Potts model on any lattice
is insufficiently constrained and is disordered at all temperatures. In
restricting our considerations to Archimedean and Laves lattices, one of
the key qualitative properties of a lattice is whether it is regular (all
sites equivalent) or irregular, meaning that it has different types of
site with different local coordination numbers. While both types of system
may possess a large number of Potts configurations minimizing the total
energy, the irregular lattices have nontrivial entropies, which in a
number of cases drive only a partial ordering transition on some of the
inequivalent sublattices.

The majority of our results are obtained for three particular Laves lattices,
the diced, Union-Jack, and centered diced lattices, which have integer average
coordination number ${\bar z}$ but behavior rather different from regular
lattices with $z = {\bar z}$. Specifically, for the value $q = q_c$ where
the regular lattice is critical, they all show finite-temperature transitions
to states of partial order. Thus the irregular lattice geometry leads quite
generally to high values of $q_c$ (indeed, $q_c > 5$ for ${\bar z} = 6$
on the centered diced lattice). The entropic selection mechanism is such
that the partial order is always on the site of highest local coordination,
which creates a high number of satisfied bonds while imposing a $q-1$ Potts
degeneracy on all of the connected sites.

The finite-temperature transitions from partially ordered states to disordered
states may in different cases be very obvious or extremely subtle. We have
analyzed both types and shown that this is a function of the number of
competing partially ordered states; when there is no unique sublattice
with the largest connectivity, then the system must resolve this competition
and the result can be a very weak transition. Indeed, the existence of
inequivalent sites in a lattice cannot on its own guarantee a partially
ordered ground state, because sufficiently large $q$ will always cause
disorder, and so the existence of order must be tested in every case.
However, the effectiveness of thermal fluctuations in suppressing the
partial order parameter is determined not only by the number of degenerate
states in the manifold but also by the nature of the competition between
partially ordered states.

Another factor in this competition can be the nature of the entropic driving
forces. In most of the models we have studied, the entropy of the ground
manifold of minimum-energy states is extensive, scaling with the volume of
the system. However, we have also discovered some situations where the
balance of connectivity (${\bar z}$), $q$, and the interactions ($J$) is such
that the ground state has one-dimensional correlations and the degeneracy
scales only with the linear dimension of the system. Despite the resulting
sub-extensive entropy, the ground manifold remains highly degenerate, and
this is sufficient to preserve the physics of partial order.

We comment here that we have largely avoided considering systems with
frustrated interactions. Frustration is regarded as a very general driving
force for (complete) suppression of order parameters and in certain systems
for the existence of only partially ordered ground states. The family of
AF Potts models can largely be categorized into three regimes, one with
$q > {\bar z}$, which exhibits disorder at all temperatures, one with
(crudely) $q \sim {\bar z}$, which exhibits phenomena including
zero-temperature criticality and entropy-driven partial order, and
one with $q \ll {\bar z}$ where the ground state usually has complete
order. Frustration is the key factor affecting the nature and extent
of order in the last of these categories. This paper concerns almost
exclusively the intermediate category, where entropic effects dominate
and frustration is absent -- quite simply, a triangle is not a frustrated
unit when $q \ge 3$.

In addition to partially ordered ground states, we have also investigated
the formation of partially ordered states at intermediate temperatures.
In the models we consider, these occur in systems with conventional, fully
ordered ground states, which are also states of low (or zero) degeneracy.
As the temperature is increased, the huge entropic preference for partially
ordered states of high degeneracy can drive an additional phase transition.
The typical regime for this type of behavior is where $q$ is slightly smaller
than in the systems with partially ordered ground states. The consequence is
a system with multiple phase transitions, from order to partial order at low
temperatures and then from partial order to complete disorder, which we are
able to characterize completely by computing the magnetizations on every
sublattice. Potts models possess both the symmetry of their lattice and
the $q$-fold permutation symmetry of the Potts variable, and models showing
separate phase transitions give a very clear example of sequential breaking
or restoration of partial symmetries, which can be used to classify the
transition type (universality class)\cite{Wu1982_RMP54-235}.

One of the key features of our calculations is that they yield
quantitative thermodynamic information about Potts models, and in particular
about the partially ordered states at low temperatures. Highly accurate values
for the entropies and magnetizations can be compared with expectations for
different, competing, partially ordered model states, to deduce their
contributions to the true ground state. In this study, we have exposed
a number of models where the thermodynamic properties do not match with naive
expectations based on perfect order on the highest-coordinated sublattice.
For reasons of space, we have not dwelled on the development of scenarios
for improving the analytical description of Potts states, although our data
allow direct comparisons with models for the leading ``defect'' configurations
within a state of perfect partial order. When these defects are present at
finite densities due to their entropic contributions, their effects will be
observable in the entropy and sublattice magnetization we compute. Our results
also permit detailed comparisons with different models in classical statistical
mechanics, further developing cross-links within the field.

Finally, we have demonstrated the power of tensor-based numerical methods
for problems in classical statistical mechanics. The ability to express
the partition function as a tensor product, to renormalize systematically,
and to truncate in the tensor dimension, gives unprecedented access to
accurate thermodynamic information. The method is completely general in
that calculations can be performed for all lattice geometries and all
values of $q$, with no restrictions to special cases. We have exploited
this power to find and characterize previously unknown phase transitions,
to quantify thermodynamic properties both at the transition and at low
temperatures, and thus to gain extra insight into the physics of partial
ordering processes. We close by noting that the development of tensor-based
numerical techniques remains in its relative infancy, and that significant
improvements in size (tensor dimension) and accuracy may still be expected.
This would make possible a new level of quantitative discussion for topics
such as scaling exponents and universality at phase transitions, which are
currently still at the frontiers of our capabilities.

\begin{acknowledgments}
The first two authors contributed equally to this work.
We thank Y.-J. Deng, W.-A. Guo, M.-X. Liu, Z.-C. Wei, and F.-Y. Wu for
helpful discussions. This work was supported
by the National Science Foundation of China under Grants
No. 10934008, No. 10874215, and No. 11174365 and by the
National Basic Research Program of China under
Grants No. 2012CB921704 and No. 2011CB309703.

\end{acknowledgments}

\end{document}